\documentclass{aa}  

\usepackage{natbib}
\usepackage{graphicx}
\usepackage{txfonts}
\usepackage{hyperref}
\usepackage{color}
\usepackage{upquote} 
\usepackage{xspace}
\usepackage{ulem}  

\makeatletter
\renewcommand*\aa@pageof{, page \thepage{} of \pageref*{LastPage}}
\makeatother

\def\Gaia{\textit{Gaia}\xspace}
\def\gmag{\ensuremath{G}\xspace}
\def\gbp{\ensuremath{G_\mathrm{BP}}\xspace}
\def\grp{\ensuremath{G_\mathrm{RP}}\xspace}
\def\rangeG{\ensuremath{\mathrm{range}(\gmag)}\xspace}
\def\absMeanG{\ensuremath{\mathrm{M}_G\xspace}}

\def\NobsG{\ensuremath{N_\mathrm{G}}\xspace}
\def\NobsBP{\ensuremath{N_\mathrm{BP}}\xspace}
\def\NobsRP{\ensuremath{N_\mathrm{RP}}\xspace}

\def\NobsRPandBP{\ensuremath{N_\mathrm{RP \cap BP}}\xspace}
\def\BPminusRP{\ensuremath{G_\mathrm{BP} - G_\mathrm{RP}}\xspace}

\def\bandBP{{BP}\xspace}
\def\bandRP{{RP}\xspace}
\def\bandBPplusRP{BP+RP\xspace}
\def\fluxexcess{BP and RP flux excess\xspace}
\def\fluxexcesses{BP and RP flux excesses\xspace}
\def\epochFluxG{\ensuremath{f_\mathrm{G}}\xspace}
\def\epochFluxBP{\ensuremath{f_\mathrm{BP}}\xspace}
\def\epochFluxRP{\ensuremath{f_\mathrm{RP}}\xspace}
\def\fluxG{\ensuremath{I_G}\xspace}
\def\fluxBP{\ensuremath{I_{BP}}\xspace}
\def\fluxRP{\ensuremath{I_{RP}}\xspace}
\def\fluxErrorG{\ensuremath{\varepsilon(\fluxG)}\xspace}
\def\fluxErrorBP{\ensuremath{\varepsilon(\fluxBP)}\xspace}
\def\fluxErrorRP{\ensuremath{\varepsilon(\fluxRP)}\xspace}
\def\varProxyG{\ensuremath{A_\mathrm{proxy,G}}\xspace}
\def\varProxyBP{\ensuremath{A_\mathrm{proxy,BP}}\xspace}
\def\varProxyRP{\ensuremath{A_\mathrm{proxy,RP}}\xspace}
\def\varProxyBPplusRP{\ensuremath{A_\mathrm{proxy,BP+RP}}\xspace}
\def\varProxyBPplusRPnullCov{\ensuremath{A'_\mathrm{proxy,BP+RP}}\xspace}
\def\varProxyGsquare{\ensuremath{A^2_\mathrm{proxy,G}}\xspace}
\def\varProxyBPsquare{\ensuremath{A^2_\mathrm{proxy,BP}}\xspace}
\def\varProxyRPsquare{\ensuremath{A^2_\mathrm{proxy,RP}}\xspace}
\def\varProxyBPplusRPsquare{\ensuremath{A^2_\mathrm{proxy,BP+RP}}\xspace}
\def\varProxyCovBPplusRP{\ensuremath{C_\mathrm{proxy,Cov(BP,RP)}}\xspace}
\def\excessFactor{\ensuremath{C}\xspace}
\def\excessFactorFiducial{\ensuremath{C_\mathrm{fid}}\xspace}
\def\excessFactorNormalized{\ensuremath{C'}\xspace}
\def\excessFactorNormalizedLimit{\ensuremath{C'_\mathrm{lim}}\xspace}

\begin{document} 
  \title{Large-amplitude variables in \Gaia Data Release 2%
         \thanks{Table D.1 is only available in electronic form at the CDS via anonymous ftp to cdsarc.u-strasbg.fr (130.79.128.5) or via http://cdsweb.u-strasbg.fr/cgi-bin/qcat?J/A+A/}
        }
  \subtitle{Multi-band variability characterization}
  
  \author{N.~Mowlavi\inst{\ref{Sauverny},\ref{Ecogia}}\fnmsep\thanks{Corresponding author: N. Mowlavi
(\href{mailto:Nami.Mowlavi@unige.ch}{\tt Nami.Mowlavi@unige.ch})},
          L.~Rimoldini\inst{\ref{Ecogia}},
          D.~W.~Evans\inst{\ref{Cambridge}},
          M.~Riello\inst{\ref{Cambridge}},
          F.~De~Angeli\inst{\ref{Cambridge}},
          L.~Palaversa\inst{\ref{Zagreb},\ref{Cambridge}},
          M.~Audard\inst{\ref{Sauverny},\ref{Ecogia}},
          L.~Eyer\inst{\ref{Sauverny},\ref{Ecogia}},
          P.~Garcia-Lario\inst{\ref{ESAC}},
          P.~Gavras\inst{\ref{ESAC}},
          B.~Holl\inst{\ref{Sauverny},\ref{Ecogia}},
          G.~Jevardat de Fombelle\inst{\ref{Ecogia},\ref{SixSq}},
          I.~Lec\oe ur-Ta\"ibi\inst{\ref{Ecogia}},
          K.~Nienartowicz\inst{\ref{Ecogia},\ref{Sednai}}
         }
         
  \authorrunning{Mowlavi et al.}
  
  \institute{Department of Astronomy, University of Geneva, Chemin Pegasi 51, 1290 Versoix, Switzerland \label{Sauverny}
             \and
             Department of Astronomy, University of Geneva, Chemin d'Ecogia 16, 1290 Versoix, Switzerland\label{Ecogia}
             \and
             Institute of Astronomy, University of Cambridge, Madingley Road,
Cambridge CB3 0HA, UK \label{Cambridge}
             \and
             Ru{\dj}er Bo\v{s}kovi\'{c} Institute, Bijeni\v{c}ka cesta 54, 10000
Zagreb, Croatia \label{Zagreb}
             \and
             European Space Astronomy Centre (ESA/ESAC), Villanueva de la Canada, 28692 Madrid, Spain \label{ESAC}
             \and
             SixSq, Route de Meyrin 267, 1217 Meyrin, Switzerland \label{SixSq}
             \and
             Sednai Sarl, 1204 Geneva, Switzerland \label{Sednai}
            }

  \date{Received 16 September 2020 / Accepted 20 December 2020}

  \abstract
   {Photometric variability is an essential feature that sheds light on the intrinsic properties of celestial variable sources, the more so when photometry is available in various bands.
    In this respect, the all-sky \Gaia mission is particularly attractive as it collects, among other quantities, epoch photometry measured quasi-simultaneously in three optical bands for sources ranging from a few magnitudes to fainter than magnitude 20.
   }
   {The second data release (DR2) of the mission provides mean \gmag, \gbp and \grp photometry for $\sim$1.4 billion sources, but light curves and variability properties are available for only $\sim$0.5 million of them.
    Here, we provide a census of large-amplitude variables (LAVs) with amplitudes larger than $\sim$0.2~mag in the \gmag band for objects with mean brightnesses between 5.5 and 19~mag.
   }
   {To achieve this, we rely on variability amplitude proxies in \gmag, \gbp and \grp computed from the uncertainties on the magnitudes published in DR2.
    We then apply successive filters to identify two subsets containing sources with reliable mean \gbp and \grp (for studies using colours) and sources having compatible amplitude proxies in \gmag, \gbp and \grp (for multi-band variability studies).
   }
   {The full catalogue gathers 23\,315\,874 LAV candidates, and the two subsets with increased levels of purity contain, respectively, 1\,148\,861 and 618\,966 sources.
    A multi-band variability analysis of the catalogue shows that different types of variable stars can be categorized according to their colours and blue-to-red amplitude ratios as determined from the \gmag, \gbp, and \grp amplitude proxies.
    More specifically, four groups are globally identified.
    They include:
    long-period variables in a first group with amplitudes more than twice larger in the blue than in the red;
    hot compact variables in a second group with amplitudes smaller in the blue than in the red;
    classical instability strip pulsators in a third group with amplitudes larger in the blue than in the red by 50\% to 80\%;
    and other non-pulsating variables in a fourth group, mainly achromatic, but 10\% of them still having 20\% to 50\% larger amplitudes in the blue than in the red.

   }
   {The catalogue constitutes the first census of \Gaia LAV candidates extracted from the public DR2 archive.
    The overview presented here illustrates the added value of the mission for multi-band variability studies, even at this stage when epoch photometry is not yet available for all sources.
   }


\maketitle

\section{Introduction}
\label{Sect:Introduction}

Since the end of the 20th century, the number of known variable stars has dramatically increased, boosted by the operation of large-scale surveys in the search for dark matter, such as the
MACHO \citep{AlcockAllsmanAlves97},
EROS \citep{Palanque-DelabrouilleAfonsoAlfert_etal98},
and OGLE \citep{UdalskiKubiakSzymanski97} surveys.
In the last few years, 
the Catalina survey reached $\sim$110\,000 variables \citep{DrakeDjorgovskiCatelan_etal17},
Pan-STARRS $\sim$240\,000 variables \citep{SesarHernitschekMitrovic_etal17},
ATLAS $\sim$430\,000 variables \citep{HeinzeTonnyDenneau_etal18},
\Gaia $\sim$500\,000 variables \citep{HollAudardNienartowicz_etal18},
ASAS-SN $\sim$220\,000 variables \citep{JayasingheStanekKochanek_etal20_V},
ZTF $\sim$600\,000 variables \citep{ChenWangDeng_etal20},
OGLE-IV $\sim$1\,000\,000 variables (\url{http://ogledb.astrouw.edu.pl/~ogle/OCVS/}),
and the American Association of Variable Star Observers (AAVSO) lists $\sim$1\,500\,000 variables as of June 2020 (\url{https://www.aavso.org}).

The \Gaia mission offers a unique opportunity in this field.
It provides astrometry, photometry, and spectro-photometry for stars all over the sky, in the wide brightness range of a few magnitudes to above 20~mag, as well as spectroscopy for the bright objects \citep{GaiaPrustiDeBruijne_etal16}.
And, for multi-band variability studies, the mission is unique because of the availability of quasi-simultaneous photometric measurements in three bands (\gmag, \gbp, and \grp within 50~sec, 100~sec if including the radial velocity spectrometer RVS).

In \Gaia data release 2 \citep[DR2, ][]{GaiaBrownVallenari_etal18}, variability amplitudes measured from epoch photometry are provided for a subset of $\sim$500\,000 variable stars of specific variability types.
This represents only a small fraction of the variables present in the public \Gaia archive.
For all other sources not included in these $\sim$500\,000 variables, and which hence do not have published photometric time series, their variability amplitude can still be estimated using the published photometric uncertainties.
This is due to the fact that these uncertainties are derived from the standard deviation of the light curves and hence include information about both measurement uncertainties and source variability.
We have taken advantage of this feature to build a multi-band variability catalogue of large-amplitude ($\gtrsim\!0.2$~mag) variables (LAVs) for all objects published in \Gaia DR2.

The variability amplitude proxies used to estimate the amplitudes in \gmag, \gbp and \grp are introduced in Sect.~\ref{Sect:varProxy}.
Our catalogue of \Gaia DR2 LAVs is then presented in Sect.~\ref{Sect:catalogue}, in which three datasets (Datasets~A, B, and C) are identified for different purposes.
The quality of the catalogue in terms of both completeness and purity is also addressed in that section.
Section~\ref{Sect:exploration} then illustrates the usage of the catalogue with two examples.
The first demonstrates an application of the mutli-band variability amplitudes to identify different categories of variable stars, while the second presents the sample of LAVs with good parallaxes.
The main body of the text ends with a summary and concluding remarks in Sect.~\ref{Sect:Conclusions}.

Additional material is presented in several appendices.
The extraction of LAVs from the \Gaia archive and the removal of outliers is detailed in Appendix~\ref{Appendix:DatasetA}.
The sum of the fluxes in the blue (BP) and red (RP) spectrophotometers is compared to the flux in the main \gmag band in Appendix~\ref{Appendix:BPRPexcess}, knowing that the summed transmission curve of the two spectrophotometers is close to the transmission curve of \gmag.
An amplitude proxy for \bandBPplusRP and its relation with the individual amplitude proxies for \gbp and $\grp$ are derived in Appendix~\ref{Appendix:AproxyBPplusRP}.
Finally, the electronic table of our \Gaia DR2 LAV catalogue is described in Appendix~\ref{Appendix:electronicTable}.

The notations used in this paper regarding \Gaia fluxes and magnitudes comply with the notations adopted in \citet{EvansRielloDeAngeli_etal18} \citep[see also][Sect.~5.3.5]{Busso_etal18_DR2docChapt5}:
\epochFluxG represents the epoch flux of one CCD photometric measurement in the astrometric focal plane, and \epochFluxBP and \epochFluxRP represent the wavelenth-integrated epoch flux during one transit in the blue and red spectrophotometric focal planes, respectively;
\fluxG, \fluxBP, and \fluxRP represent the (inverse-variance weighted) mean fluxes in the respective photometric bands of a given source over the 22 months of data gathered in DR2.
Finally \gmag, \gbp, and \grp are the mean magnitudes derived from \fluxG, \fluxBP, and \fluxRP, respectively.
Epoch magnitudes per se are not used in this paper, but when we mention it, we notated it $\gmag(t)$.

\section{The variability amplitude proxy}
\label{Sect:varProxy}

\subsection{Definitions}
\label{Sect:varProxy_definitions}

For constant stars, the uncertainty $\varepsilon(I)$ on the weighted mean flux $I$ can be estimated from the variance $\sigma_f^2$ of the $N$ flux measurements $f$ using $\varepsilon^2(I) = \sigma_f^2 \left(\sum_i^N w_i^2\right) / \left(\sum_i^N w_i\right)^2$, where $w_i$ denotes the weight associated with the $i$th measurement%
\footnote{
The expected variance of a weighted mean $\bar{x}$ is $\sigma^2_{\bar{x}}=\sigma^2_x\, V_2/V_1^2$, where $\sigma_x^2$ is the true variance of measurements $x_i$, $V_1=\sum w_i$ and $V_2=\sum w_i^2$, with $w_i$ denoting the weights associated with measurements $x_i$ \citep[see Eq.~A.31 in][]{Rimoldini14}.
When $\sigma^2_{\bar{x}}$ is estimated by this expression, the true variance $\sigma_x^2$ is unknown but it can be represented by the sample-size unbiased weighted variance $S^2_{x}=[V_1^2 / (V_1^2 - V_2)] \, s^2_{x}$
, where $s^2_{x}=\sum_i w_i(x_i-\bar{x})^2/V_1$ is the biased weighted variance \citep[Eq.~A.140 in Appendix~A of][]{Rimoldini14}.
It follows that $\sigma^2_{\bar{x}}\approx [V_2 / (V_1^2 - V_2)] \,s^2_{x}$, or $s^2_{x}/(N-1)$ for $N$ observations in the unweighted limit.
In our case, however, $\sigma^2_{\bar{x}}$ is published, so we can estimate the true variance $\sigma^2_x$ of the measurements from $\sigma^2_{\bar{x}}\, V_1^2 / V_2$, or $\sigma^2_{\bar{x}} N$ in the unweighted scenario.
}%
.
Since flux weights are not published in \Gaia DR2, we estimate $\sigma_f$ assuming equally weighted measurements: $\sigma_f=\varepsilon(I) \sqrt{N}$ (this form might overestimate $\sigma_f$ as the effective number of measurements is less than $N$ if weights are unequal).
To obtain a quantity independent of the flux (which depends on various parameters such as the integration time), it is convenient to express $\sigma_f$ relative to the mean flux, $\sigma_f/I$.
This ratio is also proportional to the standard deviation $\sigma_m$ in magnitude.
For $\sigma_f/I \ll 1$, we have
\begin{equation}
\sigma_m  \approx \frac{2.5}{\ln(10)}\, \frac{\sigma_f}{I} \approx 1.09\, \frac{\varepsilon(I)}{I}\, \sqrt{N} \;. 
\label{Eq:sigma_m}
\end{equation}
The value of $\sigma_m$ computed from $\sigma_f/I$ in this way may be underestimated for large $\varepsilon(I)/I$ variations because of the non-linear relation between flux and magnitude.
In practice, however, the approximation turns out to be sufficiently accurate for our purposes, even for the large variability amplitudes considered here (see Sect.~\ref{Sect:varProxy_range}).

In \Gaia DR2, the published mean flux uncertainty \fluxErrorG of a source, whether constant or variable, is computed from the standard deviation of its \epochFluxG flux curve.
Therefore, based on Eq.~\ref{Eq:sigma_m}, the quantity
\begin{equation}
  \varProxyG = \sqrt{\NobsG} \; \frac{\fluxErrorG}{\fluxG}
\label{Eq:AproxyG}
\end{equation}
can be used as a proxy for the scatter in \gmag light curves.
For constant stars, it approximates the standard deviation of \gmag light curves due to noise and uncalibrated systematic effects \citep[see Sect.~5.3.5 of the \Gaia DR2 documentation in][]{Busso_etal18_DR2docChapt5}, to a factor of 1.09 (from Eq.~\ref{Eq:sigma_m}).
For variable stars, the standard deviation is larger than it would be if the star was constant because of the additional contribution from stellar variability.
Therefore, the amplitude proxy reflects the variability amplitude of astrophysical origin if the latter dominates the variability recorded in the signal.

Equation~\ref{Eq:AproxyG} has already been applied to both DR1 and DR2 for the study of specific types of variable stars such as
Miras in the Magellanic Clouds \citep[MCs;][DR1]{DeasonBelokurovErkal_etal17},
RR Lyrae variables in the MCs \citep[][DR1]{BelokurovErkalDeason_etal17} and in the Galaxy \citep[][DR1]{IorioBelokurovErkal_etal18},
pre-main sequence (PMS) stars \citep[][DR2]{VioqueOudmaijerSchreiner_etal20},
white dwarfs \citep[][DR2]{EyerRimoldiniRohrbasser_etal20},
and cataclysmic variables \citep[CVs;][DR2]{AbrahamsBloomMowlavi_etal21}.

Similarly to Eq.~\ref{Eq:AproxyG}, we define amplitude proxies \varProxyBP and \varProxyRP for \gbp and \grp, respectively, using
\begin{equation}
  \varProxyBP = \sqrt{\NobsBP} \; \fluxErrorBP / \fluxBP \;, \label{Eq:AproxyBP}
\end{equation}
\vskip -6mm
\begin{equation}
  \varProxyRP = \sqrt{\NobsRP} \; \fluxErrorRP / \fluxRP \;, \label{Eq:AproxyRP}
\end{equation}
where \NobsBP and \NobsRP are the numbers of observations in \gbp and \grp, respectively, and \fluxErrorBP and \fluxErrorRP are the published uncertainties on \fluxBP and \fluxRP, respectively.

\subsection{Relation between amplitude proxy and range}
\label{Sect:varProxy_range}

\begin{figure*}
	\centering
	\includegraphics[trim={0 81 129 50},clip,height=97pt]{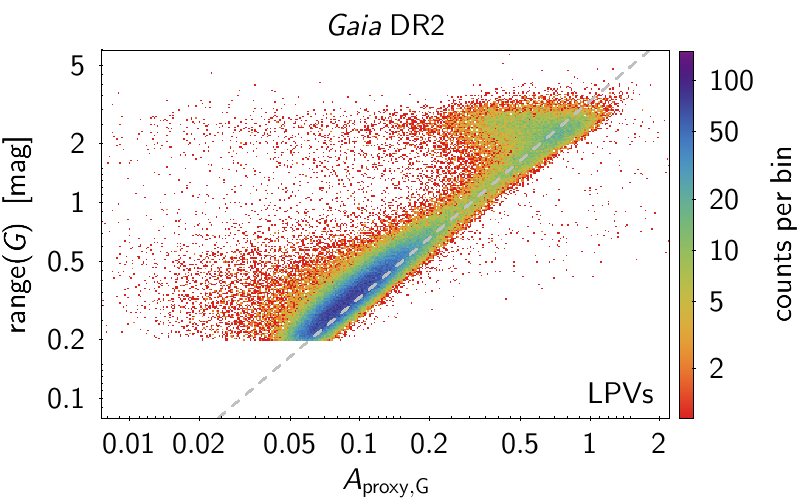}%
\includegraphics[trim={101 81 129 0},clip,,height=110pt]{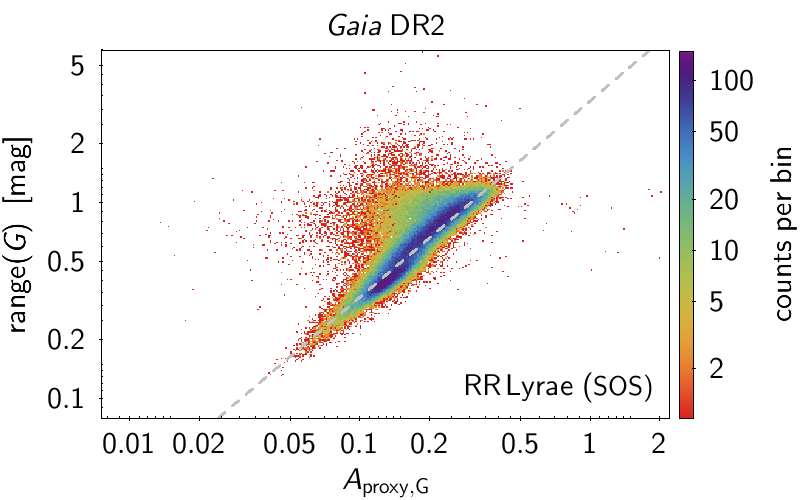}%
\includegraphics[trim={101 81 0 50},clip,,height=97pt]{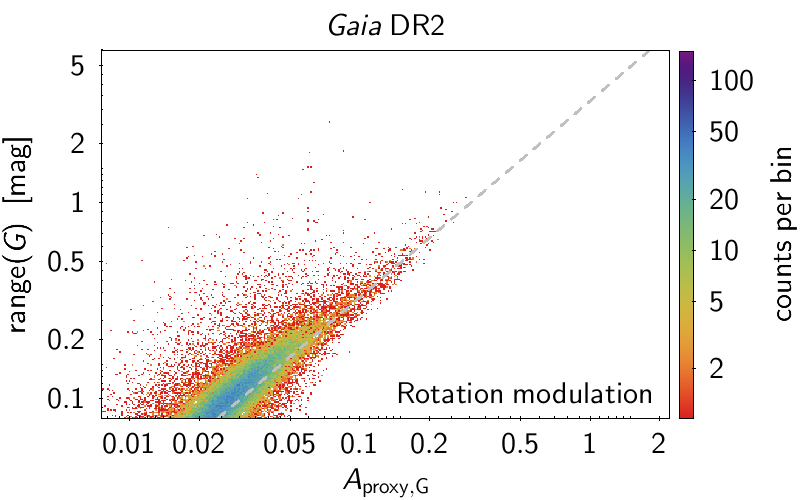}
	\vskip -0.3mm
\includegraphics[trim={0 0 129 50},clip,,height=118.2pt]{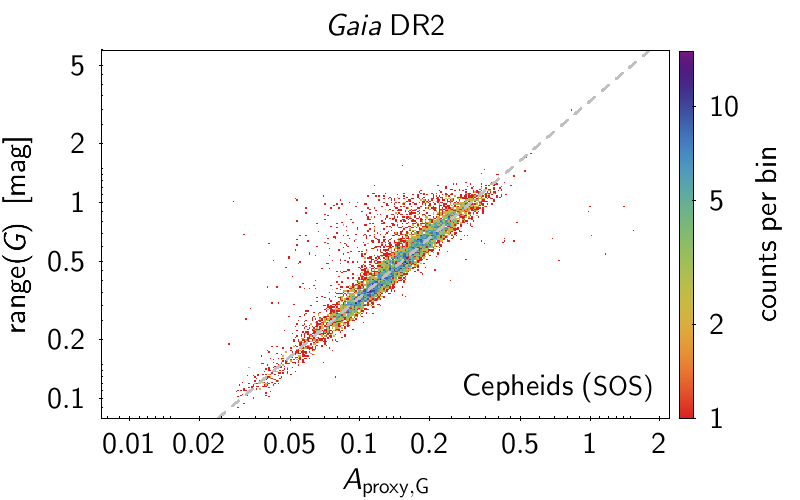}%
\includegraphics[trim={101 0 129 50},clip,,height=118.2pt]{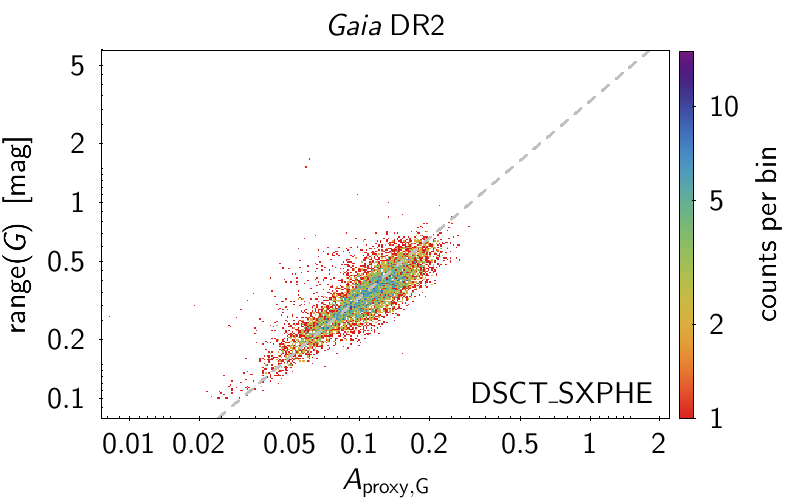}%
\includegraphics[trim={101 0 0 50},clip,,height=118.2pt]{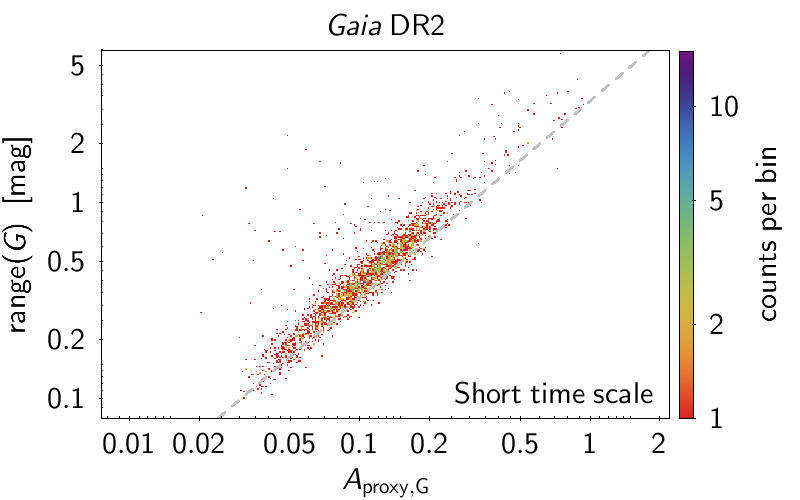}
	\caption{Density maps of the variability range of \gmag time series (in ordinate) versus amplitude proxy (in abscissa) of selected variable stars published in \Gaia DR2 for the variability types indicated in the upper-right corner of each panel.
	         The RR~Lyrae and Cepheid type candidates shown in the figure are restricted to the subset provided in the \texttt{Specific Object Study} (SOS) tables of the data release \citep[see][and more specifically their Fig.~3]{HollAudardNienartowicz_etal18}.
	         The colours of each grid cell in the maps are related to the logarithm of the density of points in the cells according to the colour scale shown on the right of each row of panels.
	         All panels in a given row share the same density colour scale.
	         The dashed diagonal line in each panel corresponds to $\rangeG = 3.3\, \varProxyG$.
	}
	\label{Fig:rangeGvsVarProxyG}
\end{figure*}

The relation between \varProxyG and \rangeG is not unique, as it depends on light curve shape and time sampling.
For a purely sinusoidal function of amplitude $A$ (i.e. peak-to-peak amplitude $2\, A$), the standard deviation is
$
\sigma_{\sin} = \sqrt{\frac{1}{\pi} \int_0^\pi A^2\sin^2(x) \;dx} = A / \sqrt{2}
$.
Therefore, for a densely and evenly sampled sine light curve $\gmag_\mathrm{sin}$, Eq.\ref{Eq:sigma_m} leads to 
$
\mathrm{range}(\gmag_\mathrm{sin}) = 2 \sqrt{2} \, \sigma_{\sin}
                                    \simeq 3.07 \sqrt{\NobsG} \, \fluxErrorG / \fluxG
$.
The proportionality constant would be different for other curve shapes.
For a triangular or a sawtooth wave, for example, $\mathrm{range} = 2\sqrt{3} \, \sigma$, and the proportionality factor would be 3.76 instead of 3.07.

The relation between \varProxyG and \rangeG for \Gaia DR2 is verified with data published in DR2%
\footnote{
We take for \rangeG the quantity \texttt{range\_mag\_g\_fov} provided in the \texttt{vari\_time\_series\_statistics} table in the \Gaia archive.
A more robust estimate of \rangeG is available in DR2 for some variability types, such as for Cepheids and RR Lyrae for which the amplitudes determined from modelled light curves are published, while the value from the statistics table has the advantage to be computed in a similar way for all published variables.
}%
.
This is shown in Fig.~\ref{Fig:rangeGvsVarProxyG} for the various variability types for which time series have been published in DR2.
They concern
151\,761 long-period variables (LPVs),
140\,784 RR~Lyrae variables,
9575 Cepheids,
147\,535 main-sequence (MS) variables induced by rotation modulation,
8882 $\delta$\,Scuti/SX\,Phoenicis type candidates,
and a sample of 3\,018 short time-scale variables.

Figure~\ref{Fig:rangeGvsVarProxyG} shows a proportionality between \varProxyG and \rangeG that is globally linear.
The relation between these two quantities is, however, not uniquely defined because of at least six reasons.
First, the variability proxy is based on the standard deviation of a time series, and its relation to the range depends on the light curve shape.
Second, it depends on the sampling of the signal, and thus on the position in the sky because of the \Gaia scanning law.
An example of this dependence is illustrated by the tail objects in Fig.~\ref{Fig:rangeGvsVarProxyG} departing from the diagonal line towards small \varProxyG values.
This is due to a succession of measurements within a short duration relative to the typical variability time scale of the source (which is not uncommon in the \Gaia scanning law).
These measurements (often within a fraction of a day), have similar fluxes for variables with larger time scales and they bias the standard deviation (and hence \varProxyG) towards small values, while the range of the full light curve remains unaffected.
Third, \varProxyG is based on fluxes, which does not linearly convert to magnitudes used for the computation of \rangeG.
Fourth, \varProxyG is derived from single CCD fluxes, while \rangeG is computed from their integration per field-of-view transit.
Fifth, different outlier-removal algorithms are used to disregard corrupt measurements in the per-ccd versus per-transit time series.
Finally, the standard deviation used in \varProxyG is more robust against outliers than the peak-to-peak amplitude that defines \rangeG.

\begin{figure}
	\centering
	\includegraphics[width=\linewidth]{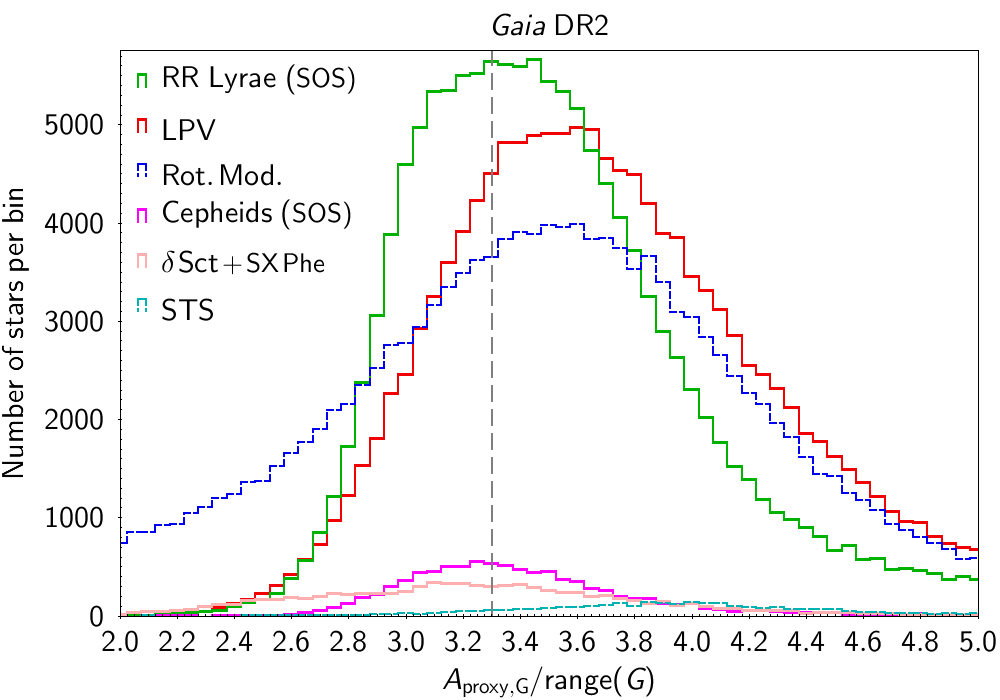}
	\caption{Histograms of \rangeG/\varProxyG ratio for the samples of various variability types shown in Fig.~\ref{Fig:rangeGvsVarProxyG}.
	The variability type corresponding to each histogram is written in the top left of the panel in the same colour as the histogram, in decreasing order of the histogram maximum.
	Pulsating stars are shown in continuous thick lines, while non pulsators, that is MS rotation modulation variables (Rot. Mod.) and short time-scale variables, are shown in dashed thin lines.
	A dashed vertical line is plotted at $\rangeG/\varProxyG=3.3$.
	}
	\label{Fig:histo_rangeGoverVarProxyG}
\end{figure}

The \rangeG/\varProxyG ratio is displayed in Fig.~\ref{Fig:histo_rangeGoverVarProxyG} for the various variability types displayed in Fig.~\ref{Fig:rangeGvsVarProxyG}.
It is seen that this ratio comprises between $\sim$3.2 for $\delta$~Sct-type variables and $\sim$3.5 for LPVs
and rotation modulation MS stars, with a value of $\sim$3.3 for Cepheids and RR Lyrae variables.
Only the small sample of short time-scale variables has a distribution peaked at a higher value around four.
The relation found for LPVs is consistent with the relation $QR_5(\gmag) \simeq 3.3 \, \varProxyG$ found in \citet{MowlaviTrabucchiLebzelter_2019} for the same set of \Gaia DR2 LPVs, $QR_5(\gmag)$ being the 5-95\% quantile range.

Given the above considerations, the relation 
\begin{equation}
    \rangeG \simeq 3.3 \, \varProxyG \;
\label{Eq:rangeGvsVarProxyG}
\end{equation}
was chosen for the LAVs studied in this paper.
The proportionality factor 3.3 in Eq.~\ref{Eq:rangeGvsVarProxyG} is of course approximate, as shown above, but it provides a useful relation to estimate the magnitude variability range, which is not available in the \Gaia DR2 archive, from the amplitude proxy.

\section{The catalogue}
\label{Sect:catalogue}

The distribution of the amplitude proxy \varProxyG defined by Eq.~\ref{Eq:AproxyG} is shown in Fig.~\ref{Fig:varProxyGsG_randomSample} versus magnitude \gmag for a random sample of 100 million \Gaia sources brighter than 19.5~mag.
The lower envelope of higher density sources represents constant stars.
Sources with amplitude proxies larger than the values characterizing constant stars are potentially variable.
Limits at $\varProxyG\!=\!0.06$, 0.15 and 0.3 are shown in the figure by solid, dashed and dotted red lines,  respectively, corresponding to estimated peak-to-peak \gmag variability amplitudes of $\sim$0.2, $\sim$0.5 and $\sim$1 mag, respectively.
To avoid contamination by constant stars, we restrict our catalogue to sources with

\begin{equation}
\left\{
  \begin{array}{l}
     \varProxyG > 0.06 \; , 
     \vspace{2mm}\\
     5.5 < \gmag/\mathrm{mag}< 19 \; .
  \end{array}
\right.
 \label{Eq:datasetLimits}
\end{equation}
Figure~\ref{Fig:varProxyGsG_randomSample} shows that the contribution of intrinsic stellar variability should dominate that of data noise in these parameter ranges.
Caution, however, must be taken at the faintest ($\gmag\!\gtrsim\! 18.5$~mag) end where data noise may provide a larger contribution to \varProxyG, and around $\gmag\!=\!11$ and 13~mag where the photometric data reduction pipeline changes calibration regimes \citep[at 13~mag due to a change of window class, and at 11~mag due to gate activation, see][ in particular their Fig.~9]{EvansRielloDeAngeli_etal18}.
The second condition in Eq.~\ref{Eq:datasetLimits} intends to stay clear of the faintest and brightest ends of \gmag where noise (at the faint side) and systematics due to poor handling of saturation in DR2 (at the bright side) become significant relative to intrinsic variability.

\subsection{Datasets A, B, and C}
\label{Sect:catalogue_datasets}

\begin{figure}
	\centering
	\includegraphics[width=\linewidth]{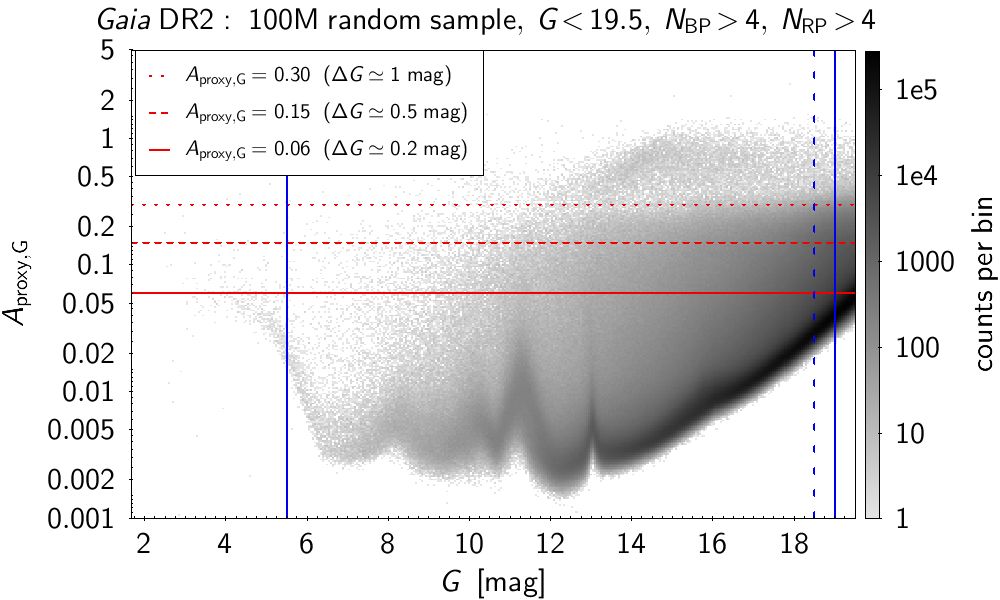}
	\caption{Density map of the variability amplitude proxy (ordinate) versus \gmag magnitude (abscissa) for a random sample of 100 million \Gaia DR2 sources that have at least five measurements in \gbp and \grp.
	A dotted, dashed and solid horizontal red line is plotted as eye-guides at $\varProxyG=0.30$, 0.15 and 0.06, respectively.
	They correspond approximately to peak-to-peak amplitudes in \gmag of 1~mag, 0.5~mag and 0.02~mag, respectively.
	Vertical continuous blue lines are plotted at $\gmag=5.5$~mag and 19~mag, which define the magnitude limits of the sample studied in this paper.
	Additionally, a vertical dashed blue line is plotted at 18.5~mag.
	}
	\label{Fig:varProxyGsG_randomSample}
\end{figure}

We provide three datasets, called Datasets~A, B, and C\footnote{
The catalogue of LAVs is available for download, see Appendix~\ref{Appendix:electronicTable}.
}%
.
Each dataset is a subset of the previous one, with Dataset~A being the full catalogue of LAVs.
The number of sources in each dataset, and the filtering conditions that lead to their definitions are summarized in 
Table~\ref{Tab:datasetsSummary}.
The datasets are characterized as follow.

\paragraph{\textbf{Dataset~A:}} This dataset contains all LAVs that satisfy Eq.~\ref{Eq:datasetLimits} and are cleaned from sources whose light curves appear to be affected by instrumental arteafacts at specific times of the mission (filter a1 in Table~\ref{Tab:datasetsSummary}, see Sect.~\ref{Appendix:DatasetA_Sky} in Appendix~\ref{Appendix:DatasetA} for more information).
Sources that may potentially contain \gmag epoch magnitudes fainter than 20.5~mag are also excluded (filter a2 in Table~\ref{Tab:datasetsSummary}, see Appendix~\ref{Appendix:DatasetA_faintG}).

The procedure used to import the data from the \Gaia DR2 archive and details on the filtering criteria are given in Appendix~\ref{Appendix:DatasetA}.

\paragraph{\textbf{Dataset~B:}} This subset of Dataset~A is to be preferentially used if reliable \gbp and \grp magnitudes are needed (such as for colour-magnitude diagrams).
The selection relies on the fact that $\fluxBP+\fluxRP$ must be close to \fluxG as a result of the wavelength transmission bands of \gmag, \gbp and \grp \citep{EvansRielloDeAngeli_etal18}.
A source with a larger-than-expected summed flux $\fluxBP+\fluxRP$ relative to \fluxG is therefore suspected to have inconsistent \gmag, \gbp and \grp measurements.
While unreliable \fluxexcesses are, in DR2, due in many cases to BP/RP integrated fluxes of poorer quality, similar problems can also affect \gmag-band measurements.
We refer to Appendix~\ref{Appendix:BPRPexcess} for a discussion on this (see in particular Sect.~\ref{Appendix:BPRPexcess_summary}).

The \fluxexcess $(\fluxBP+\fluxRP)/\fluxG$ depends on the spectral type, and thus on \BPminusRP colour.
We derive in Appendix~\ref{Appendix:BPRPexcess} a normalized \fluxexcess, notated \excessFactorNormalized (Eq.~\ref{Eq:normalizedExcessFactor} in the Appendix), which should be close to one at all \BPminusRP colours for typical stars, and apply the filtering criteria b2 and b3 listed in Table~\ref{Tab:datasetsSummary} to derive Dataset~B.
This can be done only if the source has \fluxBP and \fluxRP values in \Gaia DR2, which imposes the additional selection criterion b1 listed in Table~\ref{Tab:datasetsSummary}.

\paragraph{\textbf{Dataset~C:}} This subset of Dataset~B is to be preferentially used if reliable \varProxyBP and \varProxyRP are needed (such as for multi-band variability studies in \gmag, \gbp and \grp).
The selection relies on the fact that the variability in \bandBPplusRP must be consistent with the variability in \gmag given the wavelength transmission bands.
A variability in \gmag that is not present in \bandBPplusRP is suspicious (note, however, that this could happen in the case of an anti-correlated variability in the blue and in the red%
\footnote{
An example of anti-correlated blue/red variability is given by Ap stars.
The variability of these stars is due to the presence of spots caused by the migration of certain chemical elements modifying the opacity of the star's surface.
The presence of spots may induce two effects in the star's atmosphere: a blocking effect and then a back-warming effect.
For cold Ap stars ($<10\,000$~K), the blocking effect is in the blue part of the stellar spectra, and the re-emission in the red.
The variations in different filters can be anti-phased \citep{MuciekNorthRufener_etal85}, thus strongly attenuating the variations integrated in a large filter.
This was highlighted for the wide Hipparcos $H_\mathrm{p}$ band by \citet{Eyer98} (Sects.~10.2.2 and 13.3).
It can also be seen in the motion of variable stars in the Hertzsprung-Russell diagram \citep[such as in Fig.~11 of][]{GaiaEyerRimoldini_etal19}), where Ap stars ($\alpha^2$~Canum Venaticorum) have horizontal motions, that is noticeable variations in \BPminusRP with little change in the \gmag band.
}%
).
Likewise, a variability observed in \bandBPplusRP but not in \gmag may indicate additional noise in \gbp and/or \grp that would make \varProxyBP and/or \varProxyRP unreliable (note, however, that, in such a case, \varProxyG may still be reliable).

\begin{table}
\caption{Summary of the number of sources in Datasets~A, B, and C, and of the number of sources removed by the successive filtering criteria that lead from the public \Gaia DR2 archive (first line in the table) to each dataset.
The subsets in these datasets that have parallax uncertainties better than 10\% are shown in parenthesis.
        }
\centering
\begin{tabular}{l r}
\hline\hline
\rule{0pt}{2.0ex}Criterion & Nbr of sources \\
\hline
\rule{-4pt}{3.0ex}$5.5 < \gmag < 19$, $\varProxyG>0.06$          & 23\,830\,345 \\
\hline

\rule{0pt}{3.0ex} a1)~ {In sky stripes: $\varProxyG\!>\!0.1$ or $\gmag\!<\!18.3$} &   $-514\,084$ \\
\rule{0pt}{1.0ex} a2)~ {$\gmag + 1.65 \, \varProxyG < 20.5$} &       $-387$ \\
\rule{-5pt}{2.5ex} \textbf{Dataset A}                                & \textbf{23\,315\,874}  \\
\rule{0pt}{2.0ex} ($\varpi/\epsilon(\varpi)>10$)                     & (401\,480)  \\
\hline

\rule{0pt}{3.0ex} b1)~ {has \gbp and \grp}                           & $-5\,535\,102$ \\
\rule{0pt}{1.0ex} b2)~ {$\excessFactorNormalized < 1.04 + 0.001 \, (\BPminusRP-1)^3$}\!\!\!\!\!\!\!\! & $-16\,626\,421$ \\
\rule{0pt}{1.0ex} b3)~ {$\excessFactorNormalized > 0.9$}             &     $-5490$ \\
\rule{-5pt}{2.5ex} \textbf{Dataset B}                                & \textbf{1\,148\,861} \\
\rule{0pt}{2.0ex} ($\varpi/\epsilon(\varpi)>10$)     & (110\,521) \\
\hline
\rule{0pt}{3.0ex} c1)~ {$\varProxyG < 1.5 \, \varProxyBPplusRPnullCov$}   &    $-66\,364$ \\
\rule{0pt}{1.0ex} c2)~ {$\varProxyG > 0.8 \, \varProxyBPplusRPnullCov$}   &   $-121\,237$ \\
\rule{0pt}{1.0ex} c3)~ {$N_\mathrm{BP,RP}\ge10$}                     &    $-25\,755$ \\
\rule{0pt}{1.0ex} c4)~ {$|\NobsBP-\NobsRP| \le 1$}                   &   $-211\,607$ \\
\rule{0pt}{1.0ex} c5)~ {$7.8 < \NobsG\,/\,\NobsRP < 10.2 $}          &    $-93\,323$ \\
\rule{0pt}{1.0ex} c6)~ {$\gbp + 1.65 \, \varProxyBP < 20.5$}         &    $-11\,609$ \\
\rule{-5pt}{2.5ex} \textbf{Dataset C}                                & \textbf{618\,966} \\
\rule{0pt}{2.0ex} ($\varpi/\epsilon(\varpi)>10$)                     & (85\,046)  \\
\hline
\end{tabular}
\label{Tab:datasetsSummary}
\end{table}

\begin{figure}
	\centering
	\includegraphics[width=\linewidth]{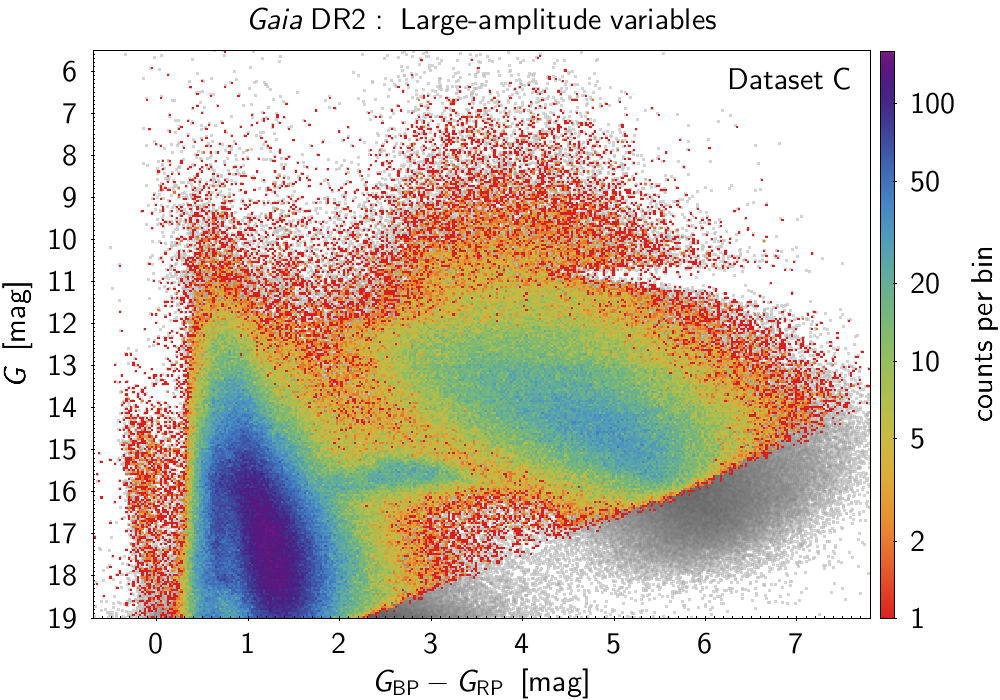}
	\caption{Density map of the colour-magnitude diagram of Dataset~C.
	         Dataset~B is plotted in grey in the background.
	         The axes range have been limited for better visibility.
	         The lack of very red sources at $\gmag \simeq 11$~mag in Datasets~B and C is due to limitations in the DR2 processing leading to too low \fluxexcesses (see text).
	}
	\label{Fig:CM_density_DatasetC}
\end{figure}

\begin{figure*}
	\centering
	\includegraphics[trim={0 81 146 0},clip,width=0.46\linewidth]{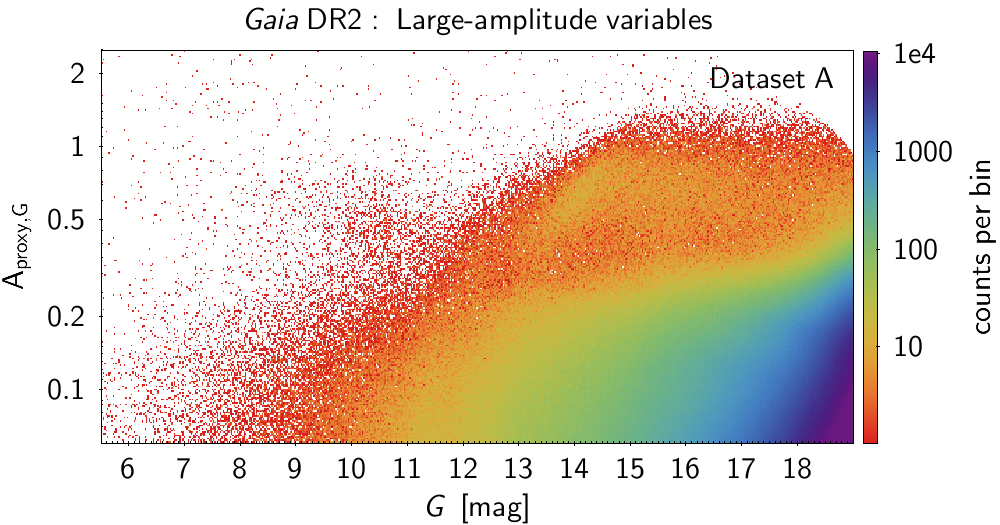}%
\includegraphics[trim={102 81 0 0},clip,width=0.483\linewidth]{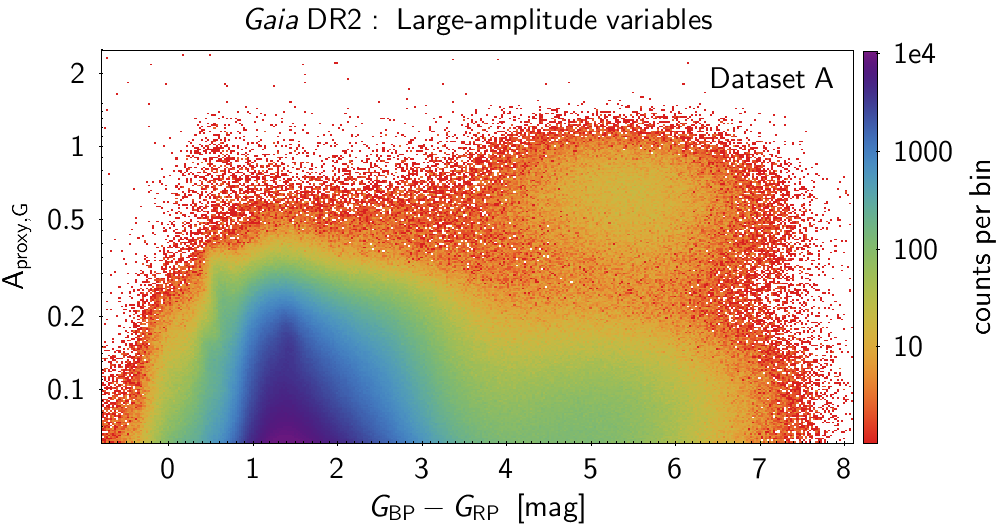}
	\vskip -0.5mm
	\includegraphics[trim={0 81 146 50},clip,width=0.46\linewidth]{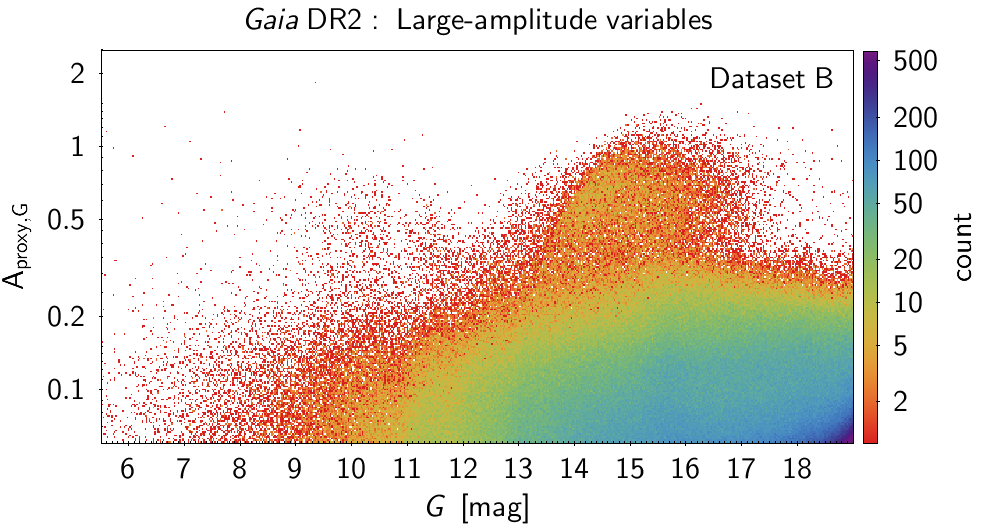}%
\includegraphics[trim={102 81 0 50},clip,width=0.483\linewidth]{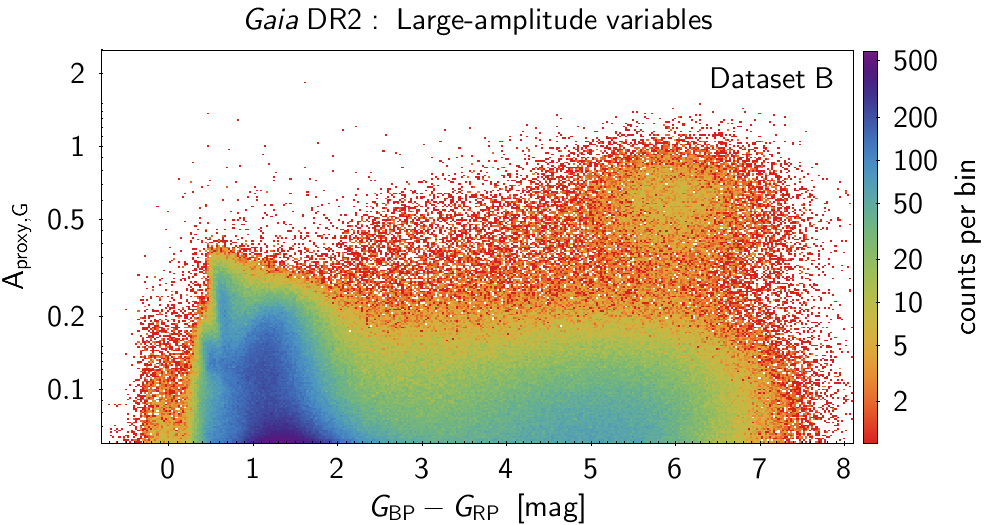}
	\vskip -0.5mm
	\includegraphics[trim={0 0 146 50},clip,width=0.46\linewidth]{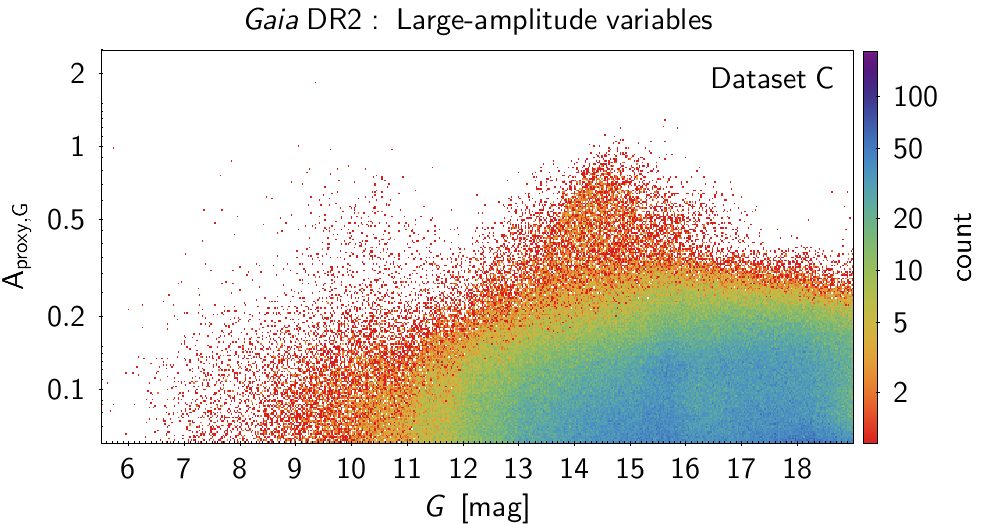}%
\includegraphics[trim={102 0 0 50},clip,width=0.483\linewidth]{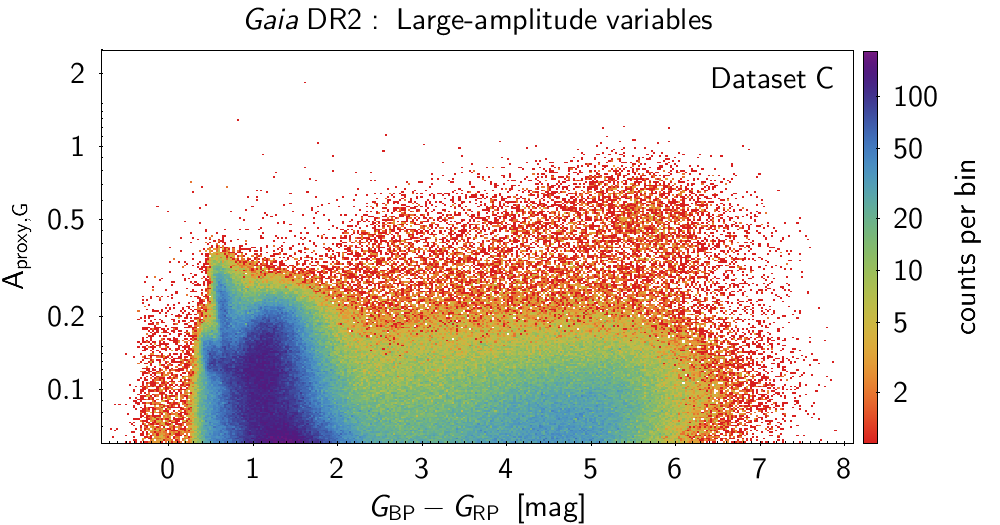}
	\caption{Density maps of the \gmag amplitude proxy versus \gmag magnitude (left panels) and versus \BPminusRP colour (right panels) for Datasets~A (top panels) B (middle panels) and C (bottom panels).
	         The colours of each grid cell in the maps are related to the logarithm of the density of points in the cells according to the colour scale shown on the right of each row of panels.
	         The two panels in a given row share the same density colour scale.
	         We note that the top-right panel of Dataset~A should be analysed with caution, as it contains sources with unreliable \gbp and/or \grp.
	}
	\label{Fig:varProxyGvsGAndColour}
\end{figure*}

The amplitude proxy \varProxyBPplusRP of the summed \bandBPplusRP is not available in \Gaia DR2, and cannot be computed with the available DR2 quantities.
This would require flux time series in order to evaluate the covariance term between \gbp and \grp.
Therefore, we derive in Appendix~\ref{Appendix:AproxyBPplusRP} an approximation to \varProxyBPplusRP , notated \varProxyBPplusRPnullCov, that neglects the covariance term but is computable with the available DR2 data (Eq.~\ref{Eq:varProxyBPplusRPwithoutCov}).
The filtering conditions c1 and c2 listed in Table~\ref{Tab:datasetsSummary} use this quantity to select sources for Dataset~C, based on the analysis performed in Appendix~\ref{Appendix:AproxyBPplusRP} on the conditions expected to be satisfied by \varProxyBPplusRPnullCov (Eq.~\ref{Eq:conditionVarProxyGBPRPRatio}).
In addition, we require the sources to have at least ten measurements in \gbp and \grp (condition c3), and to have similar numbers of field-of-view transits in \gmag, \gbp and \grp (conditions c4 and c5).
These extra conditions are meant to ensure similar time distributions between the three photometric time series, a condition that is essential for useful comparison of their variability properties given the large amplitudes considered here.
Finally, sources that may potentially contain \gbp epoch magnitudes fainter than 20.5~mag are also excluded (filter c6 in Table~\ref{Tab:datasetsSummary}).
We note that the equivalent condition for \grp is always satisfied.

\paragraph{Colour-magnitude diagram.}
\label{Sect:catalogue_datasets_CM}

The colour-magnitude (CM) diagram of Dataset~C is shown in Fig.~\ref{Fig:CM_density_DatasetC}.
It reveals a lack of very red sources ($\BPminusRP \gtrsim 4.5$~mag) at $\gmag\simeq11$~mag.
This is due to limitations in the DR2 processing, as shown in Appendix~\ref{Appendix:BPRPexcess_lowFactors}, which lead to too low \fluxexcesses for very red stars at these magnitudes (see in particular Fig.~\ref{Fig:CM_withExcessFactorNormalized_datasetA}).
The feature is present in Dataset~B as well (shown in grey in the background of Fig.~\ref{Fig:CM_density_DatasetC}) as the exclusion of sources with too small \fluxexcesses is performed with filter b3 listed in Table~\ref{Tab:datasetsSummary}.
The excess of sources around $\gmag \simeq 16$~mag with $\BPminusRP$ from 2~mag to 3.5~mag in Fig.~\ref{Fig:CM_density_DatasetC} is linked to the population of LPVs in the Magellanic Clouds.

\paragraph{\gmag-band variability.}
\label{Sect:catalogue_datasets_varProxyG}

The distributions of \varProxyG versus \gmag and versus \BPminusRP are shown in Fig.~\ref{Fig:varProxyGvsGAndColour} for the three datasets (we stress however that Dataset~A should in principle not be used when \BPminusRP colour is required).
We highlight here two features seen in these diagrams to illustrate the pros and cons of the various datasets.
The first concerns the presence of a population with very large amplitudes in all three datasets, with  $\varProxyG \gtrsim 0.3$.
The great majority of them are Miras, as suggested by their red colours in the right panels of Fig.~\ref{Fig:varProxyGvsGAndColour}.
They are relatively numerous in Dataset~A, but their number decreases significantly in Datasets~B and C.
This loss of completeness from Dataset~A to B and C is addressed in Sect.~\ref{Sect:catalogue_completeness}.

The second feature is the presence of a large number of faint LAV candidates ($\gmag\gtrsim 18$~mag) in Dataset~A (top-left panel in Fig.~\ref{Fig:varProxyGvsGAndColour}) with amplitudes close to the lower limit of $\varProxyG=0.06$ considered here.
These are most probably contaminants due to increasing noise level when \gmag approaches 19~mag, as seen in Fig.~\ref{Fig:varProxyGsG_randomSample}.
This excess of faint LAVs is much smaller in Dataset~B, and basically absent in Dataset~C (Fig.~\ref{Fig:histo_magG}).
The purity of the datasets will be addressed in Sect.~\ref{Sect:catalogue_purity}.

\paragraph{Multi-band variability.}
\label{Sect:catalogue_datasets_multiband}

Figure~\ref{Fig:varProxyRatiosBPGvsRPG_DatasetC} displays $\varProxyRP/\varProxyG$ versus $\varProxyBP/\varProxyG$.
High densities of sources are observed in specific regions of the diagram, revealing distinct multi-band variability properties.
The densest region contains quasi-achromatic variables with $\varProxyG \simeq \varProxyBP \simeq \varProxyRP$.
The next two densest groups are both observed in the region where variability amplitude is larger in the blue than in the red (i.e. on the right side of the uppermost dashed line in Fig.~\ref{Fig:varProxyRatiosBPGvsRPG_DatasetC}).
This is seen more clearly in Fig.~\ref{Fig:varProxyBPRPvsBPmRP_DatasetC} that displays $\varProxyBP/\varProxyRP$ versus \BPminusRP, where the two groups are observed at $\varProxyBP/\varProxyRP \simeq 1.63$ and 2.2, respectively.
The multi-band variability properties of Dataset~C is analysed in Sect.~\ref{Sect:exploration_multiband}.

\subsection{Completeness}
\label{Sect:catalogue_completeness}

The completeness of the three datasets is difficult to assess in absolute terms.
Estimates are attempted in this section based on data from the \Gaia DR2 catalogues of variable stars (Sect.~\ref{Sect:catalogue_completeness_DR2}), from the ASAS-SN survey (Sect.~\ref{Sect:catalogue_completeness_ASASSN}), and from the ZTF survey (Sect.~\ref{Sect:catalogue_completeness_ZTF}).

\subsubsection{Completeness estimate based on \Gaia DR2 variables}
\label{Sect:catalogue_completeness_DR2}

The completeness relative to \Gaia DR2 variables is given in Table~\ref{Tab:completeness} for the six variability types published in DR2, that is LPVs, RR~Lyrae stars, Cepheids, MS rotation modulation variables, $\delta$\,Scuti/SX\,Phoenicis stars, and short time-scale variables.
For RR~Lyrae and Cepheid variables, we consider both \Gaia DR2 samples provided in the classification and Specific Object Study (SOS) tables \citep[see][]{HollAudardNienartowicz_etal18}.
The completeness is estimated by checking the fraction of these variables that are recovered in Datasets~A, B, and C.
To achieve this, we first restrict the DR2 samples to the conditions defining our datasets, that is $5.5 \!<\! \gmag/\mathrm{mag} \!<\! 19$ and $\varProxyG\!>\!0.06$.
The number of variables satisfying these conditions for each variability type are given in the third column of Table~\ref{Tab:completeness}.
The fraction of these variables that are present in Datasets~A, B, and C are then provided in the fourth to sixth columns, respectively, with their percentages given in the row below the numbers.

Table~\ref{Tab:completeness} shows that practically all variables published in DR2 are present in Dataset A.
For dataset~B, a difference is observed between pulsating and non-pulsating stars.
Pulsating stars have completeness levels between 58\% and 96\% in Dataset~B.
This is excellent considering that Dataset~B contains only $\sim$5\% of Dataset~A.
For non strictly periodic stars, the completeness levels are much lower, being of 33\% for the DR2 rotation modulation variables, and only 5\% for short time-scale variables.
We note that rotation modulation candidates are not expected to have variability amplitudes larger than $\sim$0.2~mag, a statement supported by the very small fraction of the DR2 rotation modulation candidates that have $\varProxyG>0.06$ (2181 out of 147\,535 candidates, see Table~\ref{Tab:completeness}).
The situation is different for short time-scale candidates.
Contrary to the case of rotation modulation candidates, the majority of them do have large amplitudes in DR2 (2641 out of 3018 candidates, see Table~\ref{Tab:completeness}), and these are all, except one, in Dataset~A.
However, only 5\% of them remain in Dataset~B, a reduction factor that is similar to the overall reduction from Dataset~A to B (4.8\%, see Table~\ref{Tab:datasetsSummary}).
We remind that the short time-scale candidates published in DR2 were identified from their variability in the \gmag-band CCD timeseries, while we are dealing here with \gmag-band transit photometry.
That difference may explain their relative numbers in Datasets~A and B.

For Dataset~C, the fraction of LAVs kept from Dataset~B is around 60\% to 70\% for all variability types, except for short time-scale candidates that have a reduction factor of $\sim$50\% from Dataset~B to C.
We stress that the completeness numbers given in Table~\ref{Tab:completeness} are upper limits, because 
the catalogues of variables published in \Gaia DR2 are themselves not complete \citep[their completeness varies greatly with variability type and sky location, see Tables~2 and 3 of][]{HollAudardNienartowicz_etal18}.

\begin{figure}
	\centering
	\includegraphics[width=\linewidth]{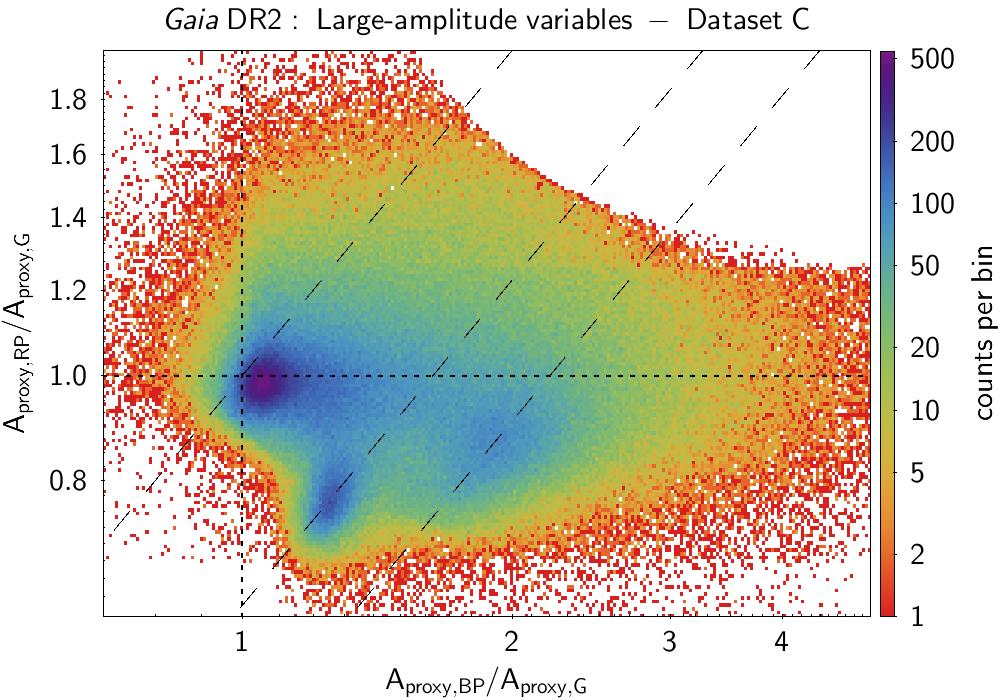}
	\caption{$\varProxyRP/\varProxyG$ versus $\varProxyBP/\varProxyG$ for all LAV candidates in Dataset~C.
	         Dashed lines are drawn at $\varProxyBP/\varProxyRP=1$, 1.63 and 2.2.
	         Dotted lines are further added at $\varProxyRP=\varProxyG$ and $\varProxyBP=\varProxyG$ to guide the eyes.
	         The axes ranges have been limited for better visibility.
	}
	\label{Fig:varProxyRatiosBPGvsRPG_DatasetC}
\end{figure}

\begin{figure}
	\centering
	\includegraphics[width=\linewidth]{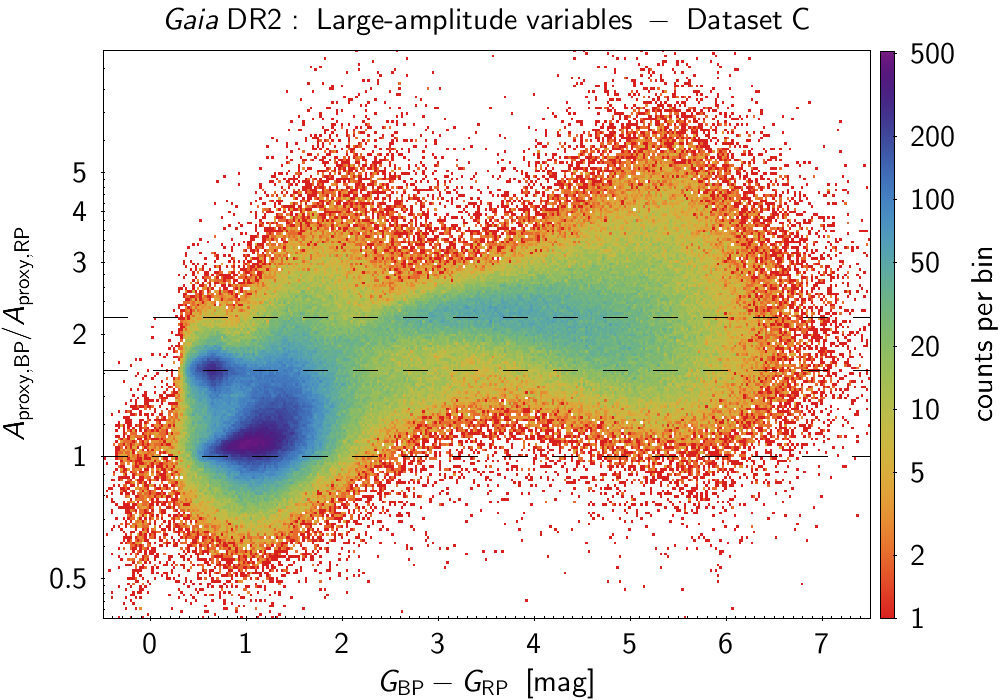}
	\caption{Same as Fig.~\ref{Fig:varProxyRatiosBPGvsRPG_DatasetC}, but for $\varProxyBP/\varProxyRP$ versus \BPminusRP.
	}
	\label{Fig:varProxyBPRPvsBPmRP_DatasetC}
\end{figure}

\subsubsection{Completeness estimate based on ASAS-SN survey}
\label{Sect:catalogue_completeness_ASASSN}

\begin{table*}[t]
\caption{Completeness of Datasets~A, B, and C with respect to the samples of variable stars published in dedicated \Gaia DR2 catalogues of variable stars.
         The origin of the samples in DR2 \citep[i.e. either classification table or SOS table, see][in particular their Fig.~3]{HollAudardNienartowicz_etal18} is indicated in parenthesis.
         Reference to the paper describing the specific catalogue is given in the last column.
         The numbers in italics denote the percentages of sources present in Datasets~B and C compared to Datasets~A and B, respectively.
        }
\centering
\begin{tabular}{l r | r| r r r | l}
\hline\hline
\rule{0pt}{2.5ex}  Variability type     &   \multicolumn{1}{c|}{Total} & \multicolumn{4}{c}{$5.5\!<\!G /\mathrm{mag}\!<\!19, \; \varProxyG\!>\!0.06$} &  \multicolumn{1}{|c}{Reference} \\
\rule{0pt}{2.2ex}     &  in DR2 &  in DR2 & Dataset~A & Dataset~B & Dataset~C & \\
\hline
\hline
\rule{0pt}{2.2ex} LPV (SOS)                & 151\,761 & 138\,193 & 138\,150 &  111\,941 &  76\,527 &  \citet{MowlaviLecoeurLebzelter_etal18}\\
                                           &         &         &  \multicolumn{3}{c|}{ \hspace{0.5cm} \small{\textit{81\%}} \hspace{1cm} \small{\textit{68\%}}} \\
\hline
\rule{0pt}{2.2ex} RR Lyr (classif)         & 195\,780 & 127\,123 & 127\,103 &  74\,189 &  48\,891 & \citet{RimoldiniHollAudard_etal19} \\
                                           &         &         &  \multicolumn{3}{c|}{ \hspace{0.5cm} \small{\textit{58\%}} \hspace{1cm} \small{\textit{66\%}}} \\
\hline
\rule{0pt}{2.2ex} RR Lyr (SOS)             & 140\,784 &  90\,024 &  89\,976 &  59\,824 &  39\,977 & \citet{ClementiniRipepiMolinaro_etal19} \\
                                           &         &         &  \multicolumn{3}{c|}{ \hspace{0.5cm} \small{\textit{66\%}} \hspace{1cm} \small{\textit{67\%}}} \\
\hline
\rule{0pt}{2.2ex} Cep (classif)            &   8550 &   8367 &   8362 &  5403 &  3887 & \citet{RimoldiniHollAudard_etal19} \\
                                           &         &         &  \multicolumn{3}{c|}{ \hspace{0.5cm} \small{\textit{65\%}} \hspace{1cm} \small{\textit{72\%}}} \\
\hline
\rule{0pt}{2.2ex} Cep (SOS)                &   9575 &   9197 &   9194 &   5685 &   4131 & \citet{ClementiniRipepiMolinaro_etal19} \\
                                           &         &         &  \multicolumn{3}{c|}{ \hspace{0.5cm} \small{\textit{62\%}} \hspace{1cm} \small{\textit{73\%}}} \\
\hline
\rule{0pt}{2.2ex} $\delta$~Sct, SX Phe (classif) &   8882 &   3328 &   3310 &   3183 &   2016 & \citet{RimoldiniHollAudard_etal19} \\
                                           &         &         &  \multicolumn{3}{c|}{ \hspace{0.5cm} \small{\textit{96\%}} \hspace{1cm} \small{\textit{63\%}}} \\
\hline
\rule{0pt}{2.2ex} Rot. modul. (SOS)        & 147535 &   2181 &   2129 &     692 &     463 & \citet{lanzafameDistefanoMessina_etal18} \\
                                           &         &         &  \multicolumn{3}{c|}{ \hspace{0.5cm} \small{\textit{33\%}} \hspace{1cm} \small{\textit{67\%}}} \\
\hline
\rule{0pt}{2.2ex} Short time-scale (SOS)                &   3018 &   2641 &   2640 &     129 &      60 & \citet{RoelensEyerMowlavi_etal18} \\
                                           &         &         &  \multicolumn{3}{c|}{ \hspace{0.5cm} \small{\textit{5\%}} \hspace{1cm} \small{\textit{47\%}}} \\
\hline
\end{tabular}
\label{Tab:completeness}
\end{table*}

\begin{table*}[t]
\caption{Completeness estimates of Datasets~A, B, and C based on ASAS-SN \citep{JayasingheKochanekStanek_etal18} and ZTF \citep{ChenWangDeng_etal20} surveys.
        }
\centering
\begin{tabular}{l | l r | l r r r r }
\hline\hline
\rule{-8pt}{2.5ex}  Survey    &      \multicolumn{2}{l|}{Amplitude-limited sample}  & Crossmatch            &           Cleaned & Dataset~A & Dataset~B & Dataset~C \\\hline
\hline
\rule{-8pt}{2.2ex} ASAS-SN    &  \small{$Amp(V)\!>\!0.2$~mag} : &           468\,527    & \small{from ASAS-SN}        &           328\,249 &   59\,641 &    54\,713 &  40\,772 \\
                   666\,502    &                                  &              &                        &         & \multicolumn{3}{c}{ \hspace{0.5cm} \small{\textit{92~\%}} \hspace{1cm} \small{\textit{75~\%}}} \\
\rule{0pt}{2.2ex}             &                                  &              &  \small{4" XM on Dataset A} &           198\,591 &  198\,591 &   181\,999 & 135\,196 \\
                              &                                  &              &                        &         & \multicolumn{3}{c}{ \hspace{0.5cm} \small{\textit{92~\%}} \hspace{1cm} \small{\textit{74~\%}}} \\
\cline{4-8}
\rule{0pt}{2.5ex}             &  \small{$Amp(V)\!>\!0.4$~mag} : &           284\,277    &  \small{from ASAS-SN}        &           195\,440 &   43\,353 &    39\,314 &  28\,648 \\
                              &                                  &              &                        &         & \multicolumn{3}{c}{ \hspace{0.5cm} \small{\textit{90~\%}} \hspace{1cm} \small{\textit{73~\%}}} \\
\rule{0pt}{2.2ex}             &                                  &              &  \small{4" XM on Dataset A} &           146\,287 &  146\,287 &   132\,518 &  96\,293 \\
                              &                                  &              &                        &         & \multicolumn{3}{c}{ \hspace{0.5cm} \small{\textit{91~\%}} \hspace{1cm} \small{\textit{73~\%}}} \\
\hline\hline
\rule{-8pt}{2.2ex} ZTF     &  \small{$Amp(r)\!>\!0.2$~mag} : &           484\,306    & \small{2" XM on Dataset A} &           311\,726 &  311\,726 &   225\,458 & 146\,746 \\
                   781\,602 &                                  &              &                        &         & \multicolumn{3}{c}{ \hspace{0.5cm} \small{\textit{72~\%}} \hspace{1cm} \small{\textit{65~\%}}} \\
\rule{-8pt}{2.2ex}         &  \small{$Amp(r)\!>\!0.4$~mag} : &           250\,821    & \small{2" XM on Dataset A} &           160\,602 &  160\,602 &   113\,288 & 73\,673 \\
                           &                                  &              &                        &         & \multicolumn{3}{c}{ \hspace{0.5cm} \small{\textit{71~\%}} \hspace{1cm} \small{\textit{65~\%}}} \\
\hline
\end{tabular}
\label{Tab:completeness_otherSurveys}
\end{table*}

The ASAS-SN survey of variable stars \citep{ShappeePrietoStanek_etal14,JayasingheKochanekStanek_etal18} has published 666\,502 sources\footnote{
We use in this paper the ASAS-SN catalogue downloaded from \url{https://asas-sn.osu.edu/variables} on June 2, 2020.
}%
, of which 646\,027 have a \Gaia crossmatch ID identified in their catalogue.
To compare with our sample of \Gaia LAVs, we apply two filters to the initial ASAS-SN dataset.
The first filter consists in considering only ASAS-SN sources that have $Amp(V)$ variability amplitudes larger than 0.2~mag to comply with the lower \gmag amplitude limit in our datasets.
The resulting ASAS-SN sample contains 468\,527 sources, of which 454\,910 sources have a \Gaia DR2 ID in their catalogue.

\begin{figure}
	\centering
	\includegraphics[width=\linewidth]{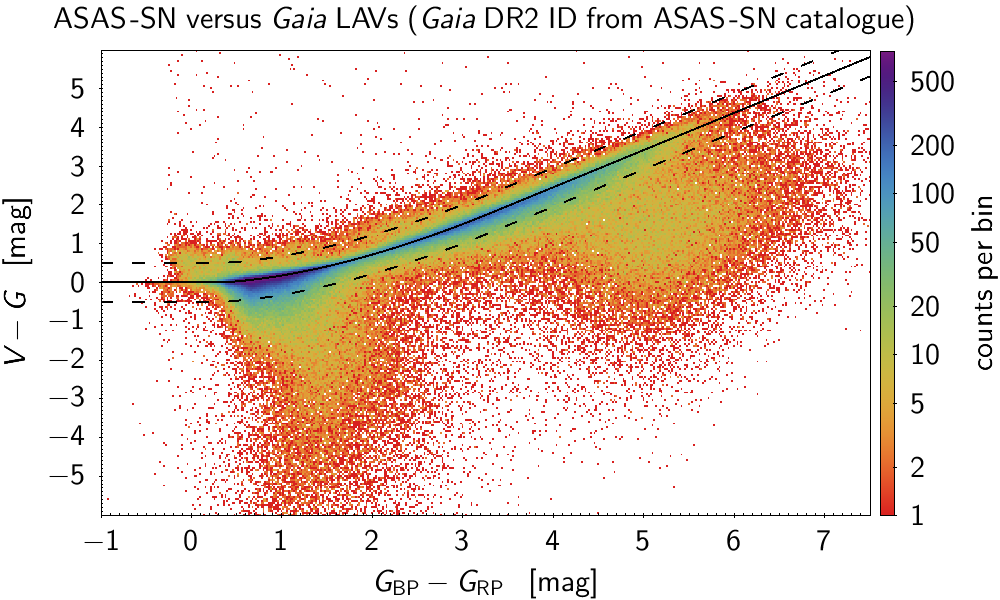}
	\caption{Density map of the difference between mean ASAS-SN $V$ and mean \Gaia \gmag magnitudes versus \Gaia $\gbp - \grp$ colour for all ASAS-SN sources with a \Gaia DR2 ID provided in the ASAS-SN catalogue.
	         A random number between $-5$ and $+5$~mmag has been added to $V$ to smooth out the two-digit precision magnitude published in the ASAS-SN catalogue.
	         This is not necessary for the \Gaia magnitudes, which are reported with four digits in that catalogue.
	         The solid line represents Eq.~\ref{Eq:Vasas_vs_G} as a function of $\gbp - \grp$, and the two dashed lines represent deviations of 0.5 magnitude above and below this function.
	         The axis ranges have been limited for better visibility.
	}
	\label{Fig:VminusGvsBPminusRP}
\end{figure}

The second filter aims at limiting the number of ASAS-SN -- \Gaia mismatched identifications.
Mismatches could result from, among other reasons, the poorer sky resolution of ASAS-SN ($\sim$8") compared to that of \Gaia ($\sim$0.4"). 
We therefore exclude sources that have ASAS-SN $V$ magnitudes potentially incompatible with the \gmag magnitudes of their \Gaia crossmatch candidates.
The compatibility must take into account the two different photometric filter responses.
The distribution of $V - \gmag$ for all ASAS-SN sources with \Gaia DR2 IDs is shown in Fig.~\ref{Fig:VminusGvsBPminusRP} versus $\gbp-\grp$.
The relation
{\small
\begin{align}
  V & = \gmag + \left\{ 
                      \begin{array}{lc}
                         0                             &        \small{\gbp-\grp < 0.25}\\
                         0.29\,(\gbp-\grp-0.25)^{1.63} & 0.25 < \gbp-\grp < 3.25\\
                         0.96\,(\gbp-\grp)-1.3818      &        \gbp-\grp > 3.25
                       \end{array}
                       \right.
\label{Eq:Vasas_vs_G}
\end{align}
}
\!\!is found to describe well the transformation from \gmag to $V$ as a function of $\gbp-\grp$ for the LAVs.
It is also compatible with the $G-V$ relation provided by \cite{EvansRielloDeAngeli_etal18} within their colour validity range. 
We restrict our final ASAS-SN sample to sources that have a maximum deviation of 0.5~mag between the observed ZTF $V$ and the value that would be obtained from $G$ with relation (\ref{Eq:Vasas_vs_G}) (i.e. that are located between the two dashed lines in Fig.~\ref{Fig:VminusGvsBPminusRP}).

\begin{figure}
	\centering
	\includegraphics[width=\linewidth]{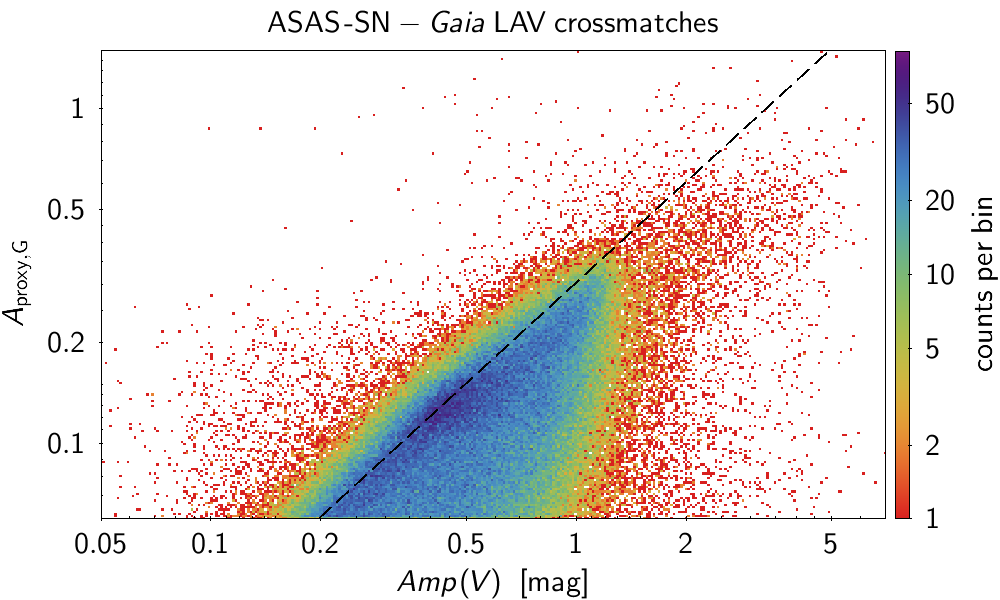}
	\caption{Density map of the \Gaia \gmag amplitude proxy versus ASAS-SN $V$ amplitude of all crossmatches found using a cone search of 4" on the sky and that have $V$ values between the two dashed lines in Fig.~\ref{Fig:VminusGvsBPminusRP}.
	         A random number between $-5$ and $+5$~mmag has been added to the $V$ amplitude to smooth out the two-digit precision numbers published in the ASAS-SN catalogue for this quantities.
	         The dashed line shows the relation $Amp(V)=3.3\,\varProxyG$.
	         The axis ranges have been limited for better visibility.
	}
	\label{Fig:ASAS_varProxyVsAmplV}
\end{figure}

The final `cleaned' sample of sources satisfying both conditions, on $Amp(V)$ and on $V-\gmag$, depends on \Gaia crossmatch identification.
Using the \Gaia DR2 IDs listed in the ASAS-SN catalogue, we get 328\,249 sources, with only 59\,641 of them present in our Dataset~A (see Table~\ref{Tab:completeness_otherSurveys}).
If we perform a sky crossmatch between ASAS-SN and our \Gaia LAVs using a search cone radius of 4", we find 198\,591 crossmatches.
This is about three times more than the above mentioned number of sources with a \Gaia DR2 IDs in the ASAS-SN catalogue for this sample.
The condition $Amp(V)>0.2$~mag is not at the origin of this discrepancy since we get similar conclusions with $Amp(V)>0.4$~mag (see Table~\ref{Tab:completeness_otherSurveys}).
The $V$ variability amplitudes in the final samples are also globally compatible with the values of \varProxyG of their \Gaia crossmatched counterparts, as shown in Fig.~\ref{Fig:ASAS_varProxyVsAmplV} for the sample using sky crossmatches (a similar diagram is obtained using the ASAS-SN crossmatches).
We thus do not know the reason for the discrepancy between the number of crossmatches reported in the ASAS-SN catalogue and the number found with a direct sky crossmatch.
Therefore, we cannot use the sample of ASAS-SN LAVs to estimate the completeness of Dataset~A.
However, we can check the fraction of crossmatches that remain from Dataset~A to B, and from B to C.
It amounts to $\sim$90\% from Dataset~A to B, and to $\sim$75\% from Dataset~B to C, irrespective of the initial sample (among the four cases listed in Table~\ref{Tab:completeness_otherSurveys}).
This shows that the majority of ASAS-SN LAVs that are present in Database~A have relatively good \Gaia multi-band photometry.

\subsubsection{Completeness estimate based on ZTF survey}
\label{Sect:catalogue_completeness_ZTF}

\begin{figure}
	\centering
	\includegraphics[width=\linewidth]{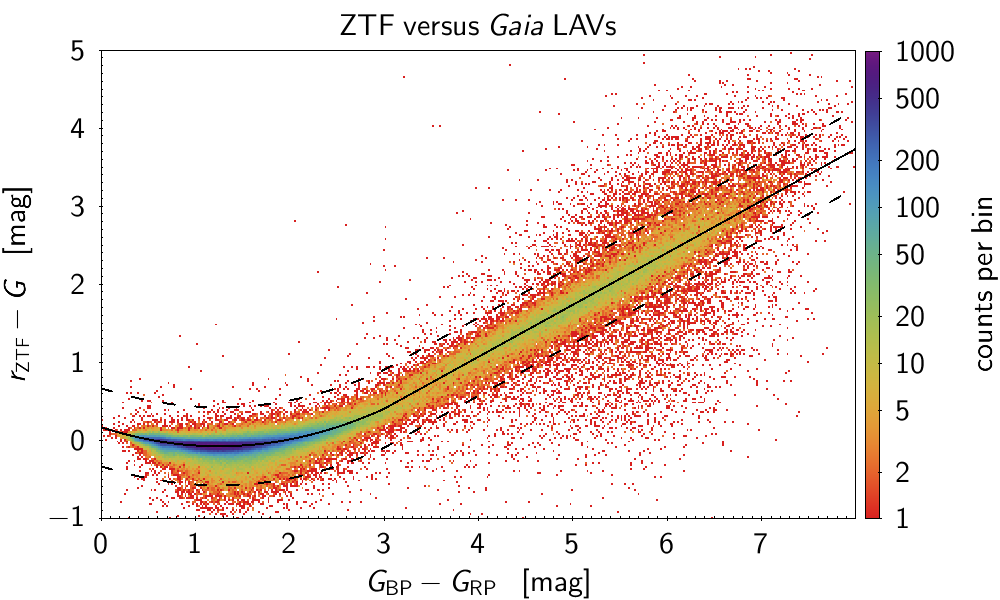}
	\caption{Same as Fig.~\ref{Fig:VminusGvsBPminusRP}, but for ZTF $r_\mathrm{ZTF}$ magnitudes of all ZTF sources crossmatched with \Gaia LAVs of Dataset~A using a 2" cone search on the sky.
	         The solid line represents Eq.~\ref{Eq:rZTF_vs_G} as a function of $\gbp - \grp$, and the two dashed lines represent deviations of 0.5 magnitude above and below this function.
	}
	\label{Fig:rminusGvsBPminusRP}
\end{figure}

We consider the ZTF catalogue of periodic variables published by \citet{ChenWangDeng_etal20} that contains 781\,602 sources.
We follow the same procedure as is done for ASAS-SN in Sect.~\ref{Sect:catalogue_completeness_ASASSN}, here applied to the $r_\mathrm{ZTF}$-band photometry of ZTF.
We first restrict the ZTF sample to sources with $Amp(r_\mathrm{ZTF})>0.2$~mag.
This selects 484\,306 sources, of which 67\% (324\,646 sources) have a 2" sky crossmatch with our Dataset~A (the numbers are summarized in Table~\ref{Tab:completeness_otherSurveys}).
We then adopt the following relation to convert from \gmag to $r_\mathrm{ZTF}$ for our LAVs:
{\small
\begin{align}
  r_\mathrm{ZTF}  & = \gmag + \left\{ 
                      \begin{array}{lc}
                         -0.08 + 0.15673\,(\gbp-\grp-1.25)^2 & \gbp-\grp < 3\\
                         (2/3)\,(\gbp-\grp) - 1.6      &  \gbp-\grp > 3
                       \end{array}
                       \right.
\label{Eq:rZTF_vs_G}
\end{align}
}
\!\!which is shown by the solid line in Fig.~\ref{Fig:rminusGvsBPminusRP}.
The dispersion of $r_\mathrm{ZTF}-\gmag$ in the ZTF sample around the fiducial relation (\ref{Eq:rZTF_vs_G}) is much smaller than it was the case for ASAS-SN (compare Figs.~\ref{Fig:VminusGvsBPminusRP} and \ref{Fig:rminusGvsBPminusRP}).
We still keep a filtering condition of a maximum of 0.5~mag dispersion of $r_\mathrm{ZTF}$ with respect to $r_\mathrm{ZTF}(\gmag)$ for ZTF sources, which leads to a final ZTF sample of 311\,726 sources.
They are all by construction in Dataset~A.
225\,458 of them are present in Dataset~B and 146\,746 sources are in Dataset~C.
These reduction factors of 72\% from Dataset~A to B and of 65\% from Dataset B to C are comparable to the factors obtained in Sect.~\ref{Sect:catalogue_completeness_DR2} using \Gaia DR2 variables.

\begin{figure}
	\centering
	\includegraphics[width=\linewidth]{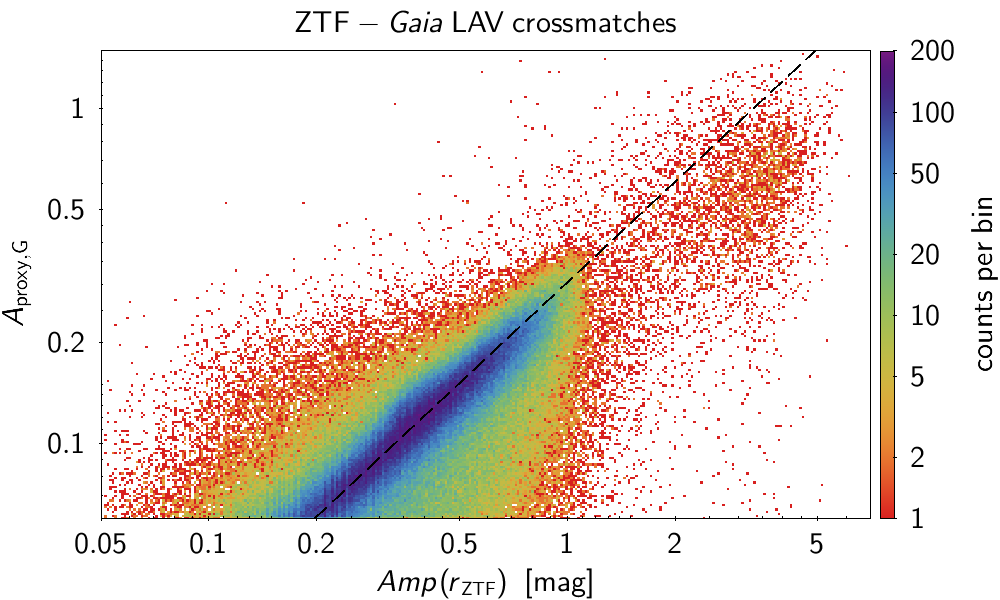}
	\caption{Same as Fig.~\ref{Fig:ASAS_varProxyVsAmplV}, but for \Gaia \gmag amplitude proxy versus ZTF $r_\mathrm{ZTF}$ amplitude of all crossmatches found using a cone search of 2" on the sky and that have $r_\mathrm{ZTF}$ values between the two dashed lines in Fig.~\ref{Fig:rminusGvsBPminusRP}.
	}
	\label{Fig:ZTF_varProxyVsAmplr}
\end{figure}

The sky resolution of ZTF is very good, with 60\% of the 324\,646 ZTF -- Dataset~A crossmatches having an angular separation less than 0.1", and 95\% less than 0.2".
Given the magnitude depth of ZTF ($r_\mathrm{ZTF}$ up to 21~mag), we may expect that all ZTF sources are present in \Gaia.
The $r_\mathrm{ZTF}$ amplitude from ZTF is compared to \varProxyG in Fig.~\ref{Fig:ZTF_varProxyVsAmplr}.
It confirms the $\varProxyG = 3.3 \; Amp(r_\mathrm{ZTF})$ relation established in Sect.~\ref{Sect:varProxy_range} (Eq.~\ref{Eq:rangeGvsVarProxyG}).
It also confirms an inevitable dispersion around this relation due to survey properties (such as time sampling and photometric precision) and stellar variability and spectral properties (such as photometric filter responses).
In particular, a non-negligible fraction of ZTF sources with $Amp(r_\mathrm{ZTF})>0.2$~mag have $\varProxyG<0.06$ and are missed in Dataset~A (see Fig.~\ref{Fig:ZTF_varProxyVsAmplr}).
This observation enables us to estimate a completeness factor of 67\% of Dataset~A relative to the ZTF catalogue of periodic variables.

\subsection{Purity}
\label{Sect:catalogue_purity}

We address the  question of the purity of Datasets~A, B, and C in two different ways.
The first method is based on the consistency of variability amplitudes in the three \Gaia photometric bands (Sect.~\ref{Sect:catalogue_purity_multiband}).
The second method makes use of magnitude distributions (Sect.~\ref{Sect:catalogue_purity_magnitude}).

\subsubsection{Purity estimate based on multi-band variability}
\label{Sect:catalogue_purity_multiband}

\begin{figure}
	\centering
	\includegraphics[trim={0 81pt 0 0},clip,width=\linewidth]{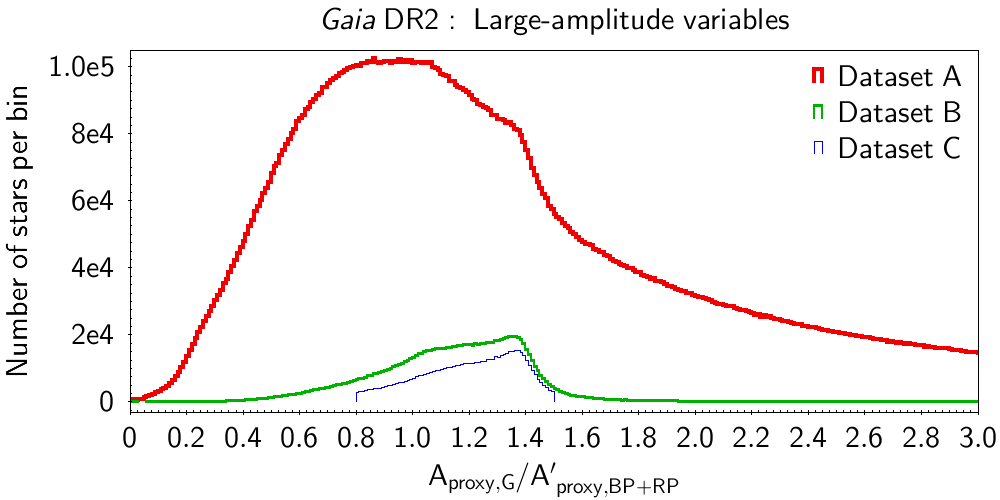}
	\vskip -0.5mm
	\includegraphics[trim={0 0 0 50pt},clip,width=\linewidth]{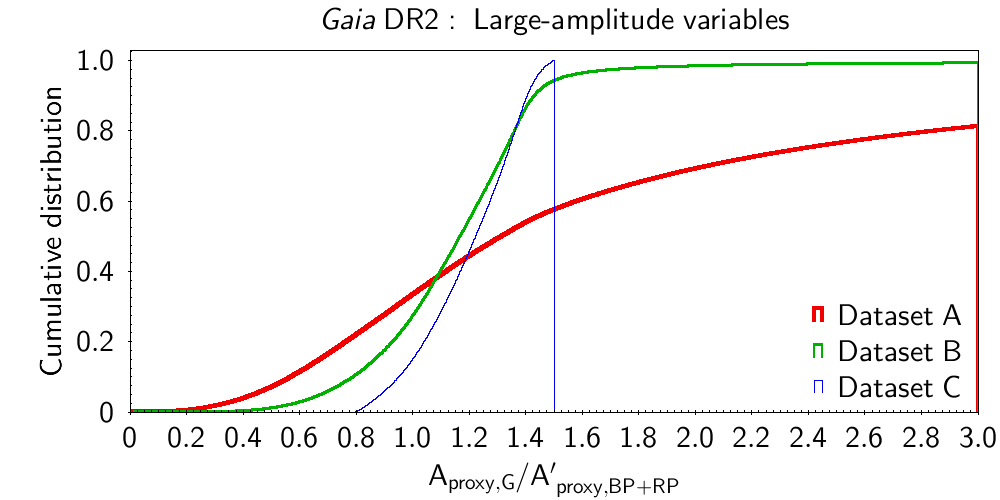}
	\caption{Distribution of $\varProxyG/\varProxyBPplusRPnullCov$ in Datasets~A (thick red line), B (green line) and C (thin blue line), displayed as histograms in the top panel and as cumulative histograms in the bottom panel.
	}
	\label{Fig:histo_varProxyGoverVarProxyBPRP}
\end{figure}

\begin{figure}
	\centering
	\includegraphics[trim={0 81pt 0 0},clip,width=\linewidth]{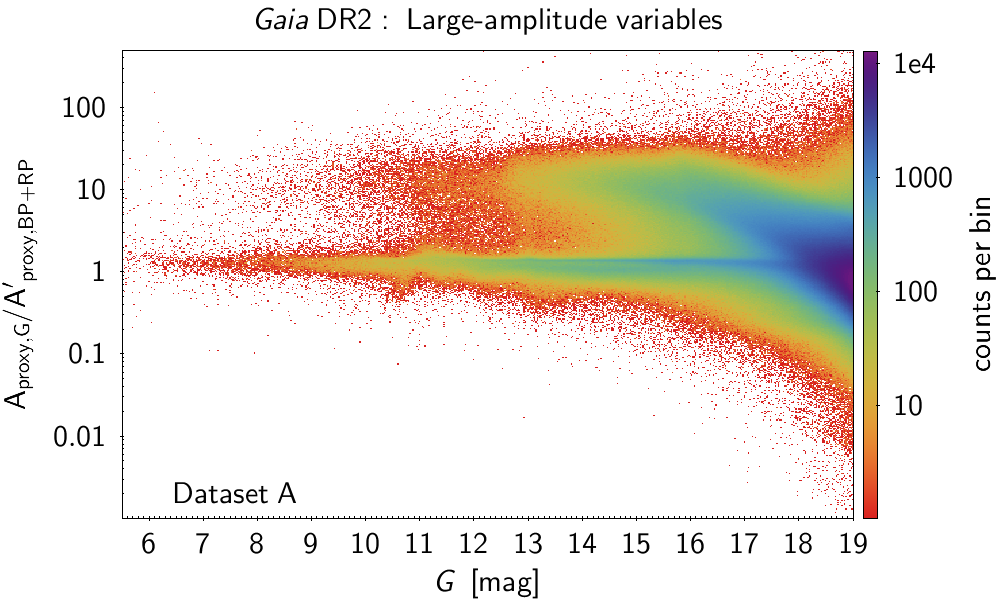}
	\vskip -0.5mm
	\includegraphics[trim={0 0 0 50pt},clip,width=\linewidth]{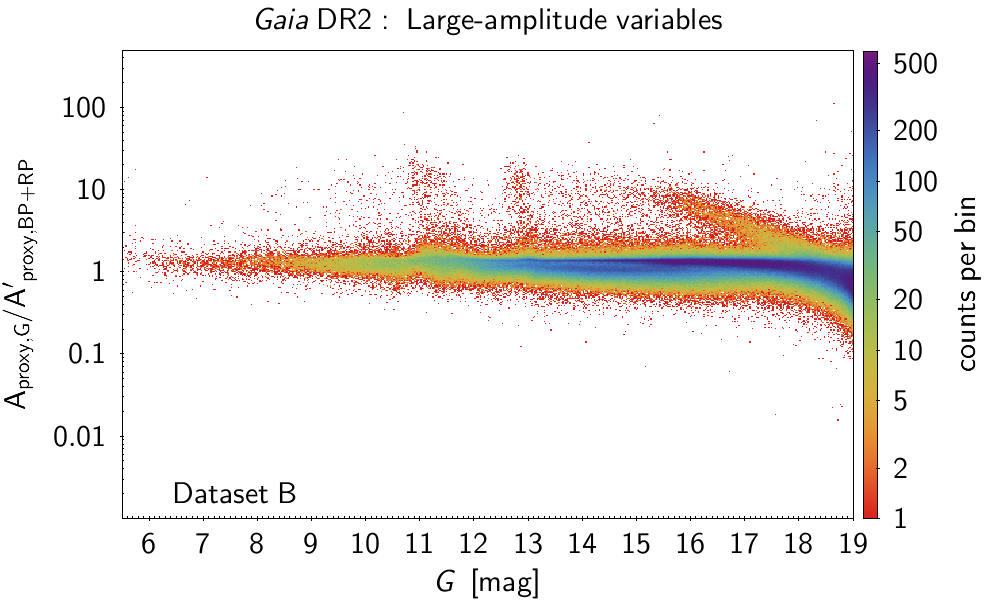}
	\caption{Density map of the ratio $\varProxyG/\varProxyBPplusRPnullCov$ versus \gmag for Dataset~A (top panel) and B (bottom panel).
	         The abscissa range has been limited for better visibility.
	         The ordinate scales are kept identical in the two panels.
	}
	\label{Fig:varProxyGOverBPplusRPvsG_DatasetsABC}
\end{figure}

Any variability detected in \gmag should be present in \bandBPplusRP.
Therefore, we can estimate an upper limit for the purity level by checking the consistency between \gmag-band amplitude (\varProxyG) and combined \bandBPplusRP-band amplitude (\varProxyBPplusRPnullCov).
The analysis presented in Appendix~\ref{Appendix:AproxyBPplusRP} concludes that $\varProxyG/\varProxyBPplusRPnullCov$ should lie between $\sim$1 and $\sim$1.5, the exact value depending on variability type.
However, the histograms of $\varProxyG/\varProxyBPplusRPnullCov$ for the three datasets, shown in Fig.~\ref{Fig:histo_varProxyGoverVarProxyBPRP} (top panel), reveals a large fraction of sources with ratios outside this range.
This is especially true for Dataset~A.

The first case to consider is $\varProxyG < \varProxyBPplusRPnullCov$, that is when the variability amplitude detected in \gmag is smaller than the one detected in \bandBPplusRP.
The cumulative histogram of $\varProxyG/\varProxyBPplusRPnullCov$, displayed in the bottom panel of Fig.~\ref{Fig:histo_varProxyGoverVarProxyBPRP}, shows that about one quarter of LAV candidates in Datasets~A and B have $\varProxyG<\varProxyBPplusRPnullCov$, and 14\% in Dataset~C.
This can be the case if, for example, the noise in \gbp and \grp is larger than the noise in \gmag due to, among other reasons, increased residual astrophysical background, fewer CCD transits for \gbp and \grp than for \gmag, or blending effects in \bandBP and \bandRP spectra.
Figure~\ref{Fig:varProxyGOverBPplusRPvsG_DatasetsABC}, which displays $\varProxyG/\varProxyBPplusRPnullCov$ versus \gmag for both Dataset~A (top panel) and Dataset~B (bottom panel), tends to support this latter explanation, as the number of cases with $\varProxyG/\varProxyBPplusRPnullCov<1$ increases with increasing magnitude, especially for Dataset~A.
We therefore cannot, in general, use the criterion $\varProxyG < \varProxyBPplusRPnullCov$ to identify spurious \varProxyG values.
Rather, it would point to an overestimation of \varProxyBPplusRPnullCov, and hence to unreliable \varProxyBP and/or \varProxyRP values.
A condition based on this conclusion, but using the less restrictive condition $\varProxyG/\varProxyBPplusRPnullCov<0.8$, was actually used to filter out such cases in Dataset~C (filter c2 in Table~\ref{Tab:datasetsSummary}).

In the second case, when $\varProxyG > 1.5\;\varProxyBPplusRPnullCov$, the amplitude is unexpectedly larger in the \gmag band than in the \bandBPplusRP band.
This would point to a spurious value of \varProxyG.
It represents 33\% of sources in Dataset~A, but only 6\% in Dataset~B (bottom panel of Fig.~\ref{Fig:histo_varProxyGoverVarProxyBPRP}).
These sources were removed in Dataset~C (filter c1 in Table~\ref{Tab:datasetsSummary})-

If we were to consider that \varProxyG is reliable if $1.0 \!<\! \varProxyG/\varProxyBPplusRPnullCov \!<\! 1.5$ (the first condition being restrictive if interpreted as being due to the unreliability of \gmag rather than of \gbp and/or \grp, see above), we would conclude from the above estimates that the purity level with respect to \varProxyG could be around 40\% for Dataset~A, 70\% for Dataset~B, and 85\% for Dataset~C.
These numbers, however, must be taken with caution.

\subsubsection{Purity estimate based on magnitude distribution}
\label{Sect:catalogue_purity_magnitude}

\begin{figure}
	\centering
	\includegraphics[width=\linewidth]{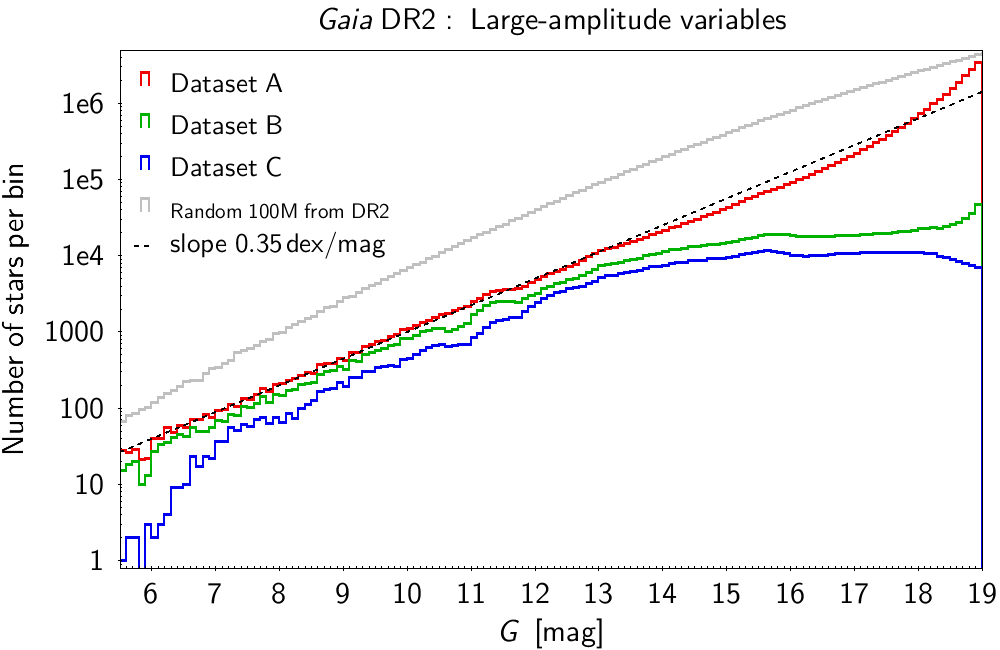}
	\caption{\gmag magnitude histograms for, from top to bottom lines, a sample of 100 million sources randomly taken from \Gaia DR2 (in grey), Dataset~A (in red), Dataset~B (in green), and Dataset~C (in blue).
	         Bins are 0.1~mag wide.
	         On the log scale of the ordinate, a dotted straight line with slope 0.35~dex/mag is adjusted to the histogram of Dataset~A.
	}
	\label{Fig:histo_magG}
\end{figure}

The \gmag magnitude distributions of the three datasets are shown in Fig.~\ref{Fig:histo_magG}.
The magnitude distribution of Dataset~A (red line) basically follows an exponential increase as a function of magnitude up to $\gmag \simeq 13$~mag (a power-ten function with a slope of $0.35\,$dex/mag is plotted in dotted line in the figure).
Above that magnitude, the slope slightly decreases up to $\gmag\simeq 16$~mag, before strongly increasing at magnitudes above $\sim$17~mag.
For comparison, the magnitude distribution of 100 million sources randomly selected from \Gaia DR2 is shown in grey in Fig.~\ref{Fig:histo_magG}.
It reveals a continuous decrease in the dex/mag slope as a function of magnitude.
If we assume a spatial distribution of LAVs in the Galaxy similar to that of all stars, the comparison of magnitude distributions of Dataset~A with the random sample suggests the presence of a significant fraction of contaminants in Dataset~A at magnitudes above $\sim$17~mag.
The number of faint contaminants is much reduced in Dataset~B, whose magnitude distribution flattens above $\sim$16~mag (green line in Fig.~\ref{Fig:histo_magG}).
Yet, the increase observed at $\gmag \gtrsim 18$~mag still indicates the presence of contaminants at the faintest end of this sample.
This is no longer the case for Dataset~C, whose magnitude distribution even decreases above $\sim$18~mag.

In conclusion, comparison of the magnitude distributions of the three datasets with that of a random \Gaia DR2 sample suggests the presence of a non-negligible fraction of contaminants at the faint side ($\sim$17~mag) in Dataset~A.
It supports a similar conclusion obtained in Sect.~\ref{Sect:catalogue_purity_multiband}, and confirms the higher purity levels estimated for Datasets~B and C.
Contaminants are still expected to pollute Dataset~B at magnitudes fainter than $\sim$18~mag, while Dataset~C is the purest of the three datasets.

\section{Catalogue exploration: Two examples}
\label{Sect:exploration}

We provide in this section two examples that illustrate the content of the catalogue and its usage.
The first case investigates the exploitation of multi-band variability amplitudes to disentangle and study different types of variable stars (Sect.~\ref{Sect:exploration_multiband}).
Section~\ref{Sect:exploration_goodparallaxes} then presents the sample of LAVs with parallax uncertainties better than 10\%.

\subsection{Multi-band variability studies}
\label{Sect:exploration_multiband}

\begin{figure*}
	\centering
	\includegraphics[trim={0 67 0 0},clip,width=\linewidth]{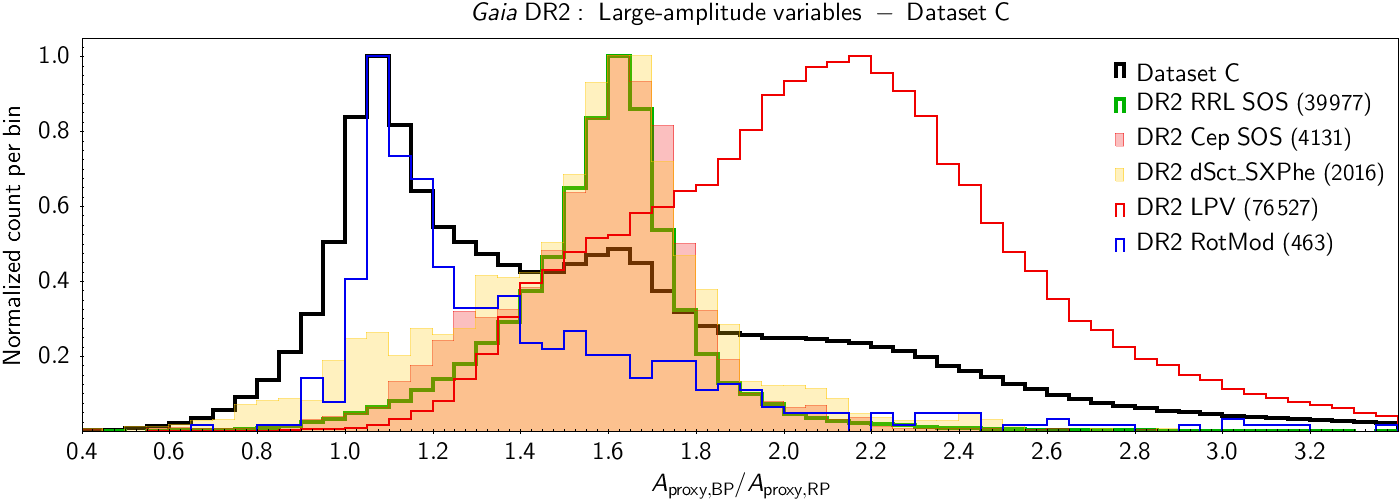}
	\vskip -0.5mm
	\includegraphics[trim={0 67 0 38},clip,width=\linewidth]{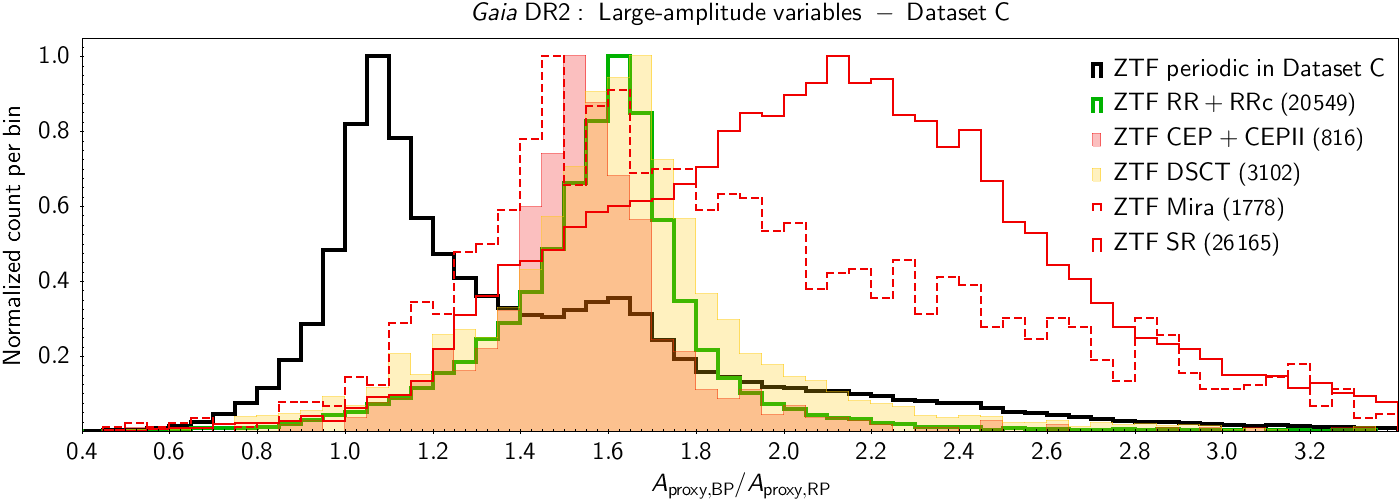}
	\vskip -0.5mm
	\includegraphics[trim={0 67 0 38},clip,width=\linewidth]{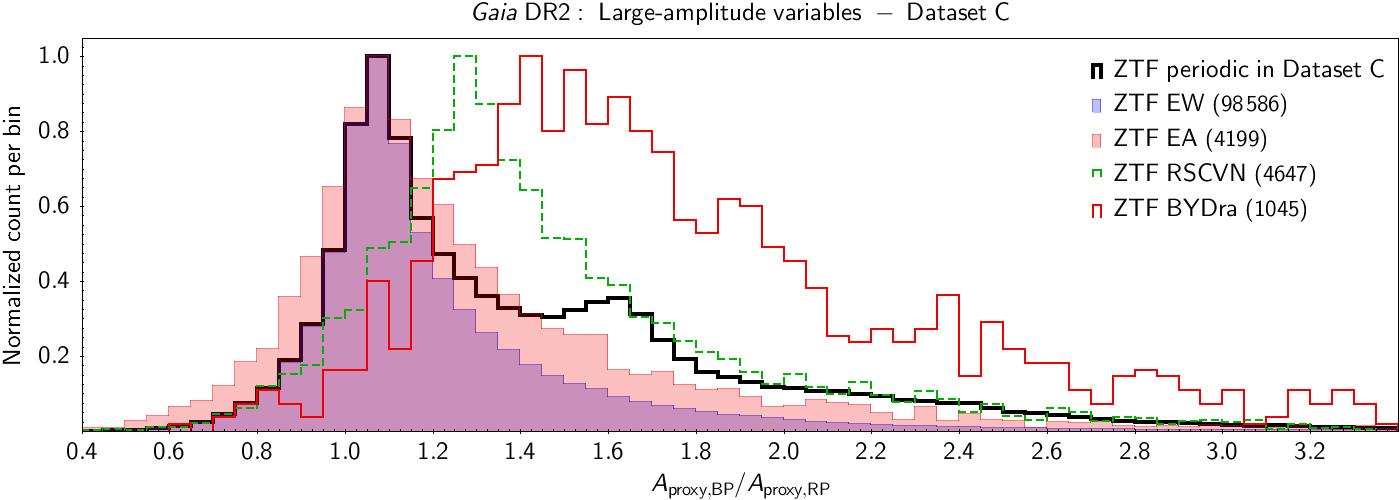}
	\vskip -0.5mm
	\includegraphics[trim={0 0 0 38},clip,width=\linewidth]{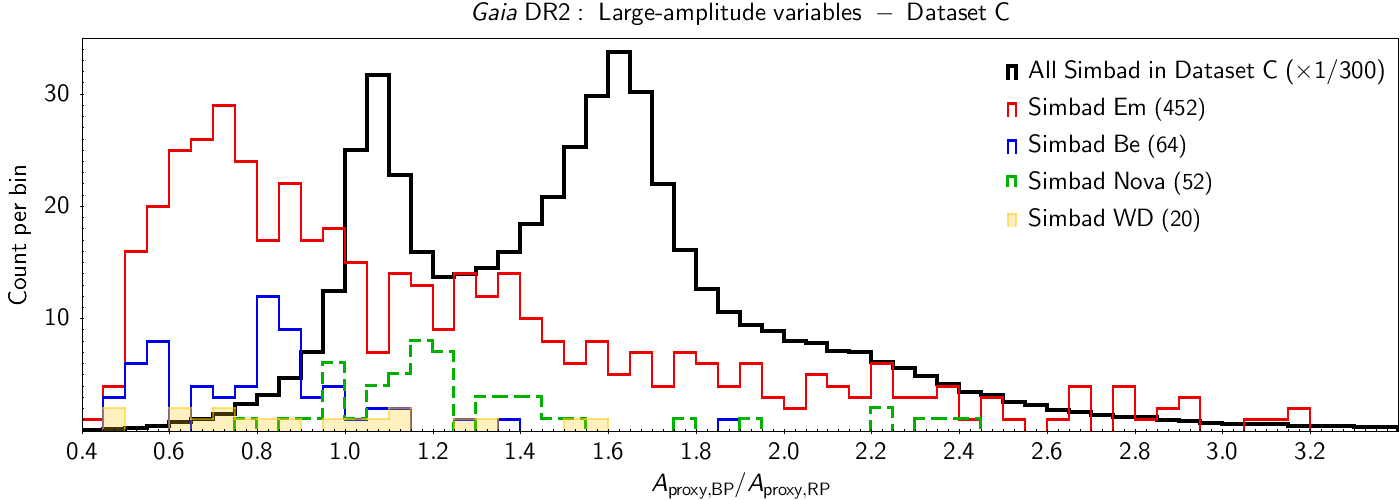}
	\caption{Histograms of $\varProxyBP/\varProxyRP$ for various types of variable candidates identified in the catalogues of \Gaia DR2 variables (top panel),  in the ZTF catalogue of periodic variables (middle panels) and in Simbad (bottom panel; `Em' are emission-line stars).
	         Only crossmatches with Dataset~C are considered.
	         The variability type of each histogram is indicated in the upper-right corner of each panel, with the number of crossmatches available in Dataset~C indicated in parenthesis next to the variability type.
	         The thick black line in each panel represents the histogram of the full sample of crossmatches in Dataset~C of the relevant catalogue.
	         The histograms are normalized to maximum count in the top three panels, and the actual counts per bin in the bottom panel.
	         The counts in the histogram of the full sample of Simbad crossmatches (thick black line in the bottom panel) have been divided by 300 for better visibility.
	         The abscissa range has been limited for better visibility.
	}
	\label{Fig:histo_varProxyBPoverRP_literature}
\end{figure*}

\begin{figure}
	\centering
	\includegraphics[trim={0 81 0 0},clip,width=\linewidth]{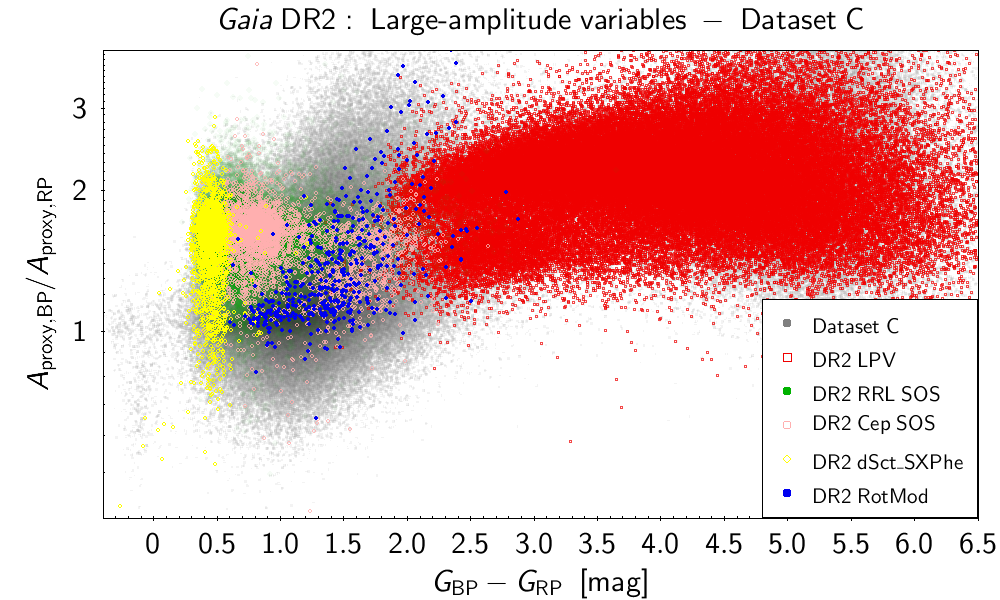}
	\vskip -0.5mm
	\includegraphics[trim={0 81 0 50},clip,width=\linewidth]{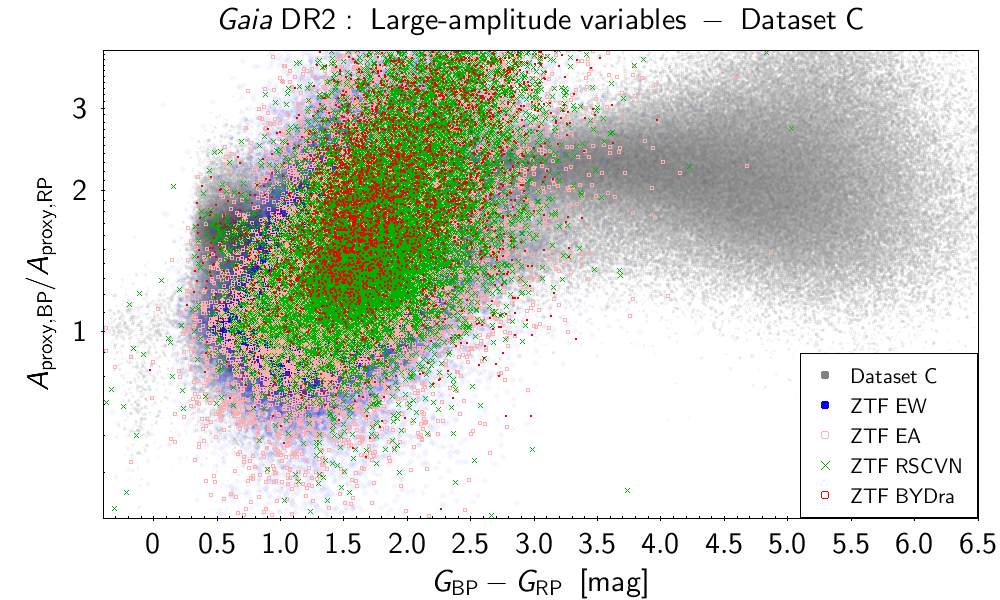}
	\vskip -0.5mm
	\includegraphics[trim={0 0 0 50},clip,width=\linewidth]{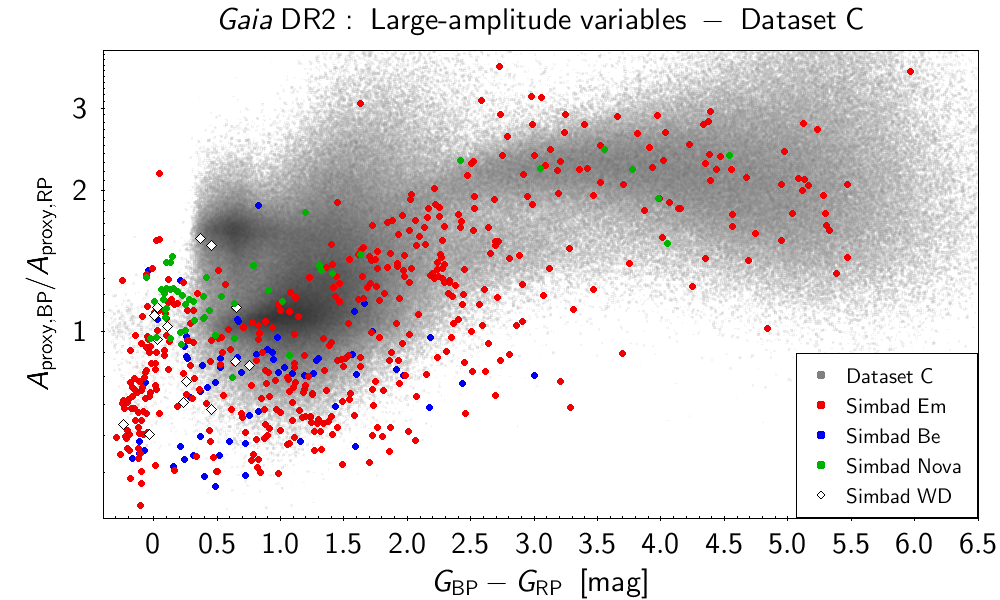}
	\caption{Distribution in the $\varProxyBP/\varProxyRP$ versus \BPminusRP diagram of Dataset~C LAVs crossmatched with some of the \Gaia DR2 catalogues of variables (top panel), with the sample of non-pulsating variables from the ZTF catalogue of periodic variables (middle panel), and with a selection of variability types crossmatched in the Simbad databse.
	         The colours of the symbols used to represent them are the same as the ones in Fig.~\ref{Fig:histo_varProxyBPoverRP_literature}.
	         The background grey points represent the full Dataset~C shown in Fig.~\ref{Fig:varProxyBPRPvsBPmRP_DatasetC}.
	         The axes ranges have been limited for better visibility.
	}
	\label{Fig:varProxyGvsBPminusRP_DatasetC}
\end{figure}

The availability of quasi-simultaneous photometric measurements in three bands confers to the \Gaia mission an invaluable advantage for variability studies.
This is obviously the case when analyzing light curves.
However, it is also an advantage for studies using variability proxies because quasi-simultaneous observations lead to consistent variability amplitude proxies in the different bands.
Multi-band measurements should be taken within time intervals that are short compared to the expected variability time-scale if compatible amplitudes are required in the different bands.
And this time-scale can be short for such variables as EA-type eclipsing binaries with short-duration deep eclipses, flare stars, or transient objects, to cite only a few types.
Therefore, in general, quasi-simultaneous photometric measurements ensure coherent variability proxies in different bands.
In \Gaia, simultaneous photometry within less than one minute is ensured in \gmag, \gbp and \grp.
Dataset~C has been defined with conditions enforcing, as much as possible, similar epoch measurements in these three bands for a given source, and is therefore best suited for multi-band variability analyses.
We therefore restrict our study in this section to Dataset~C.

We introduced two diagrams in Sect.~\ref{Sect:catalogue_datasets_multiband} that evidenced the dependence of the blue-to-red amplitude ratio ($\varProxyBP / \varProxyRP$) on stellar variability type (Figs.~\ref{Fig:varProxyRatiosBPGvsRPG_DatasetC} and \ref{Fig:varProxyBPRPvsBPmRP_DatasetC}), with at least three categories of variables highlighted in the figures.
Here, we further investigate this property based on literature data (Sect.~\ref{Sect:exploration_multiband_literature}), and identify four broad groups of variables from their multi-band variability properties (Sect.~\ref{Sect:exploration_multiband_groups}).

\subsubsection{$\varProxyBP/\varProxyRP$ ratios of known variability types}
\label{Sect:exploration_multiband_literature}

We identify the variability types of the LAVs in Dataset~C from three sources in the literature: the \Gaia DR2 catalogues of variables (already used in Sect.~\ref{Sect:catalogue_completeness_DR2}), the ZTF catalogue of periodic variables (already used in Sect.~\ref{Sect:catalogue_completeness_ZTF}), and the Simbad database \citep{WengertOchsenbeinEgret_etal00} from which crossmatches with Dataset~C are extracted using a 2" cone search on the sky.
The $\varProxyBP / \varProxyRP$ distributions of the crossmatches are shown in Fig.~\ref{Fig:histo_varProxyBPoverRP_literature}, in the top panel for \Gaia DR2 variables, in the two middle panels for the ZTF variables, and in the bottom panel for variables extracted from Simbad among a selection of variability types.
The histograms are plotted normalized to the maximum count-per-bin for better visibility in the top three panels, but kept as total count per bin for the Simbad crossmatches in the bottom panel due to the small number of sources per variability type.

Several categories of variable stars are identified from the distributions shown in Fig.~\ref{Fig:histo_varProxyBPoverRP_literature}.
First, the \Gaia DR2 samples of RR Lyrae variables (green histogram in the top panel of Fig.~\ref{Fig:histo_varProxyBPoverRP_literature}), Cepheids (filled pink), and $\delta$~Scuti/SX~Phoenicis stars (filled yellow) are seen to be distributed around $\varProxyBP / \varProxyRP \simeq 1.6$.
      These distributions are confirmed by the ZTF samples of the same variability types, shown in the second panel from top.
      They represent pulsating stars in the classical instability strip (which we denote hereafter as `classical pulsators').
      Their \BPminusRP colours and $\varProxyBP / \varProxyRP$ ratios are shown in the top panel of Fig.~\ref{Fig:varProxyGvsBPminusRP_DatasetC}.

Second, the distribution of the \Gaia DR2 sample of LPVs, shown in red in the top panel of Fig.~\ref{Fig:histo_varProxyBPoverRP_literature}, are seen to pulsate with amplitudes about twice larger in the blue than in the red, with a peak of the distribution at $\varProxyBP / \varProxyRP \simeq 2.1$.
      However, a relatively large dispersion around this peak value is observed, extending from $\varProxyBP / \varProxyRP \simeq 1.2$ to above 3.
      The ZTF sample shown in the second panel from top distinguishes between semi-regular variables (SRVs) and Miras.
      Interestingly, Miras (dashed red histogram) are seen to peak at blue-to-red amplitude ratios similar to those of classical pulsators, between 1.4 and 1.7, while SRVs (solid red line) have, on the mean, ratios above 1.8.
      While the $\varProxyBP / \varProxyRP$ ratios of the LPVs overlap with those of classical pulsators at ratios below $\sim$1.8, they are nevertheless easily identified from their red colours, as seen in Fig.~\ref{Fig:varProxyGvsBPminusRP_DatasetC} (top panel).

Third, some variability types are approximately achromatic with $\varProxyBP \simeq \varProxyRP$.
      In the ZTF sample of periodic variables, EW-type eclipsing binaries are among the most abundant ones (filled blue histogram in the third panel from top in Fig.~\ref{Fig:histo_varProxyBPoverRP_literature}).
      Their $\varProxyBP / \varProxyRP$ histogram peaks at  $\simeq$1.07.
      EA-type eclipsing binaries (filled pink histogram) follow a distribution similar to the EW eclipsing binaries, though with a slightly wider dispersion around the peak value.
      These patterns are compatible with the type of binaries they represent, EA-type binaries consisting of detached systems where the two stars keep their individual characteristics in the majority of cases, while EW-type systems share a common envelope around the two stars.
      Other variable stars also display achromatic variability.
      The $\varProxyBP / \varProxyRP$ histogram of the rotation-modulation LAV candidates published in \Gaia DR2, shown in the top panel of Fig.~\ref{Fig:histo_varProxyBPoverRP_literature} (blue histogram), also peaks at $\varProxyBP / \varProxyRP \simeq 1.1$ (we must however keep in mind that a fraction of these DR2 rotation-modulation candidates are misclassified, especially the ones considered in this paper, see Sect.~\ref{Sect:catalogue_completeness}).
      Achromatic variables span a wide range of \BPminusRP colours, typically from $\sim$0.2 to $\sim$3 mag as seen in the middle panel of Fig.~\ref{Fig:varProxyGvsBPminusRP_DatasetC} for the ZTF variables and in the top panel for the DR2 rotation-modulation candidates.

Two other non-pulsating types of LAVs from the ZTF sample are shown in the third panel from top in Fig.~\ref{Fig:histo_varProxyBPoverRP_literature}.
      They are the RS Canum Venaticorum (RS~CVn) and BY Draconis (BY~Dra) variables.
      The $\varProxyBP / \varProxyRP$ distribution of the RS~CVn candidates peaks at values between 1.25 and 1.3 (dotted green histogram).
      Interestingly, the peak of the distribution is relatively well defined, and its width not much larger than the widths observed for the classical pulsators shown in the top two panels. 
      RS~CVn variables are close binary systems with chromospheric activity and large spots on the stellar surface.
      The BY~Dra candidates in the ZTF sample, on the other hand, show a much wider $\varProxyBP / \varProxyRP$ distribution, between $\sim$1.2 and $\sim$1.9 (red histogram in the third panel from top), without a clear peak value.
      These variables also have active chromospheres, but are single stars, typically K- and M-type MS stars.
      The location of RS~CVn and BY~Dra variables in the $\varProxyBP / \varProxyRP$ versus \BPminusRP diagram is shown in the middle panel of Fig.~\ref{Fig:varProxyGvsBPminusRP_DatasetC}.

Finally, Fig.~\ref{Fig:varProxyBPRPvsBPmRP_DatasetC} reveals the presence of a population of blue stars ($\BPminusRP \lesssim 0.2$~mag) with variability amplitudes larger at long than at short wavelengths ($\varProxyBP / \varProxyRP<1$).
      No such specific class is found in the samples of \Gaia DR2 variables or in the ZTF periodic variables.
      We therefore browsed the Simbad database for crossmatches in Dataset~C that have $\varProxyBP / \varProxyRP<1$, and report in the bottom panel of Fig.~\ref{Fig:histo_varProxyBPoverRP_literature} some of the variability classes to which they belong.
      The most numerous class consists of emission-line stars, the histogram of which is shown in red (labelled `Simbad Em') in the bottom panel of Fig.~\ref{Fig:histo_varProxyBPoverRP_literature}.
      Be stars (blue histogram) form another class with amplitudes larger in the red than in the blue.
      They are also a type of emission-line stars.
      Some white-dwarf (WD) variables also have $\varProxyBP / \varProxyRP<1$ (filled yellow histogram), though their number statistics is very small (only 20 crossmatches found) and their $\varProxyBP / \varProxyRP$ distribution extends up to 1.6.
      Novae do not necessarily have $\varProxyBP / \varProxyRP<1$, though some do (dashed green histogram).
      The small number statistics of these Simbad crossmatchs prevents, however, to draw firm conclusions.
      Further insight into this group of $\varProxyBP / \varProxyRP<1$ variables will be provided from the analysis of the sample with good parallaxes in Sect.~\ref{Sect:exploration_goodparallaxes}.
      The colour distribution of the variables discussed here is shown in the bottom panel of Fig.~\ref{Fig:varProxyGvsBPminusRP_DatasetC}.

The above analyses rely on variability types published in the literature, which are, however, affected by uncertainties.
\Gaia DR2 and ZTF identifications result from automated techniques, and the ones in  Simbad have a wide and non-homogeneous origin.
Broad distributions can thus be expected for parameters derived from the analysis of these catalogues.

The summed distribution of $\varProxyBP/\varProxyRP$ in a given survey is very informative of the overall stellar sample.
These distributions are shown with thick black lines in each panel of Fig.~\ref{Fig:histo_varProxyBPoverRP_literature} for each of the Dataset~C, ZTF and Simbad samples.
The distribution of Dataset~C (top panel) is very similar to that of the ZTF sample of periodic variables (middle panels), with a predominance of achromatic variables.
In the ZTF sample, the main peak at $\varProxyBP \simeq \varProxyRP$ is due to eclipsing binaries.
Extrapolating this result to Dataset~C, we could thus expect that most of the quasi-achromatic variables in Dataset~C also consist of eclipsing binaries (this will be confirmed in Sect.~\ref{Sect:exploration_goodparallaxes_groups}).
In contrast, the summed distribution for the sample of Dataset~C \!--\! Simbad crossmatches shows a predominance of $\varProxyBP/\varProxyRP$ around 1.6, typical of classical pulsators (thick black line in the bottom panel of Fig.~\ref{Fig:histo_varProxyBPoverRP_literature}).

\subsubsection{Classification of LAVs using $\varProxyBP/\varProxyRP$}
\label{Sect:exploration_multiband_groups}

Based on the results of the previous section, we schematically categorize LAVs in four groups according, mainly, to their blue-to-red $\varProxyBP/\varProxyRP$ amplitude ratio and their \BPminusRP  colour.
The four groups and their definitions are summarized in Table~\ref{Tab:groups}.

\paragraph{Group 1.}
\label{Sect:exploration_multiband_group1}

We first consider LPVs (Group~1), which are easily identified by their red colours.
We require $\BPminusRP > 1.8$~mag.
This, however, will also include MS red dwarf and PMS stars including young stellar objects (YSOs).
To restrict Group~1 to LPVs (the other types will belong to other groups), we take advantage of their very bright intrinsic luminosities.
At the large variability amplitudes considered in this paper, LPVs mainly consist of red giants on the Asymptotic Giant Branch (AGB) (as well as the even brighter, though statistically less numerous, red supergiants).
Their typical absolute \gmag magnitudes (\absMeanG) lie between $-3$ and 0~mag.
MS red dwarfs, on the other hand, are much fainter, with typical brightnesses of $\absMeanG \!\simeq\! 8$~mag at $\BPminusRP = 2$~mag, and up to $\absMeanG \!\simeq\! 11$~mag at $\BPminusRP = 3$~mag.
They are thus of the order of ten magnitudes fainter than typical LPVs.
At brightnesses between these two extremes, we find PMS stars, usually still several magnitudes fainter than LPVs.

\begin{table}
\caption{Schematic categorization of variables using colour and wavelength-dependent variability (Figs.~\ref{Fig:varProxyRatiosBPGvsRPG_DatasetC} and \ref{Fig:varProxyBPRPvsBPmRP_DatasetC}).
In each group, the listed conditions must all be satisfied (\texttt{`AND'} operator).
        }
\centering
\begin{tabular}{l}
\hline\hline
\rule{0pt}{3.0ex} \textbf{Group 1} (mainly LPVs) \\[-5pt]
\rule{0pt}{0ex}\parbox{8cm}{
      \begin{equation}
        \hspace{0.3cm}
        \left\{
        \begin{array}{l}
          \BPminusRP >  1.8 \\
          \varpi     <  0.12+\exp{\Bigl[10 + 3\,(\BPminusRP) - 1.5 \,\gmag \Bigr]}
        \end{array}
        \right. \!
      \label{Eq:conditionsGroup1}
      \end{equation}
    }\\
\hline
\rule{0pt}{3.5ex} \textbf{Group 2} (hot compact LAVs with $\displaystyle{\frac{\varProxyBP}{\varProxyRP}} < 0.9$)  \\[-5pt]
\rule{0pt}{0ex}\parbox{8cm}{
      \begin{equation}
        \hspace{0.3cm}
        \left\{
        \begin{array}{l}
          \BPminusRP <  0.2 \\
          \varProxyBP < 0.9 \;\varProxyRP \\
          \varpi     >  0.12+\exp{\Bigl(19 - 1.5 \,\gmag \Bigr)}
        \end{array}
        \right.
      \label{Eq:conditionsGroup2}
      \end{equation}
    }\\
\hline
\rule{0pt}{3.0ex} \textbf{Group 3} (mainly classical pulsators) \\[-5pt]
\rule{0pt}{0ex}\parbox{8cm}{
      \begin{equation}
        \hspace{0.3cm}
        \left\{
        \begin{array}{l}
          \mathrm{Not~in~Group~1~or~2} \\
          \varProxyBP > 1.4 \; \varProxyRP \\
          \varProxyRP < 0.85 \; \varProxyG
        \end{array}
        \right.
      \label{Eq:conditionsGroup3}
      \end{equation}
    }\\
\hline
\rule{0pt}{3.0ex} \textbf{Group 4} (mainly non-pulsating variables) \\[-5pt]
\rule{0pt}{0ex}\parbox{8cm}{
      \begin{equation}
        \hspace{0.3cm}
          \mathrm{Not~in~Group~1,~2~or~3} \\
      \label{Eq:conditionsGroup4}
      \end{equation}
    }\\
\rule{0pt}{2.0ex} \textbf{Group 4a} (mainly chromatic non-pulsating variables) \\[-5pt]
\rule{0pt}{0ex}\parbox{8cm}{
      \begin{equation}
        \hspace{0.3cm}
        \left\{
        \begin{array}{l}
          \mathrm{In~Group~4} \\
          \varProxyBP > 1.2 \; \varProxyRP  \\
          \varProxyRP < 0.92 \; \varProxyG
        \end{array}
        \right.
      \label{Eq:conditionsGroup4a}
      \end{equation}
    }\\
\hline
\end{tabular}
\label{Tab:groups}
\end{table}

The much larger intrinsic luminosities of LPVs compared to red dwarfs and PMS stars translate into much smaller parallaxes at any given observed magnitude.
To illustrate this, we will consider a $\gmag=15$~mag LPV.
If its absolute \gmag magnitude is $\absMeanG \!\simeq\! 0$~mag, the LPV would have a parallax of $\varpi = 10^{-0.2\,(\gmag-\absMeanG-10)}~\mathrm{mas} = 0.1$~mas ($d=10$~kpc).
A red clump clump star that would be reddened at a colour of $\BPminusRP=3$~mag (typical of not too evolved LPVs as shown in Fig.~\ref{Fig:varProxyBPRPvsBPmRP_DatasetC}) would have $\absMeanG \!\simeq\! 3.5$~mag.
Consequently, its parallax would be of 0.5~mas ($d=2$~kpc) if it were seen with $\gmag=15$~mag.
A MS red dwarf at the considered colour, on the other hand, with $\absMeanG \!\simeq\! 11$~mag, would need to be much closer to have $\gmag=15$~mag, at a parallax of $\varpi \simeq 15.8$~mas ($d=63$~pc).
Therefore, a star at $\gmag = 15$~mag and $\BPminusRP=3$~mag has a high probability to be a LPV if its parallax is smaller than $\sim$0.5~mas.
The upper parallax limit sketched above for a star to be a LPV decreases with increasing magnitude.
It also depends on colour.
An empiric exploration of LAVs in Dataset~C leads to the condition $\varpi \!<\! 0.12+\exp{[10 + 3\,(\BPminusRP) - 1.5 \,\gmag ]}$~mas to identify LPV candidates in the sample of red LAVs (the additive factor of 0.12~mas counts for the typical \Gaia DR2 parallax uncertainty).
The final conditions for Group~1 are given by Eqs.~\ref{Eq:conditionsGroup1} in Table~\ref{Tab:groups}.

\begin{figure}
	\centering
	\includegraphics[width=\linewidth]{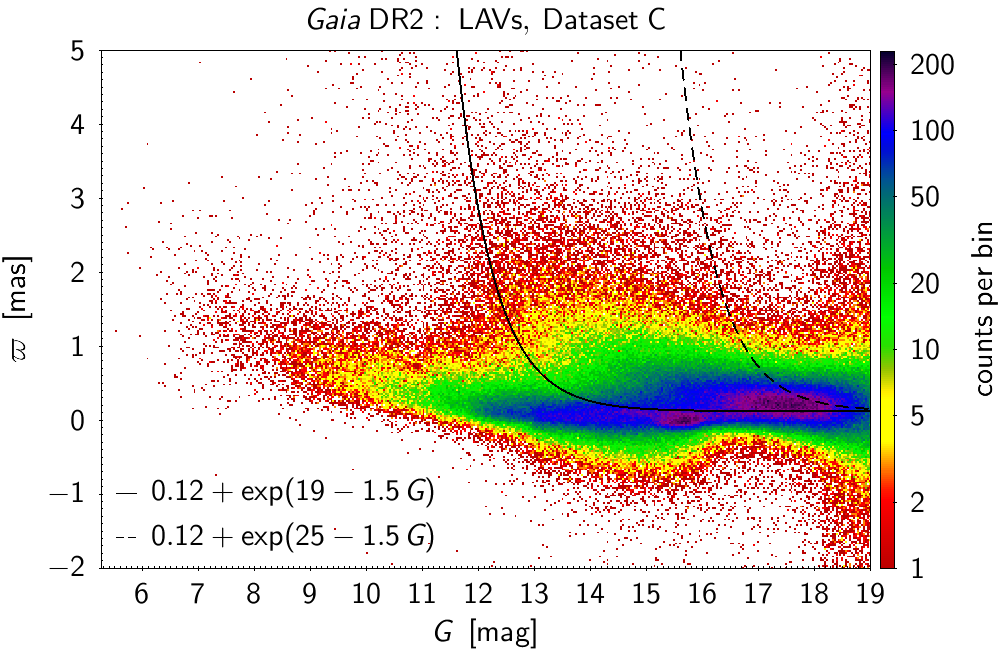}
	\caption{Density map of the parallax versus \gmag for Dataset~C.
	         The lines are examples of parallax limits below which stars are classified as LPVs (for $\BPminusRP = 3$~mag for the solid line and for for $\BPminusRP = 5$~mag for the dashed line).
	         A rainbow colour-code is used for the density map to highlight the location of the densest regions with respect to the solid and dashed lines.
	         The axes ranges have been limited for better visibility.
	}
	\label{Fig:parallaxVsG}
\end{figure}

Conditions (\ref{Eq:conditionsGroup1}) properly select LPV candidates in the sample of Dataset~C LAVs with parallax uncertainties better than 10\% (see Sect.~\ref{Sect:exploration_goodparallaxes}), but also in the full sample of Dataset~C due to their much larger brightnesses compared to other red stars.
Conditions (\ref{Eq:conditionsGroup1}) also correctly select LPVs in the Magellanic Clouds, as seen in Fig.~\ref{Fig:parallaxVsG} where they form the over-density of sources at $\varpi \simeq 0$~mas and $15 \!\lesssim\! \gmag/\mathrm{mag} \!\lesssim\! 16.2$.
We note that red LAVs other than LPVs can also be present in this group, such as R~CrB and RV~Tauri variables.

\begin{figure}
	\centering
	\includegraphics[trim={0 81pt 0 0},clip,width=\linewidth]{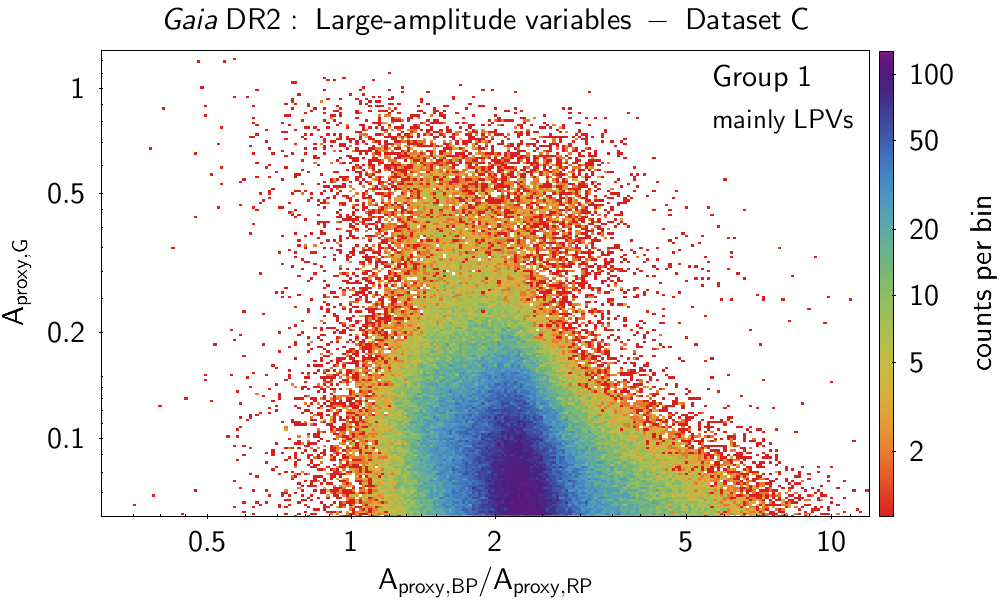}
	\vskip -0.5mm
	\includegraphics[trim={0 81pt 0 50pt},clip,width=\linewidth]{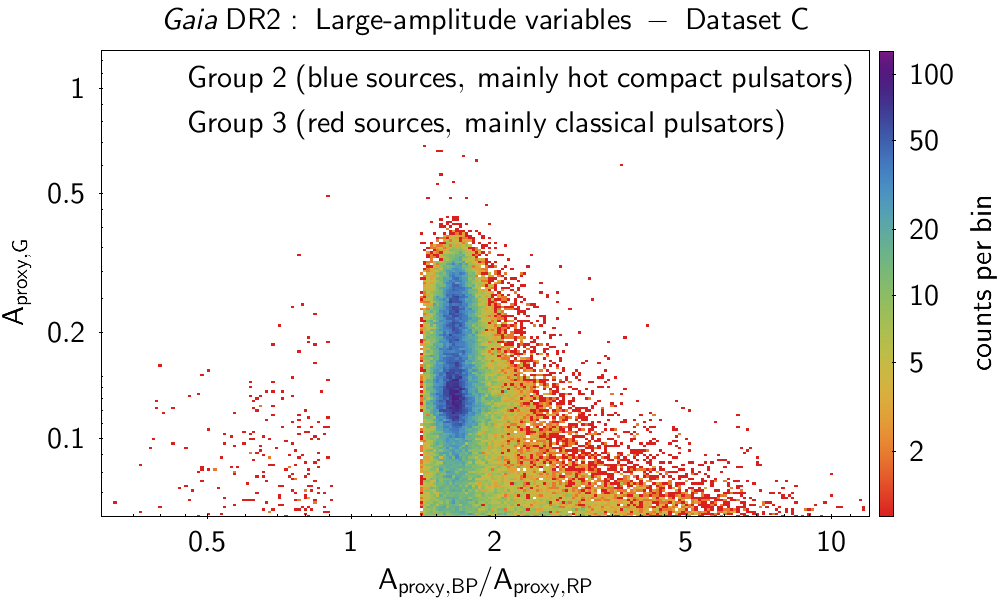}
	\vskip -0.5mm
	\includegraphics[trim={0 81pt 0 50pt},clip,width=\linewidth]{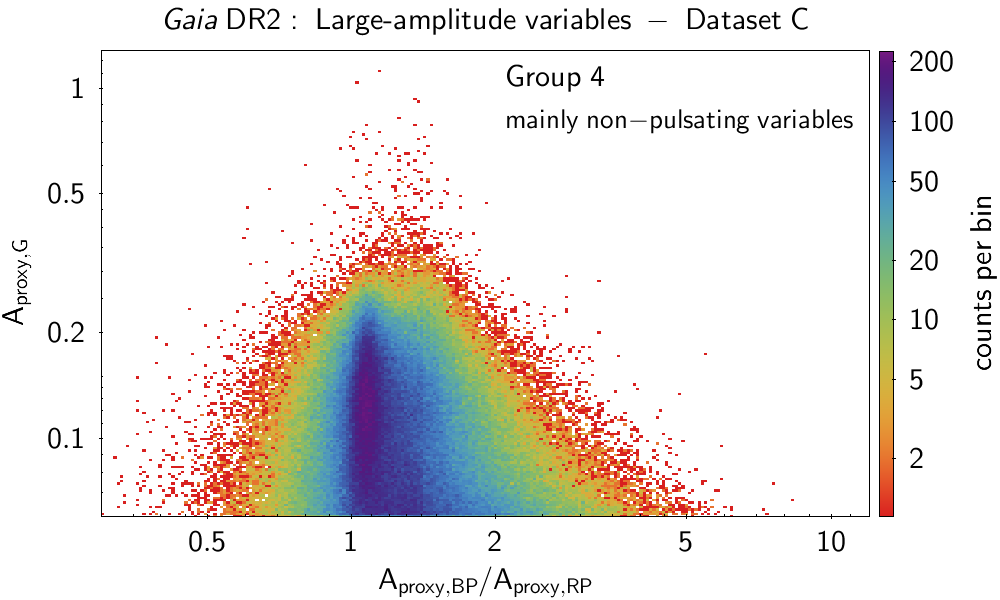}
	\vskip -0.5mm
	\includegraphics[trim={0 0 0 50pt},clip,width=\linewidth]{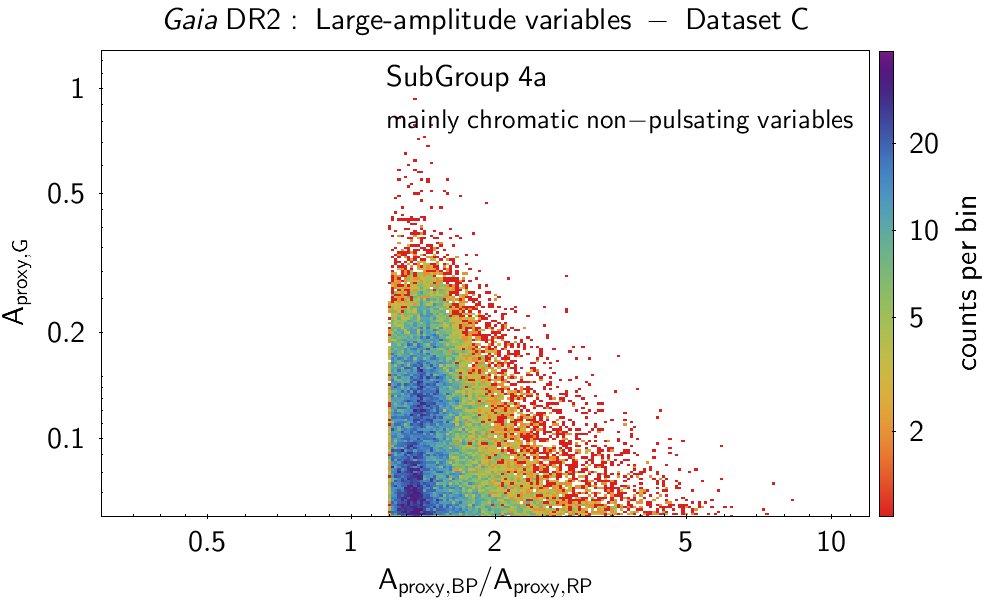}
	\caption{Density maps of \varProxyG versus $\varProxyBP/\varProxyRP$ for the different groups in Dataset~C.
	         \textbf{Top panel:} Group~1 (mainly LPVs).
	         \textbf{Second panel from top:} Group~2 for sources at $\varProxyBP/\varProxyRP<0.9$ (mainly hot compact LAVs) and Group~3 for sources at $\varProxyBP/\varProxyRP>1.4$ (mainly classical pulsators).
	         \textbf{Third panel from top:} Group~4 (mainly non-pulsating variables).
	         \textbf{Bottom panel:} Subgroup~4a (mainly chromatic non-pulsating variables).
	}
	\label{Fig:varProxyGvsVarProxyBPoverRP_DatasetC}
\end{figure}

Regarding the \gmag-band variability amplitudes of these LPVs, most of them have $\varProxyG \lesssim 0.3$ (see top panel of Fig.~\ref{Fig:varProxyGvsVarProxyBPoverRP_DatasetC}), which corresponds to peak-to-peak \gmag amplitudes of less than $\sim$1~mag.
Miras stand out at amplitudes larger than this value.
Group~1 contains one third of all LAVs in Dataset~C.

\paragraph{Group 2.}
\label{Sect:exploration_multiband_group2}

\begin{figure}
	\centering
	\includegraphics[width=\linewidth]{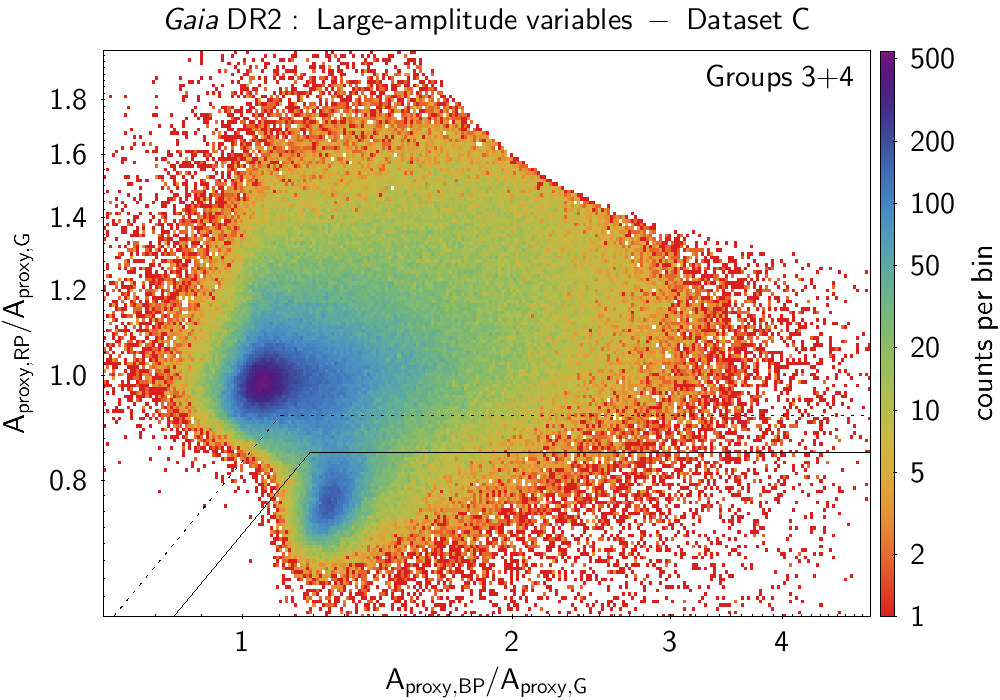}
	\caption{Same as Fig.~\ref{Fig:varProxyRatiosBPGvsRPG_DatasetC}, but for Groups~3 and 4 in Dataset~C.
	         The solid-line delineates Group~3 (below the line) and Group~4 (above the line).
	         The dashed line identifies Subgroup 4a at the small $\varProxyRP/\varProxyG$ side of Group 4 (see text).
	         The diagonal lines are given by $\varProxyBP/\varProxyRP = 1.2 $ (dashed diagonal line) and $\varProxyBP/\varProxyRP = 1.4$ (solid diagonal line).
	}
	\label{Fig:varProxyRatiosBPGvsRPG_DatasetC_Groups34}
\end{figure}

\begin{figure}
	\centering
	\includegraphics[width=\linewidth]{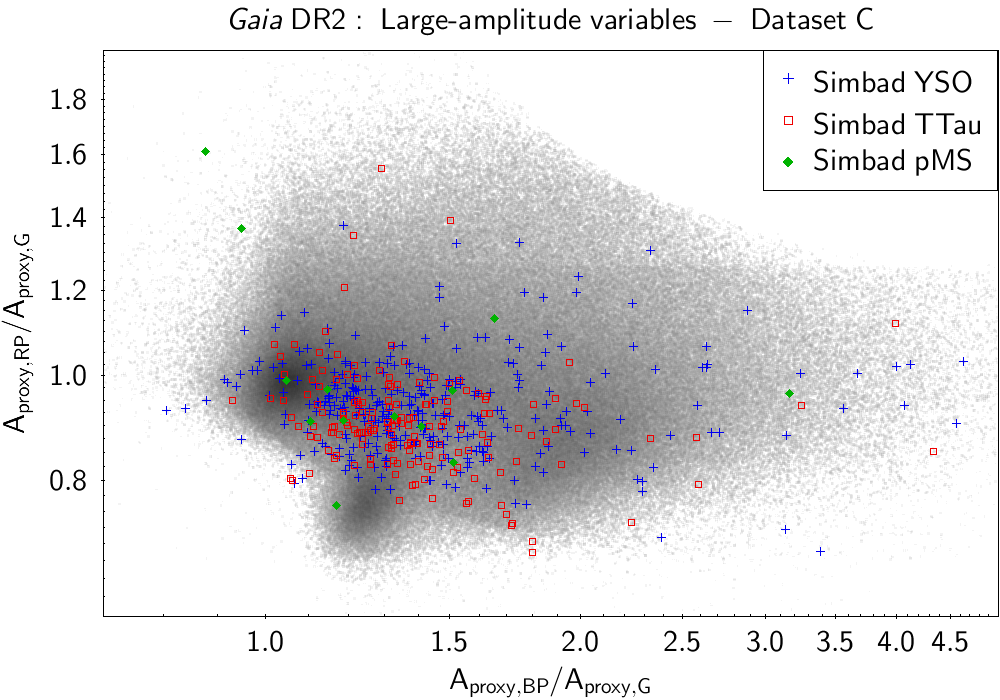}
	\caption{Same as Fig.~\ref{Fig:varProxyRatiosBPGvsRPG_DatasetC}, but with Dataset~C sources crossmatched with Simbad YSOs, T~Tauri stars and pre-MS stars shown with the markers labelled in the upper-right inset of the figure.
	}
	\label{Fig:varProxyRatiosBPGvsRPG_DatasetC_YSOs}
\end{figure}

We gather in this group blue LAVs ($\BPminusRP<0.2$~mag) with variability amplitudes larger in the red than in the blue ($\varProxyBP < 0.9 \, \varProxyRP$).
We restrict to hot stars fainter than the MS, including WDs and subdwarfs, leaving hot MS LAVs such as Ae or Be stars to Groups~3 and 4 that contain MS stars.
In order to do so, we use a magnitude-dependent limit on the parallax similar to the method used for LPVs in Group~1.
The condition is given by Eqs.~\ref{Eq:conditionsGroup3} in Table~\ref{Tab:groups}.
We note, however, that the absolute magnitude separation between hot subdwarfs and MS blue variables is only between $\sim$3 to $\sim$5 magnitudes, which will necessarily lead to some confusion for sources that do not have good parallaxes.
Group~2 contains only a small fraction of all LAVs, about 0.1\% in Dataset~C.

\paragraph{Group 3.}
\label{Sect:exploration_multiband_group3}

The multi-band amplitude ratios of the Dataset~C sample cleaned from Group~1 and 2 stars are shown in Fig~\ref{Fig:varProxyRatiosBPGvsRPG_DatasetC_Groups34}.
The removal of Group~1 stars from the sample has adequately removed the over-density of sources that was present around $\varProxyBP/\varProxyRP=2$ in the sull Dataset~C sample (see Fig.~\ref{Fig:varProxyRatiosBPGvsRPG_DatasetC}).
Two main groups remain clearly visible in Fig~\ref{Fig:varProxyRatiosBPGvsRPG_DatasetC_Groups34}, one below $\varProxyRP / \varProxyG \!\simeq\! 0.85$, and another one above that limit.
The former group contains sources that have significantly larger variability amplitudes in blue than in red, and forms our Group~3.
This group is further restricted to $\varProxyBP/\varProxyRP>1.4$, leading to the sample below the solid line in Fig~\ref{Fig:varProxyRatiosBPGvsRPG_DatasetC_Groups34}.
The conditions for Group~3 are summarized in Table~\ref{Tab:groups}.

The variables in this group are mainly the classical pulsators identified in Sect.~\ref{Sect:exploration_multiband_literature}, with $\varProxyBP \!\simeq\! 1.63 \, \varProxyRP$.
Their \gmag amplitudes are shown in the second panel from top in Fig.~\ref{Fig:varProxyGvsVarProxyBPoverRP_DatasetC} (sources with $\varProxyBP/\varProxyRP>1$).
Group~3 contains about 10\% of Dataset~C LAVs.

\paragraph{Group 4.}
\label{Sect:exploration_multiband_group4}

The fourth group contains the remaining LAVs not in Groups~1 to 3, that is at $\varProxyRP/\varProxyG$ ratios above the solid line in Fig.~\ref{Fig:varProxyRatiosBPGvsRPG_DatasetC_Groups34} (condition~\ref{Eq:conditionsGroup4} in Table~\ref{Tab:groups}).
They correspond to variables with $\varProxyBP \!\simeq\! \varProxyG \!\simeq\! \varProxyRP$.
The great majority of them are known to be non-pulsating variables.
In particular, they contain eclipsing binaries, as seen in the ZTF sample of periodic variables (see Sect.~\ref{Sect:exploration_multiband_literature}).

Figure~\ref{Fig:varProxyRatiosBPGvsRPG_DatasetC_Groups34} further reveals a small over-density of sources, within Group~4, close to the transition between Groups~3 and 4, at $1.2 \!<\! \varProxyBP/\varProxyRP \!<\! 1.4$ (between the solid and dotted lines in the figure).
With amplitudes 20\% to 40\% larger in \gbp than in \grp, the variability cannot be considered achromatic.
We therefore define Subgroup~4a with the conditions (\ref{Eq:conditionsGroup4a}) listed in Table~\ref{Tab:groups}, which select the sample between the solid and dotted lines in Fig.~\ref{Fig:varProxyRatiosBPGvsRPG_DatasetC_Groups34}.
The RS~CVn variables in the ZTF sample analysed in Sect.~\ref{Sect:exploration_multiband_literature} typically fall in this subgroup.
Pre-MS variables, YSOs and T~Tauri variables are also found in this subgroup, as shown in Fig.~\ref{Fig:varProxyRatiosBPGvsRPG_DatasetC_YSOs} where these stars identified from the Simbad database have been plotted.
The \gmag amplitude distribution of Subgroup~4a is shown in the bottom panel of Fig.~\ref{Fig:varProxyGvsVarProxyBPoverRP_DatasetC}.
It confirms the relevance of this subgroup as being distinct within Group~4, with an over-density of sources at $1.25 \lesssim \varProxyBP / \varProxyRP \lesssim 1.5$.

Group~4 is the most populated of the four groups, gathering 58\% of all LAVs in Dataset~C.
Subgroup~4a contains 10\% of Group~4.


\subsection{The sample with parallaxes better than 10\%}
\label{Sect:exploration_goodparallaxes}

\begin{figure}
	\centering
	\includegraphics[width=\linewidth]{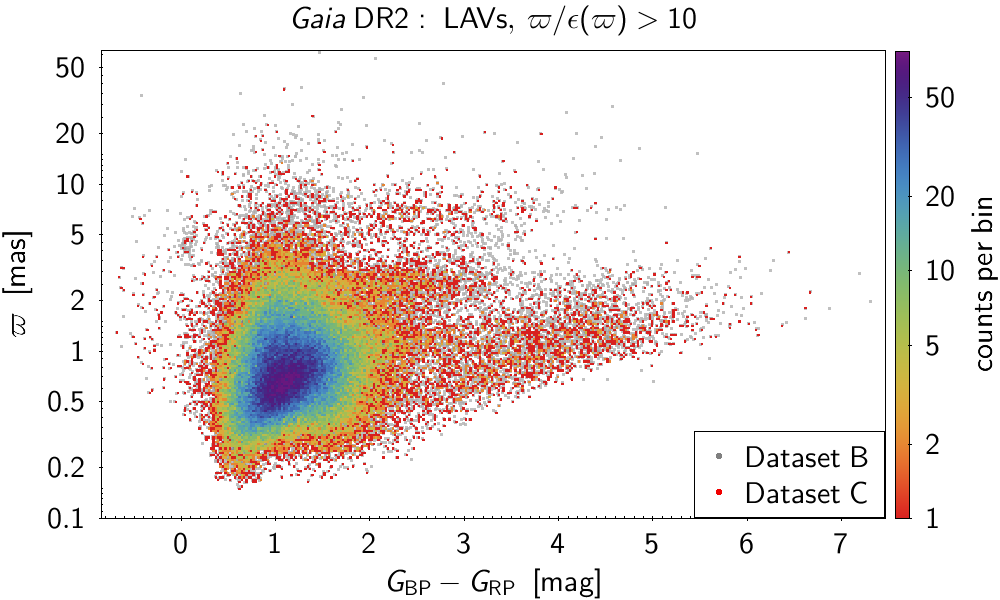}
	\caption{Parallax versus colour for sources that have relative parallax uncertainties better than 10\% in Dataset~C (coloured with the density of points according to the colour scale shown on the right of the figure).
	         Dataset~B is plotted in the background in grey. 
	}
	\label{Fig:parallaxVsBPmRP_datasetsBC}
\end{figure}

\begin{figure}
	\centering
	\includegraphics[trim={0 81pt 0 0},clip,width=\linewidth]{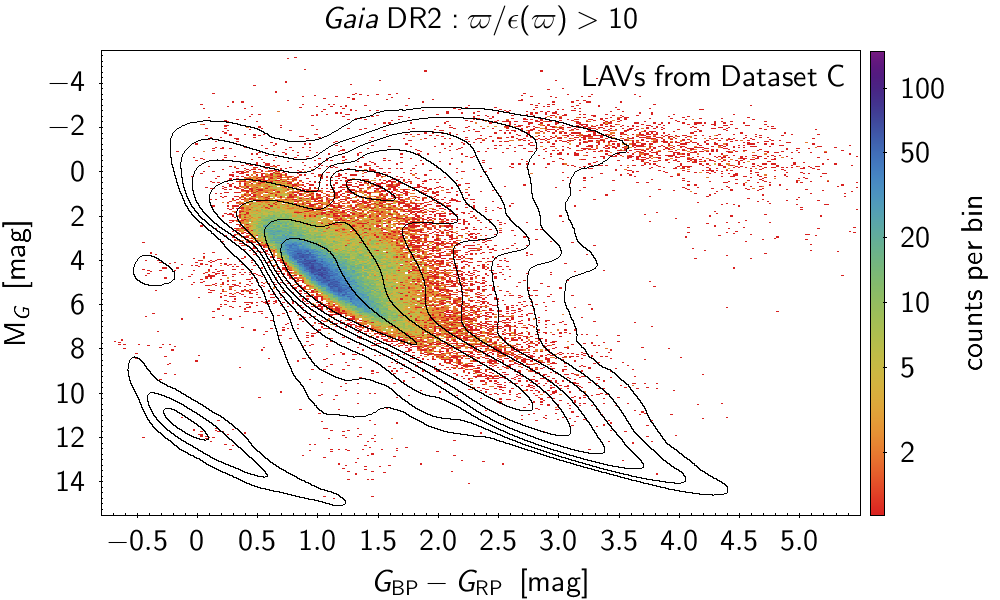}
	\vskip -0.5mm
	\includegraphics[trim={0 0 0 50pt},clip,width=\linewidth]{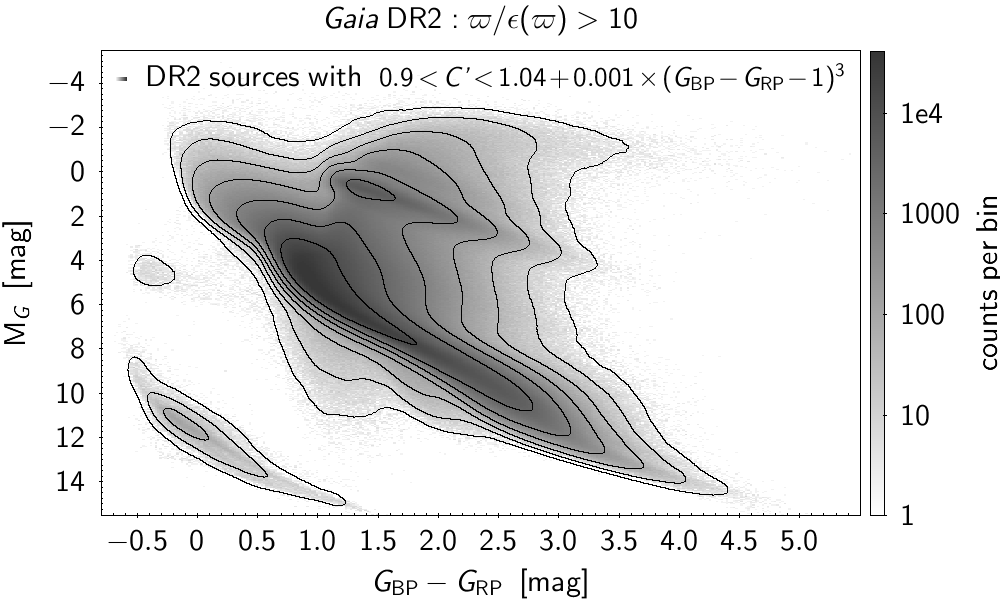}
	\caption{Observational HRDs of \Gaia DR2 sources that have relative parallax uncertainties better than 10\%.
	         The top panel shows the density map of LAV candidates from Dataset~C, while the bottom panel displays a random selection of variable+constant sources with good {\fluxexcess}.
	         The black contour lines in both panels correspond to the density lines of the sample shown in the bottom panel.
	         No correction for interstellar reddening and extinction is applied.
	}
	\label{Fig:obsHR_Cgp}
\end{figure}

\begin{figure*}
	\centering
	\includegraphics[width=0.49\linewidth]{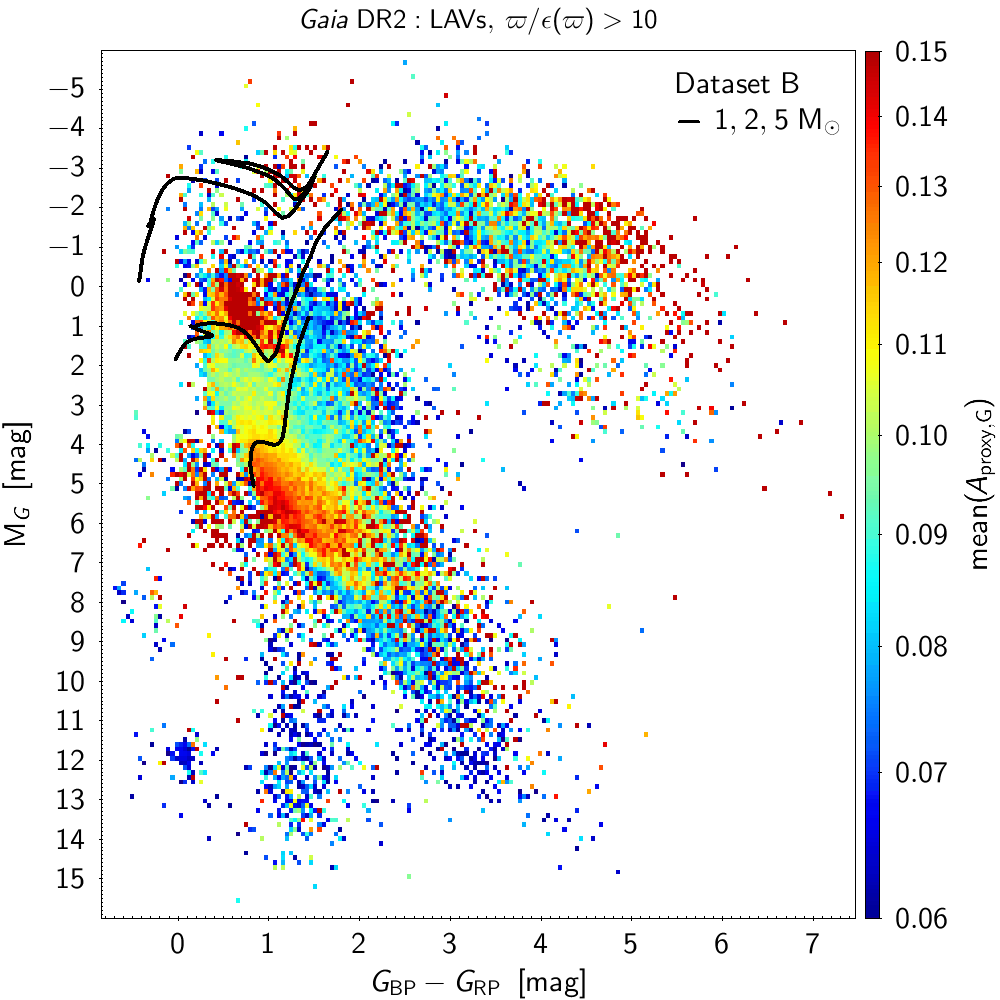}
	~~
	\includegraphics[width=0.49\linewidth]{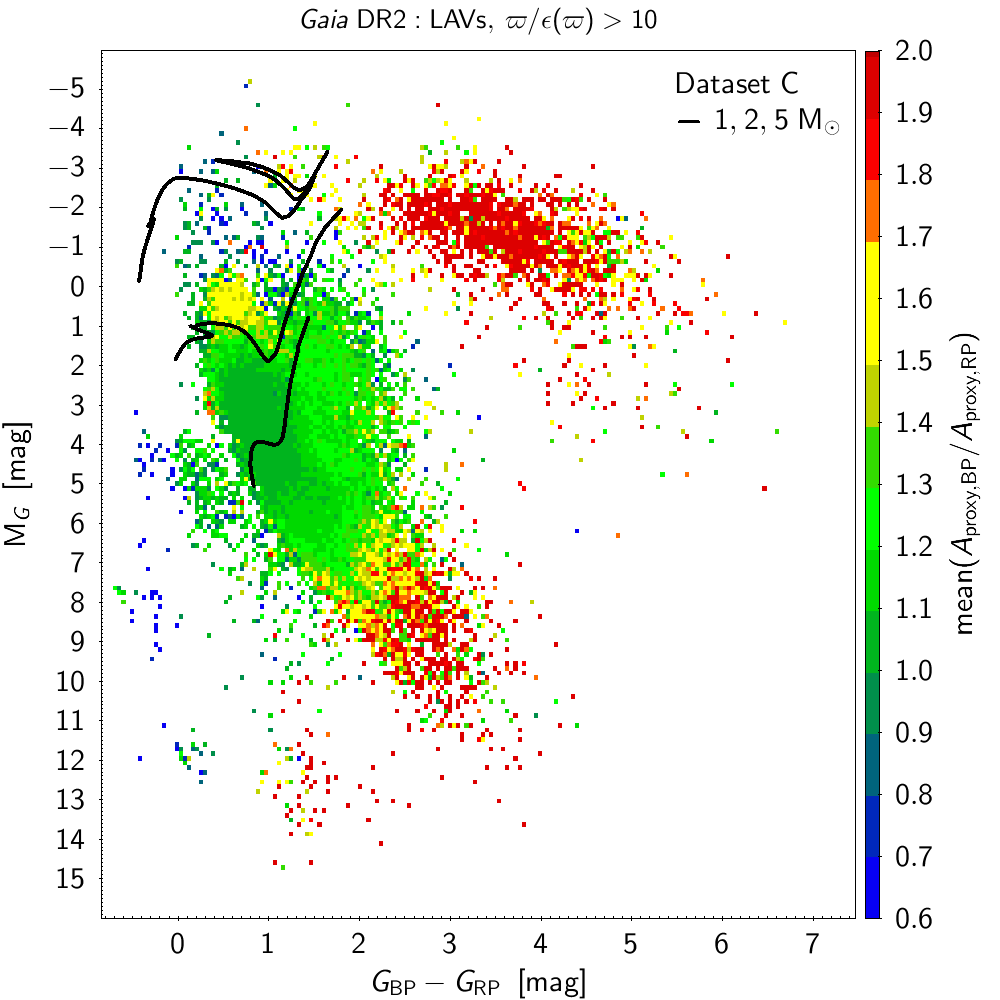}
	\caption{Observational HRD with the mean value of \varProxyG from Subset~Bgp (left panel) and of $\varProxyBP/\varProxyRP$ from Subset~Cgp (right panel) for each cell of size $[\Delta (\BPminusRP), \Delta \absMeanG] = (0.045, 0.12)$~mag, plotted in colour according to the colour-scale on the right of each panel.
	         The thin contour lines in black correspond to the density lines of the DR2 sample of constants $+$ variables shown in Fig.~\ref{Fig:obsHR_Cgp} (bottom panel).
	         The thick lines correspond to evolutionary tracks of (from bottom to top) 1, 2 and 5 M$_\odot$ solar-metallicity stellar models from \citet{EkstromGeorgyEggenberger_etal12}.
	}
	\label{Fig:obsHR_gridVarProxyGBPRP_mean}
\end{figure*}

\begin{figure}
	\centering
	\includegraphics[trim={0 81pt 0 0},clip,width=0.99\linewidth,height=148pt]{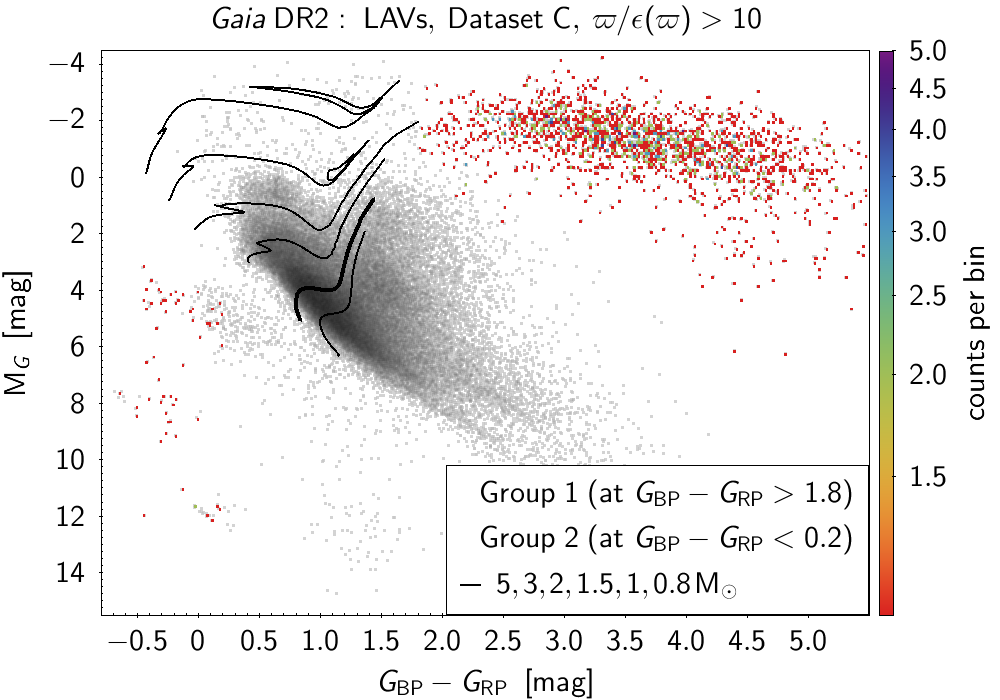}
	\vskip -0.5mm
	\includegraphics[trim={0 81pt 0 50pt},clip,width=0.99\linewidth,height=148pt]{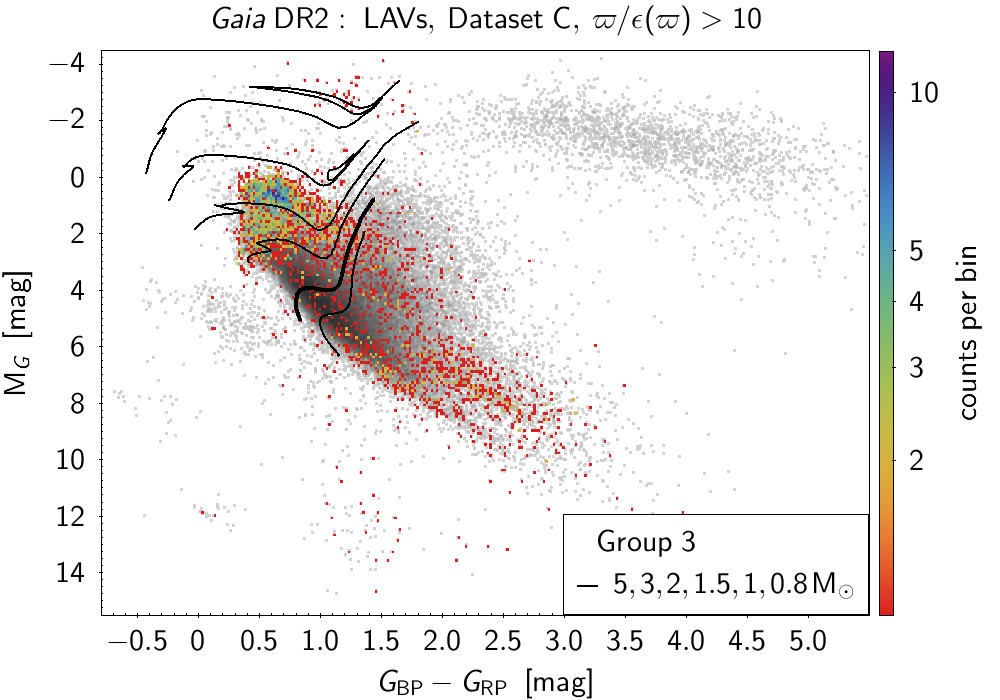}
	\vskip -0.5mm
	\includegraphics[trim={0 81pt 0 50pt},clip,width=0.99\linewidth,height=148pt]{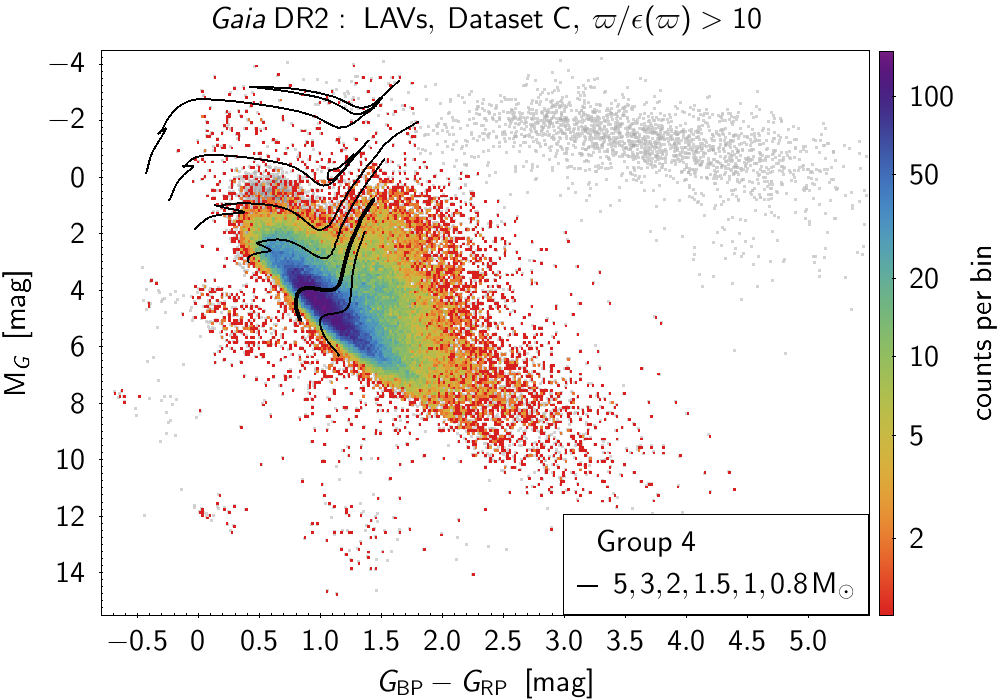}
	\vskip -0.5mm
	\includegraphics[trim={0 0 0 50pt},clip,width=0.99\linewidth,height=148pt]{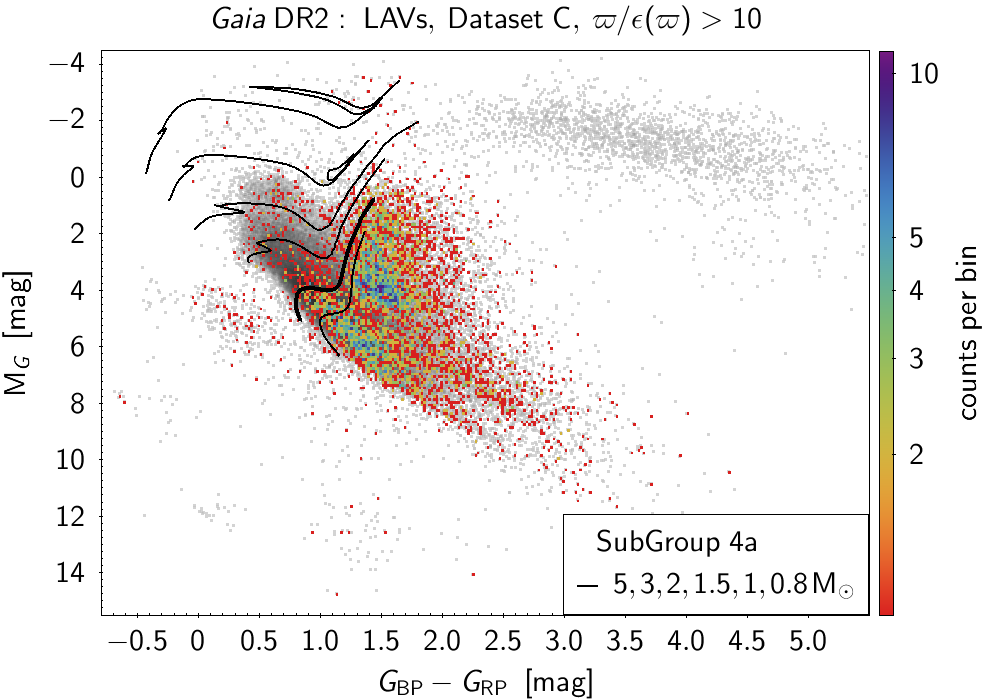}
	\caption{Density maps of observational HRDs of Group~1 (top panel, sources having $\BPminusRP>1.8$~mag), Group~2 (top panel, sources having $\BPminusRP<0.2$~mag), Group~3 (second panel from top), Group~4 (third from top) and Subgroup~4a (bottom panel) of LAVs in dataset~C with parallax uncertainties better than 10\%.
	         The background grey points show the full sample of dataset~C with $\varpi/\varepsilon(\varpi)>10$.
	         Evolutionary tracks of (from bottom to top) 0.8, 1, 1.5, 2, 3 and 5 M$_\odot$ solar-metallicity stellar models from \citet{EkstromGeorgyEggenberger_etal12} are over-plotted in black, with the 1~M$_\odot$ track rendered in thick line.
	         The axes ranges have been limited for better visibility. 
	}
	\label{Fig:obsHR_Cgp_groups}
\end{figure}

\begin{figure}
	\centering
	\includegraphics[width=\linewidth]{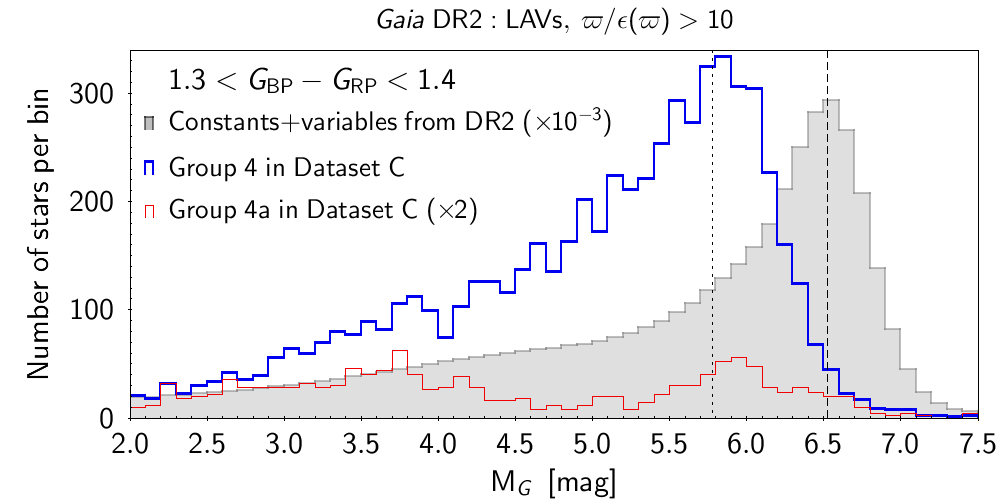}
	\caption{Histograms of the absolute \gmag magnitude of sources with parallax uncertainties better than 10\% in the colour bin $1.3 \!<\! \BPminusRP/\mathrm{mag} \!<\! 1.4$, from Group~4 (blue thick histogram) and Subgroup~4a (red thin histogram) in Dataset~C, and from the constant+variable DR2 sample (filled grey histogram).
	         Bins are 0.1~mag wide, and the numbers of sources per bin have been multiplied by two for Subgroup~3a and by $10^{-3}$ for the full DR2 sample.
	         The vertical dashed line locates the absolute magnitude at maximum of the full DR2 distribution (6.525~mag).
	         A vertical dotted line is added at an absolute magnitude 0.75~mag brighter than the dashed line.
	         The abscissa range has been limited to highlight the distribution of main-sequence stars.
	}
	\label{Fig:histo_absMagG_group4}
\end{figure}

\begin{figure}
	\centering
	\includegraphics[width=\linewidth]{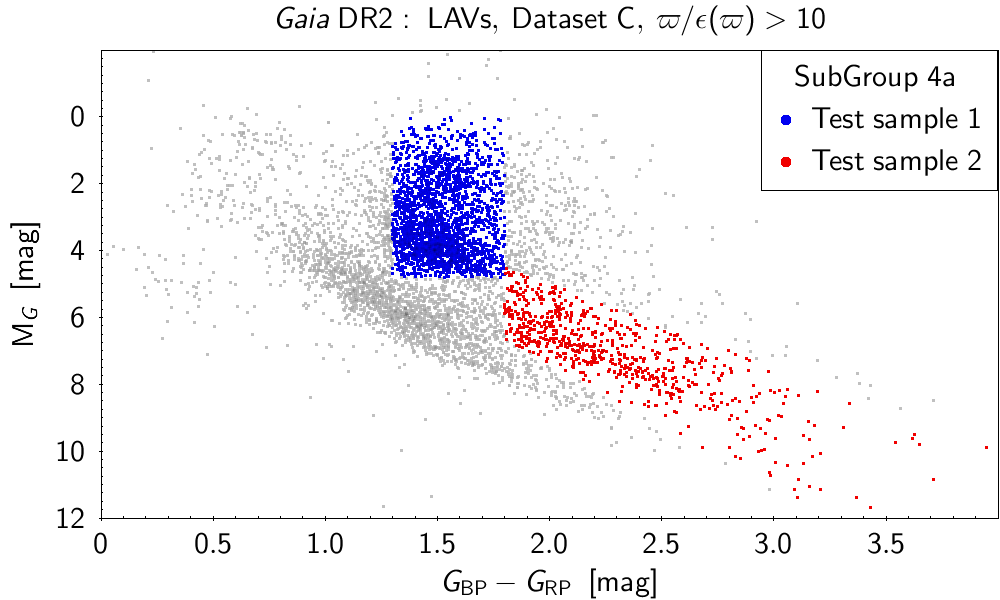}
	\includegraphics[trim={0 0 0 30pt},clip,width=\linewidth]{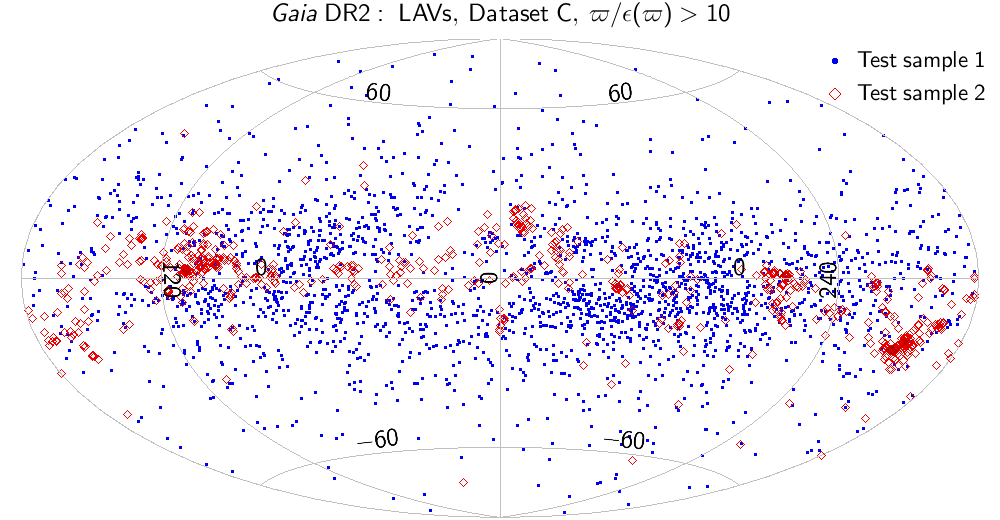}
	\includegraphics[trim={0 0 0 30pt},clip,width=\linewidth]{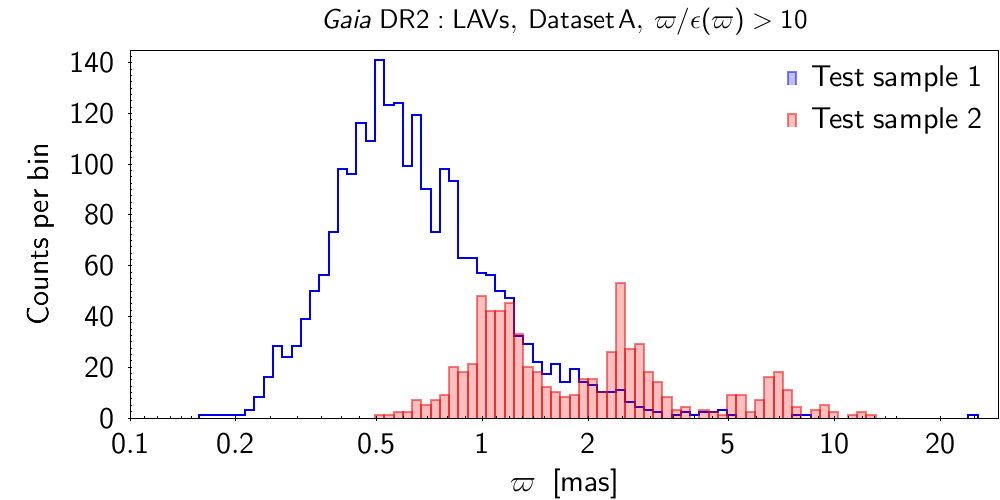}
	\caption{Properties of two Subgroup~4a samples selected from their locations in the observational HRD as shown by the blue and red samples in the top panel.
	         Their sky distributions are shown in the middle panel, and the histograms of their parallaxes are shown in the bottom panel.
	}
	\label{Fig:Subgroup4a_TTauri_YSO}
\end{figure}

We present in this section the sample of LAVs with relative parallax uncertainties better than 10\%.
We do not impose any restriction on the number of visibility periods used in the derivation of the astrometric solution as less than 1\% of Datasets~B and C that have parallax uncertainties better than 10\% have fewer than eight such periods, while 82\% of them have at least ten periods.
We will use either Dataset~B or C, as needed.
We recall from Sect.~\ref{Sect:catalogue_datasets} that Dataset~B is suitable when colours are required, while Dataset~C is required when \gbp and \grp variability amplitudes are used.
The subsets with good parallaxes in Datasets~B and C according to $\varpi/\epsilon(\varpi)>10$ are hereafter called subsets Bgp and Cgp.
The number of their sources is given in Table~\ref{Tab:datasetsSummary}.
The distribution of the parallaxes versus \BPminusRP is shown in Fig.~\ref{Fig:parallaxVsBPmRP_datasetsBC}.

We first give an overview of the datasets in the observational Hertzsprung-Russell diagram (HRD) in Sect.~\ref{Sect:exploration_goodparallaxes_overview}.
We then summarize their multi-band variability properties in Sect.~\ref{Sect:exploration_goodparallaxes_multiband}, and finally present in Sect.~\ref{Sect:exploration_goodparallaxes_groups} the distribution in this diagram of the four variability groups identified in the previous section.

\subsubsection{Overview}
\label{Sect:exploration_goodparallaxes_overview}

The observational HRD of Subset~Cgp is shown in Fig.~\ref{Fig:obsHR_Cgp} (top panel), to be compared with the distribution of a random sample of constant+variable DR2 sources in the bottom panel that also have parallaxes better than 10\% and good {\fluxexcess} (conditions b2 and b3 in Table~\ref{Tab:datasetsSummary}) \citep[the reader is referred to][for a detail presentation of this diagram]{GaiaBabusiauxVanLeeuwen_etal18}.
To guide the eyes, contour lines from the DR2 distribution in the bottom panel are reported on the Cgp distribution in the top panel.
Absolute magnitude in this and following observational HRDs is computed with $\mathrm{M}_\gmag = \gmag + 5 - 5\,\log_{10}(1000/\varpi)$.
The comparison between the two panels shows a potential shortage of LAV sources from Dataset~Cgp in some parts of the diagram.
This can be due to a real shortage of LAVs in a specific region of the diagram, such as for stars in the red clump around $(\BPminusRP, \absMeanG) \simeq (1.4, 1)$~mag.
Or it can be due to smaller statistics in Subset~Cgp ($\sim$85\,000 sources) compared to the DR2 sample ($\sim$67 million sources), combined with the parallax-limited selection.
This may explain the shortage of blue MS LAVs.
However, it can also be a selection effect resulting from the filters leading to Dataset~C (like filters c3 to c5 in Table~\ref{Tab:datasetsSummary}).
This is expected to be the case for the shortage of LAVs among low-mass M-type MS stars (red dwarfs at $\absMeanG \simeq 9-14$~mag).
M0-M5.5 dwarf stars are known to be photometrically variable with flare amplitudes that can reach the order of 1~mag \citep[e.g.][]{GuntherZhanSeager_etal20}, which fall in the amplitude range considered in this paper.
A fraction of these stars could be detected with \Gaia \citep[see][for a candidate identified in DR2]{DistefanoLanzafame20}.
However, the faint magnitudes of these stars combined with their red colours lead to the exclusion of the majority of them from Datasets~B and C.
Most of these excluded sources also have Renormalized Unit Weight Errors (RUWE) larger than 1.4 (see Appendix~\ref{Appendix:BPRPexcess}, Fig.~\ref{Fig:CMabs_parallaxOverErrGT10_Enormalized_withRuwe}).

Except for these faint sources, subsets Cgp and Bgp provide a reliable picture of LAVs in the sample with parallax uncertainties better than 10\%.
These sources reach distances up to 5~kpc at $\BPminusRP \simeq 0.6$~mag (see Fig.~\ref{Fig:parallaxVsBPmRP_datasetsBC}), while limited to $\sim$1~kpc for the reddest and bluest LAVs in the samples.

\subsubsection{Multi-band variability properties in the observational HRD}
\label{Sect:exploration_goodparallaxes_multiband}

A summary picture of the variability of LAVs across the observational HRD is shown in Fig.~\ref{Fig:obsHR_gridVarProxyGBPRP_mean}, where each cell of size $[\Delta (\BPminusRP), \Delta \absMeanG] = (0.045, 0.12)$~mag has been colour-coded according to either the mean value of \varProxyG (left panel, using Dataset~B), or the mean value of $\varProxyBP/\varProxyRP$ (right panel, using Dataset~C).
The largest variability amplitudes in \gmag (red areas in the left panel) are mainly observed for LPVs (bright red side of the diagram), Cepheids (bright side of the classical instability strip), RR Lyrae variables (in the instability strip close to the MS), eclipsing binaries (on the MS), and variables in the hot subdwarf region of the HRD (bluewards of the MS).
The regions of pre-main sequence stars redwards of the MS and between the MS and WD sequence also contain cells with $\mathrm{mean}(\varProxyG)>0.13$.

The left panel of Fig.~\ref{Fig:obsHR_gridVarProxyGBPRP_mean} is advantageously put in perspective with Fig.~9 of \cite{GaiaEyerRimoldini_etal19}, which plots the distribution of DR2 variability amplitudes across the observational HRD.
The selection criteria of the sources displayed in that paper are, however, not the same as here, leading to different patterns when comparing the two figures.
In particular, the selection used by \cite{GaiaEyerRimoldini_etal19} excludes the largest-amplitude variables with $\rangeG\!\gtrsim\! 0.75$~mag (their exclusion filter $\fluxErrorG / \fluxG \!>\! 0.02$ is equivalent, with a mean number of 130 CCD measurements, to an exclusion of sources with $\varProxyG \!\gtrsim\! 0.23$).
Another difference is the absence, in their Fig.~9, of LAV subdwarfs observed in our Fig.~\ref{Fig:obsHR_gridVarProxyGBPRP_mean} around $(\gbp-\grp, \absMeanG) = (0.5, 5.5)$~mag.
This absence in their figure is due to their additional selection criteria, mainly on astrometry.

Large-amplitude variables that have the largest blue-to-red amplitude ratios (red areas in the right panel of Fig.~\ref{Fig:obsHR_gridVarProxyGBPRP_mean}) mainly consist of LPVs.
Two other areas in the observational HRD also display large mean $\varProxyBP/\varProxyRP$ ratios, one at the faint side of the MS and another one in the faint region between the MS and the WD sequence.
Caution must however be taken for these faint red sources, as the \varProxyBP values most probably result from noise in the \gbp light curves and are thus not reliable.
The second largest $\varProxyBP/\varProxyRP$ ratios in the right panel of Fig.~\ref{Fig:obsHR_gridVarProxyGBPRP_mean} are found for classical pulsators in the instability strip (yellow concentrations in the upper MS and Cepheid region of the diagram).
White dwarfs and hot subdwarf variables, on the other hand, have the smallest blue-to-red amplitude ratios, with $\varProxyBP/\varProxyRP<1$ (blue areas in Fig.~\ref{Fig:obsHR_gridVarProxyGBPRP_mean}, right panel).
Few bright LAV candidates in the upper MS, or redwards of it, also show $\varProxyBP/\varProxyRP<0.9$.

\subsubsection{Properties of the four classification groups in the observational HRD}
\label{Sect:exploration_goodparallaxes_groups}

The LAVs in Dataset~C have been categorized in Sect.~\ref{Sect:exploration_multiband_groups} into four groups according, mainly, to their blue-to-red amplitude ratio and their colour.
Here we further analyse their properties using the observational HRD.
Their distributions in that diagram are shown in Fig.~\ref{Fig:obsHR_Cgp_groups}.
Stellar evolutionary tracks at solar-metallicity are also added from \citet{EkstromGeorgyEggenberger_etal12}%
\footnote{
Downloaded from \url{https://www.unige.ch/sciences/astro/evolution/en/database/syclist/}.
The transformation relations used in the website to derive photometry in the \Gaia bands are taken from \citet{EvansRielloDeAngeli_etal18}.
}%
to evaluate the stellar masses and evolutionary stages.
The considered sample with good parallaxes contains 2033 LAVs in Group~1 (sources with $\BPminusRP>1.8$~mag in the top panel of the figure), 59 in Group~2 (sources with $\BPminusRP<0.2$~mag in the top panel), 3531 in Group~3 (second panel from top) and 79\,074 in Group~4 (third panel from top) LAV candidates, with 6154 sources in Subgroup~4a (bottom panel).

Group~1 LAVs populate the red part of the HRD as expected for LPVs (top panel in Fig.~\ref{Fig:obsHR_Cgp_groups}).
We note in Fig.~\ref{Fig:parallaxVsBPmRP_datasetsBC} that the redder an LPV is, the less far from the Sun it can be detected in subsets Bgp and Cgp.
This is due to the combined effect of redder LPVs being fainter, and of fainter stars having less precise parallaxes.

Group~2 LAVs are not numerous.
Their location in the observational HRD indicates that they contain hot subdwarfs and white dwarfs (top panel in Fig.~\ref{Fig:obsHR_Cgp_groups}), in agreement with the definition of the group.

Group~3 LAVs are expected from the analyses of literature data presented in Sect.~\ref{Sect:exploration_multiband} to predominantly contain pulsating stars.
This is confirmed from their distribution in the observational HRD (second panel from top in Fig.~\ref{Fig:obsHR_Cgp_groups}), where most of them are seen to gather in the region of RR~Lyrae stars around $(\BPminusRP, \mathrm{M}_G) \!\simeq\! (0.51, 0.5)$~mag.
A tail extending from that region towards the faint-red side of the HRD, down to $(\BPminusRP, \mathrm{M}_G) \!\simeq\! (1.8, 5)$~mag, is also observed, compatible with RR~Lyrae stars reddened by extinction on the line of sight.
Two other classical pulsators are also visible in the diagram: $\delta$~Sct stars extending below the bulk of RR~Lyr stars at $1 \!\lesssim\! \mathrm{M}_G \mathrm{/mag}\!\lesssim\! 3$, and Cepheids at the bright side of the HRD at $-3 \!\lesssim\! \mathrm{M}_G \mathrm{/mag}\!\lesssim\! -1$.
Group~3 also contains a small fraction of variables that are not classical pulsators, as witnessed by the fainter candidates present at $\absMeanG>5$~mag (Fig.~\ref{Fig:obsHR_Cgp_groups}, second panel from top). 
They amount to less than 15\% of Group~3.

Group~4 was shown in Sect.~\ref{Sect:exploration_multiband} to predominantly contain non-pulsating LAVs.
In particular, the analysis of ZTF periodic variables in Sect.~\ref{Sect:exploration_multiband_literature} showed the quasi-achromaticity of the majority of their eclipsing binaries (see in particular Fig.~\ref{Fig:histo_varProxyBPoverRP_literature}).
A query in the SIMBAD database confirms this expectation, with three quarters of Group~4 LAVs in Subset~Cgp being classified as eclipsing binaries.
This is also consistent with the distribution of subset~Cgp in the observational HRD (top panel of Fig.~\ref{Fig:obsHR_Cgp}) when compared to the distribution of constant+variable stars shown in the bottom panel of that figure.
They reveal (top panel) a lack of sources close to the zero-age MS, as expected if they are composed of binary stars of similar masses (required for near-achromatic variability).
Figure~\ref{Fig:histo_absMagG_group4} quantifies this observation for Group~4 stars by comparing their $\absMeanG$ histogram in a given colour range (taking $1.3 \!<\! \BPminusRP/\mathrm{mag} \!<\! 1.4$, blue histogram) with that of constant+variable DR2 stars in the same colour range (filled grey histogram).
The histogram of MS dwarf stars in the latter sample peaks at $\absMeanG = 6.525$~mag (vertical dashed line in Fig.~\ref{Fig:histo_absMagG_group4}), while that of Group~4 LAVs peaks at a magnitude almost 0.75~mag brighter than this value (dotted line in Fig.~\ref{Fig:histo_absMagG_group4}), as expected if they are composed of equal-mass eclipsing binaries.
In addition to eclipsing binaries, various other types of variables are present in Group~4, as witnessed from their distributions in the observational HRD (third panel from top in Fig.~\ref{Fig:obsHR_Cgp_groups}).
A comparison with Figs.~3 to 7 of \citet{GaiaEyerRimoldini_etal19}, derived from what is known in the literature, is most instructive for their identifications.

Subgroup~4a provides additional insight on Group~4 LAVs that display chromatic variability.
The ZTF sample of periodic variables already identified RS~CVn and BY Dra variables among non-pulsating variables with $\varProxyBP/\varProxyRP>1.2$ (Sect.~\ref{Sect:exploration_multiband_literature}).
The distribution of Subgroup~4a in the observational HRD (bottom panel of Fig.~\ref{Fig:obsHR_Cgp_groups}) provides additional clues on the content of this subgroup.
It reveals, in particular, the presence of a significant population of potential LAVs in the region of the diagram between the MS and the red clump (around $\absMeanG=4$~mag and $\BPminusRP=1.5$~mag), as well as on a sequence at $\BPminusRP \!\gtrsim\! 1.8$~mag almost parallel to the MS and about two magnitudes brighter than it.
To check their nature, we selected a sample of each of these two populations from their location in the observational HRD as shown in the top panel of Fig.~\ref{Fig:Subgroup4a_TTauri_YSO}.
The first sample, shown in blue in the figure (at $\BPminusRP \simeq 1.5$~mag), is found to be distributed preferentially along the Galactic plane but with no obvious specific pattern (second panel from top).
They lie at distances between 0.3 and 3 kpc from the Sun (bottom panel).
Querying the Simbad database returns 408 crossmatches, with the four most identified types being RS~CVn (46 candidates), rotational variables (29), eclipsing binaries (23), and LPVs (15).
These variability types should be confirmed as we do not expect to find LPVs in this region of the diagram.
We note that 212 crossmatches in this sample have unidentified or uncertain Simbad variability types.
The second sample, on the other hand, shown in red in Fig.~\ref{Fig:Subgroup4a_TTauri_YSO} (top panel), is mainly distributed on the Gould Belt with a predominance in star forming regions (middle panel).
Its parallax distribution (bottom panel) confirms that its members belong to nearby star forming regions located at specific distances from the Sun.
Among the 318 Simbad crossmatches of this sample, 56 are classified as YSOs, 42 as T Tauri stars, 38 as variables of unspecified type in Orion, and 17 as emission-line stars;
130 stars of the crossmatches have no or uncertain Simbad classification.
Finally, Fig.~\ref{Fig:obsHR_Cgp_groups} (bottom panel) shows that Subgroup~4a also contains a variety of other variability types, such as LAVs on the MS (compatible with most of them being eclipsing binaries as suggested by their $\absMeanG$ distribution shown in red in Fig.~\ref{Fig:histo_absMagG_group4}), as well as some hot subdwarfs and CVs.

\section{Summary and conclusions}
\label{Sect:Conclusions}

We have presented a catalogue of 23\,315\,874 LAVs from \Gaia DR2 having peak-to-peak \gmag amplitudes larger than about 0.2~mag, selected from their amplitude proxy $\varProxyG>0.06$ (Sect.~\ref{Sect:catalogue}).
The full catalogue of sources is called Dataset~A.

We identified two sub-samples, summarized in Table~\ref{Tab:datasetsSummary}.
Dataset~B ($\sim$5\% of Dataset~A) is suitable for studies requiring \gbp and \grp magnitudes, such as studying colour-magnitude diagrams.
Dataset~C (about half of Dataset~B) is suited for multi-band variability studies involving the amplitude proxies \varProxyBP and \varProxyRP in \gbp and \grp, respectively.

Within the magnitude and amplitude range considered in this paper, the completeness of Dataset~A relative to the variables published in dedicated catalogues in \Gaia DR2 is close to 100\% (Sect.~\ref{Sect:catalogue_completeness_DR2}), while the completeness of Datasets~B and C are $\sim$70\% and $\sim$47\%, respectively (Table~\ref{Tab:completeness}).
Comparison with the ZTF catalogue of periodic variables, on the other hand, suggests a completeness factor of 67\% for Dataset~A (Sect.\ref{Sect:catalogue_completeness_ZTF}).
It also confirms the above reduction factors from Dataset~A to Datasets~B and C.
The purity levels, on the other hand, are estimated in Sect.\ref{Sect:catalogue_purity} to increase from less than 50\% in Dataset~A to $\sim$70\% in Dataset~B and $\sim$85\% in Dataset~C.

The power of \Gaia to study variable stars using quasi-simultaneous multi-band photometry has been illustrated in Sect.~\ref{Sect:exploration} with two example cases.
The first studied the blue-to-red variability amplitude ratio for different types of variable stars based on literature source identification using \Gaia DR2 variables, ZTF, and Simbad (Fig.~\ref{Fig:histo_varProxyBPoverRP_literature}).
The full Dataset~C was then classified into four groups based on, mainly, $\varProxyBP/\varProxyRP$ and \BPminusRP.
The main types of variables in these groups are, schematically,
LPVs in Group~1 with amplitudes more than twice larger in \gbp than in \grp,
blue compact objects in Group~2 with amplitudes smaller in \gbp than in \grp,
pulsators in the classical instability strip in Group~3 with $\varProxyBP \simeq 1.63\, \varProxyRP$,
and a variety of LAVs with $\varProxyBP \simeq \varProxyRP$ in Group 4;
the last group mainly consists of non-pulsating variables, but about ten percent (Subgroup 4a) have chromatic variability with $1.2 \lesssim \varProxyBP/\varProxyRP \lesssim 1.5$.
The properties of these four groups have further been investigated in Sect.~\ref{Sect:exploration_goodparallaxes} using sub-samples having parallax uncertainties better than 10\%, and examples of additional types of variables populating each group other than the main types just mentioned have been identified from the distributions in the observational HRD (Fig.~\ref{Fig:obsHR_Cgp_groups}) complemented with type identification using the Simbad database.

The catalogue of LAVs presented here constitutes the first \Gaia catalogue of LAV candidates extracted from the full public DR2 archive.
While it inevitably contains shortcomings inherent to an intermediate data release of such a mission, it provides the opportunity to study variable objects using the samples identified in Datasets~A, B, and C, depending on the purpose of the study.

Future data releases will contain additional key \Gaia results with the provision of data collected from both its RVS and BP/RP spectrophotometers.
These will be especially relevant for the study of LAVs considering that the spectra are taken quasi-simultaneously with the \gmag measurements.
The combined photometric $+$ spectroscopic time series will offer unique opportunities to further characterize \Gaia variable stars.

\begin{acknowledgements}

We thank the anonymous referee for her/his comments that led to a significant update of the paper.
This work has made use of data from the European Space Agency (ESA) mission {\it Gaia} (\url{https://www.cosmos.esa.int/gaia}), processed by the {\it Gaia} Data Processing and Analysis Consortium (DPAC,
\url{https://www.cosmos.esa.int/web/gaia/dpac/consortium}).
Funding for the DPAC has been provided by national institutions, in particular the institutions participating in the {\it Gaia} Multilateral Agreement.\\
This publication makes use of the Starlink Tables Infrastructure Library \citep{Taylor2005} to produce the figures (STILTS and Topcat).\\
This research has made use of the SIMBAD database, operated at CDS, Strasbourg, France

\end{acknowledgements}

\bibliographystyle{aa}
\bibliography{bibTex}

\begin{appendix}

\section{Dataset~A}
\label{Appendix:DatasetA}

\begin{figure}
	\centering
	\includegraphics[width=\linewidth]{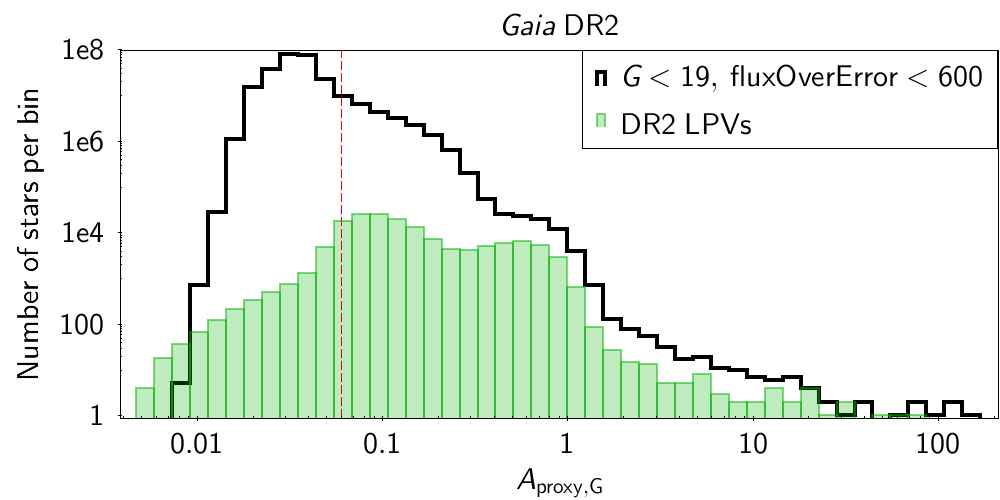}
	\caption{Distributions of the variability amplitude proxy of sources in the \Gaia DR2 archive.
	         The black histogram represents all sources brighter than $G\!=\!19$~mag with mean \gmag flux over error ratios $\fluxG/\fluxErrorG\!<\!600$.
	         The filled green histogram represents the sources published in the specific DR2 catalogue of LPVs.
	         The vertical red dashed line locates the $\varProxyG=0.06$ limit used to select large-amplitude variable candidates in this study.
	}
	\label{Fig:histo_varProxyG}
\end{figure}

We detail in this Appendix the procedure used to extract and filter the sample of large-amplitude variable candidates from the full \Gaia DR2 archive.
The extraction is described in Sect.~\ref{Appendix:DatasetA_Aproxy}.
Section~\ref{Appendix:DatasetA_Sky} then details the removal of spurious cases on specific stripes on the sky due to bad time intervals, and Sect.~\ref{Appendix:DatasetA_faintG} the removal of faint sources with large variability amplitudes.
They correspond to filters a1 and a2 mentioned in Table~\ref{Tab:datasetsSummary} of the main body of the article.
Some properties of the resulting dataset, called Dataset~A, is given in \ref{Appendix:DatasetA_Summary}.

\subsection{Extraction from the \Gaia DR2 archive}
\label{Appendix:DatasetA_Aproxy}

The quantity \varProxyG is not available in the \Gaia archive, and thus cannot be used to select sources.
Instead, we use $\fluxG / \fluxErrorG$ which is indexed in the archive.
We import all sources with $\gmag\!<\!19$~mag and $\fluxG / \fluxErrorG\!<\!600$, which amount to 256\,633\,579 sources.
The histogram of their \varProxyG is shown by the black line in Fig.~\ref{Fig:histo_varProxyG}.
Keeping only sources with $\varProxyG\!>\!0.06$, we are left with 23\,830\,862 candidates.
As as comparison, the \varProxyG distribution of the LPV candidates published in DR2, which all have the 5-95\% quantile range $QR_5(G)\!>\!0.2$~mag, is also shown in Fig.~\ref{Fig:histo_varProxyG} (green filled histrogram).
It is noted that a very small (we note the logarithmic scale of the figure) fraction of these variables have $\varProxyG\!<\!0.06$ despite their large $QR_5(G)$ amplitude, and are missed in the present sample.

The above procedure using $\fluxG / \fluxErrorG\!<\!600$ retrieves correctly all sources with $\varProxyG\!>\!0.06$ if $\NobsG\!<\!1297$.
However, sources with $\fluxG / \fluxErrorG\!>\!600$ (i.e. with very small relative uncertainties on their mean \gmag flux) may also have $\varProxyG\!>\!0.06$ if they have $\NobsG\!>\!1297$.
There are only 282 such sources, located in the north and south ecliptic poles.
Their high number of observations results from the Ecliptic Pole Scanning Law used during the \Gaia commissioning phase.
They are added to the sample separately, which reaches 23\,831\,144 candidates.

Finally, we remove from the sample all sources brighter than 5.5~mag in \gmag to comply with the limits taken for our catalogue (see Sect.~\ref{Sect:catalogue} in the main body of this article).
The total number of large-amplitude variable candidates with $\varProxyG\!>\!0.06$ in the magnitude range $5.5\mathrm{~mag} \!<\! \gmag \!<\! 19\mathrm{~mag}$ amounts to 23\,830\,345 in this initial sample.

\subsection{Filter on bad time intervals}
\label{Appendix:DatasetA_Sky}

\begin{figure}
	\centering
	\includegraphics[width=\linewidth]{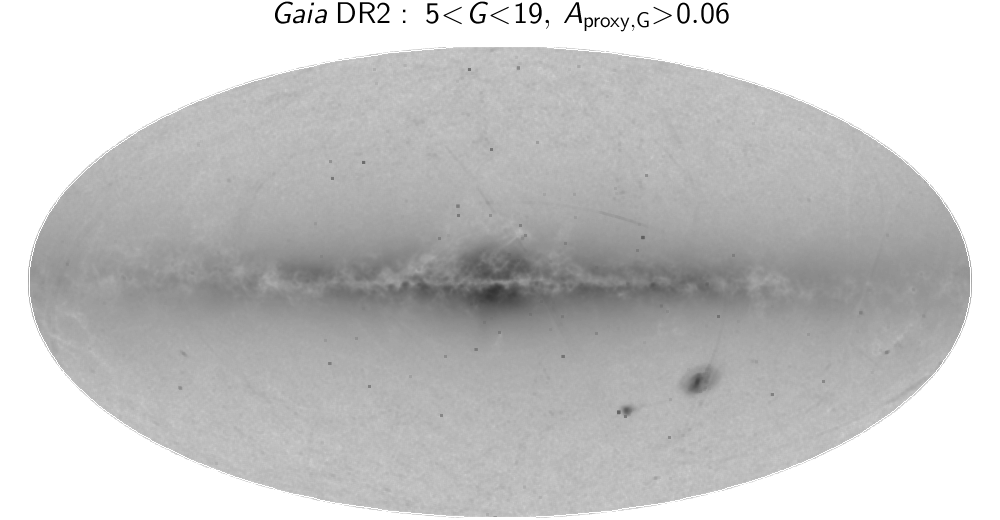}
	\caption{Sky distribution in Galactic coordinates of \Gaia DR2 sources with $\gmag\!<\!19$~mag, $\varProxyG\!>\;0.06$, non-NULL parallaxes, and which have at least five measurements in both the BP and RP bands.
	}
	\label{Fig:sky_initialDataset}
\end{figure}

\begin{figure}
	\centering
	\includegraphics[width=\linewidth]{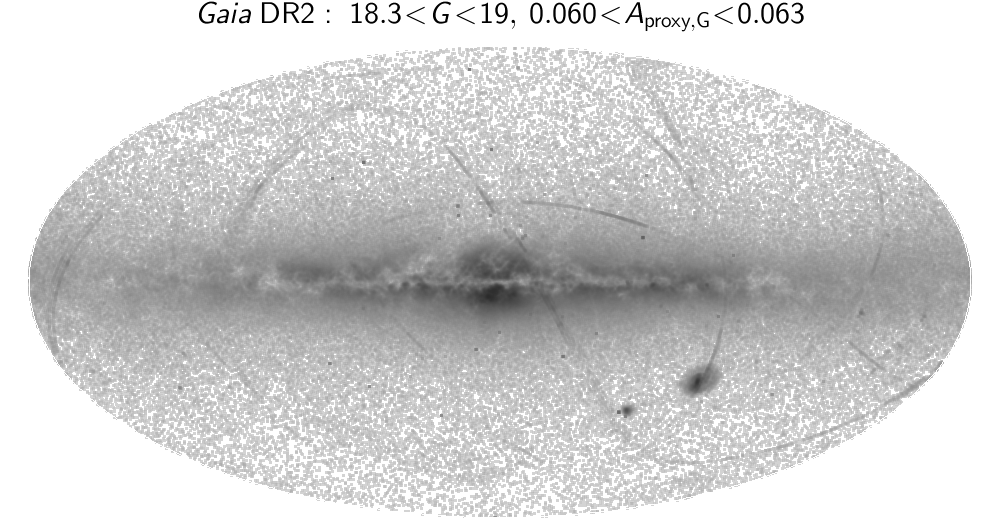}
	\includegraphics[width=\linewidth]{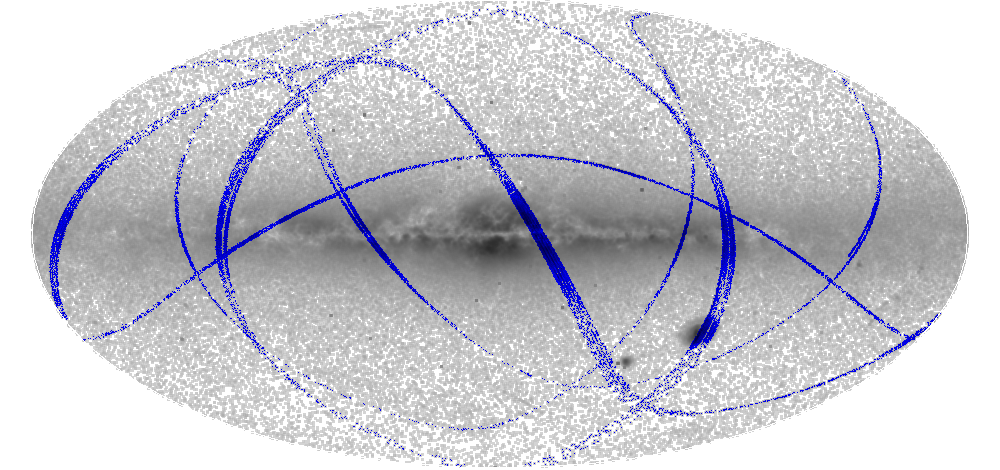}
	\caption{Same as Fig.~\ref{Fig:sky_initialDataset}, but limited to the sub-sample with $0.06\!<\!\varProxyG\!<\!0.063$ and $18.3\!<\!\gmag / \mathrm{mag}\!<\!19$.
	The time-interval-limited stripes identified with the \Gaia nominal scanning law as containing bad measurements are shown in blue in the bottom panel.
	}
	\label{Fig:sky_varProxyGLT0p063_GGT18p3}
\end{figure}

\begin{figure}
	\centering
	\includegraphics[width=\linewidth]{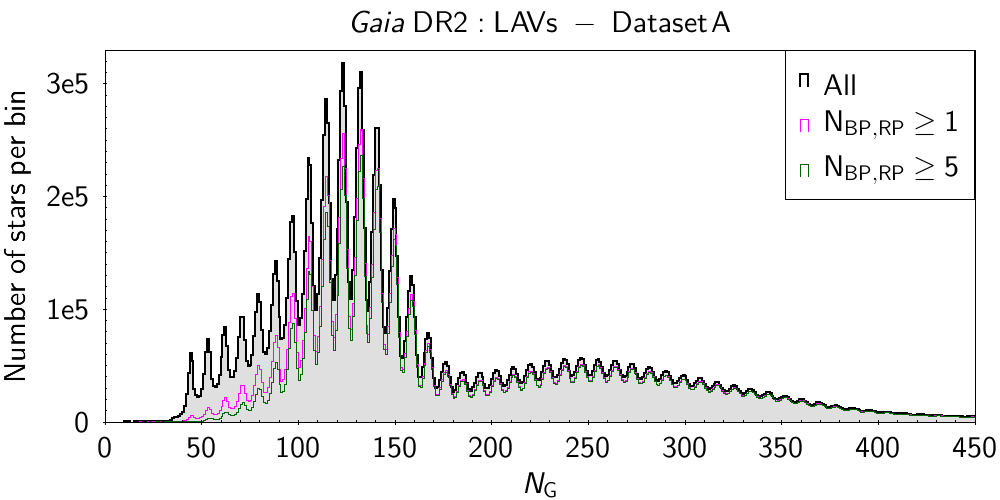}
	\caption{Number of CCD observations in \gmag of Dataset~A (filled grey histogram with black contour).
	         The histogram is limited to $\NobsG \le 450$ for better visibility, the maximum encountered number of CCD measurements being 2189, with a mean at 176.
	         The thin magenta/green histograms show the number of CCD observations in \gmag for the subsets of candidates having at least one/five measurements in both \gbp and \grp, respectively. 
	}
	\label{Fig:histo_NobsG_zoom_datasetA}
\end{figure}

\begin{figure}
	\centering
	\includegraphics[width=\linewidth]{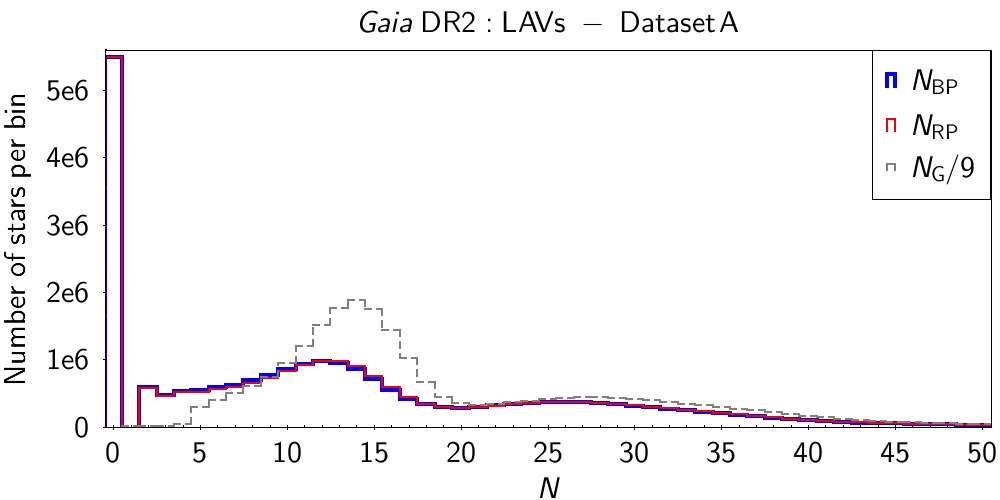}
	\caption{Number of observations in \gbp (solid blue) and \grp (solid red) of Dataset~A.
	         The histograms are limited to $N_\mathrm{BP,RP}\!\le\!50$ for better visibility, the maximum encountered number of measurements in \gbp and \grp being 235 and 235, respectively.
	         Also shown is the distribution of $\NobsG/9$ in dashed grey.
	}
	\label{Fig:histo_NobsBPRP_zoom_datasetA}
\end{figure}

The sky distribution of the sample constructed so far is displayed in Fig.~\ref{Fig:sky_initialDataset}.
It reveals stripes across the sky that are unphysical.
The stripes are clearly visible if we display the sub-sample with $\varProxyG\!<\!0.063$ and $\gmag\!>\!18.3$~mag.
This is shown in the top panel of Fig.~\ref{Fig:sky_varProxyGLT0p063_GGT18p3}.
These unphysical stripes originate from bad measurements at specific times during the mission which escaped the filters that were applied at the time of processing, sometimes coinciding with non-nominal satellite configurations but often for yet unpublished technical reasons.
They provide additional noise in the time series of faint sources during those short time intervals.
Given the \Gaia scanning law, the sources affected during these time intervals are distributed on stripes in the sky.

Checks of the sky distributions with varying \gmag and \varProxyG intervals reveal that mainly sources fainter than $\gmag=18.3$~mag are affected, and that the induced scatter in their light curves does not exceed $\varProxyG\!=\!0.1$ for the great majority of them.
Therefore, we exclude all sources in these stripes that have $\varProxyG\!<\!0.1$ and $\gmag\!>\!18.3$~mag.

To identify the time ranges that are associated with the stripes, and the sources that are in the stripes, we use the Gaia HEALPix Time Extraction tool described in Holl et al. (in preparation).
They are shown in blue in Fig.~\ref{Fig:sky_varProxyGLT0p063_GGT18p3}, bottom panel.
They contain 1\,265\,533 sources, of which 514\,084 sources have $\varProxyG\!<\!0.1$ and $\gmag\!>\!18.3$~mag.
The amplitude proxy of these latter sources are potentially dominated by noise rather than stellar variability, and are excluded from our sample.
After this last filter on sources potentially affected in the identified bad time intervals, our sample contains 23\,316\,261 large-amplitude candidates.

\subsection{Filter faint candidates with large variability amplitudes}
\label{Appendix:DatasetA_faintG}

\begin{figure}
	\centering
	\includegraphics[width=\linewidth]{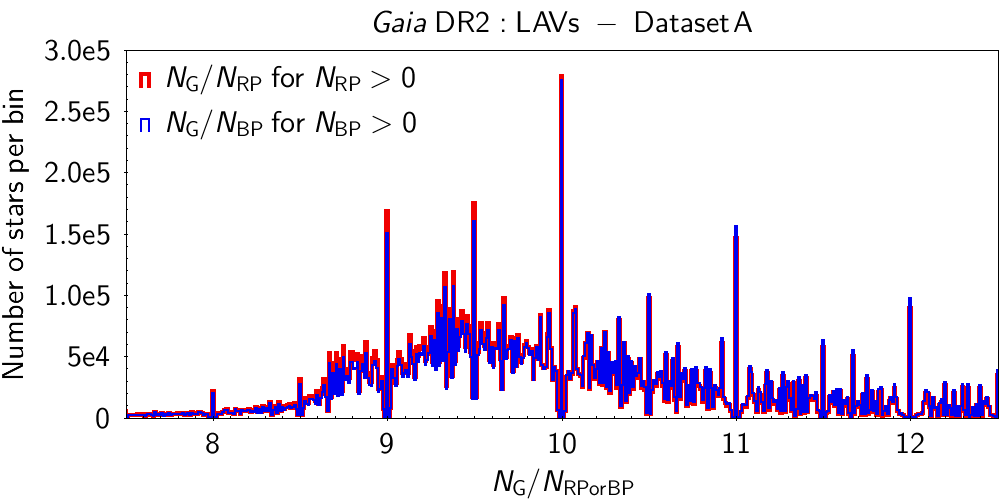}
	\caption{Ratio of the number of CCD observations in \gmag to the number of transit observations in \grp (thick red) and \gbp (thin blue) for all sources in Dataset~A that have non-zero measurements in \grp and \gbp, respectively.
	Bins are 0.01 wide.
	}
	\label{Fig:histo_ratioNobsRPBPG_zoom_datasetA}
\end{figure}

Since we are dealing with large amplitude variables, we must ensure, to the best possible way, that none of the epoch measurements in the \gmag time series becomes too faint, or else the faintest epoch measurements will be missed or have too large uncertainties, and the amplitude of the recorded light curve will be affected.

However, we do not have access to the DR2 epoch photometry for all stars.
We therefore use a trick to identify (and exclude) these large-amplitude faint sources, using \varProxyG.
If we assume $\rangeG \simeq 3.3\, \varProxyG$ (see Eq.~\ref{Eq:rangeGvsVarProxyG} in the main text), we estimate the faintest epoch measurement in the (not-availble) time series to be equal to $\gmag + 1.65 \, \varProxyG$.
We then keep only sources that have
\begin{equation}
  \gmag + 1.65 \, \varProxyG < 20.5 \; .
\label{Eq:condition_G_AproxyG}
\end{equation}
This filtering removes 387 sources from the sample determined so far, leading to a final list of 23\,315\,874 large-amplitude variable candidates in Dataset~A.

\subsection{Summary}
\label{Appendix:DatasetA_Summary}

Here we present some characteristics of Dataset~A in complement to the analysis presented in the main body of the article.

\paragraph{a) Number of measurements.}

The histograms of the number of good CCD measurements for \gmag is shown in Fig.~\ref{Fig:histo_NobsG_zoom_datasetA} for Dataset~A.
The wiggles observed in the histogram occur at multiples of nine.
They reflect the CCD distributions in the \Gaia focal plane, each transit in the astrometric field going through nine CCDs in six cases out of seven, and through eight CCDs in one case out of seven \citep{GaiaPrustiDeBruijne_etal16}.

The number of observations in \gbp and \grp are shown in Fig.~\ref{Fig:histo_NobsBPRP_zoom_datasetA}.
About one fourth of the sources have no measurement in \gbp and \grp (5\,430\,305 sources exactly, that is 23\% of Dataset~A), and another 105\,025 sources lack measurement in either one of the two bands (about half for each band).
Figure~\ref{Fig:histo_NobsBPRP_zoom_datasetA} further shows that the number of measurements are about equal in both bands.

\begin{figure}
	\centering
	\includegraphics[width=\linewidth]{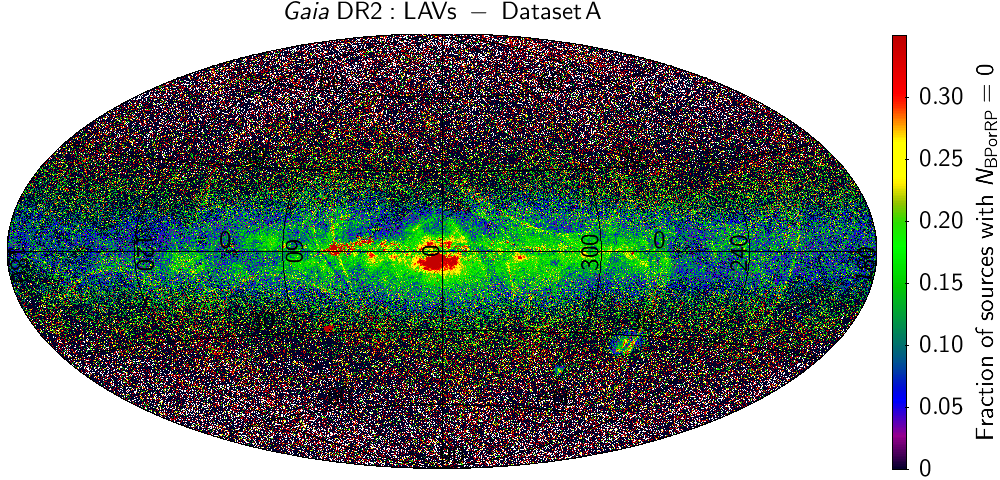}
	\caption{Fraction of sources from Dataset~A within each level 8 HEALPix tile on the sky that have no observation in either \gbp or \grp.
	         The fraction is displayed according to the linear colour scale shown on the right of the panel, sources with a fraction larger than 0.35 being rendered in red.
	         The sky is represented in Galactic coordinates.
	}
	\label{Fig:sky_datasetA_fracNobsBpRpOverG}
\end{figure}

\begin{figure}
	\centering
	\includegraphics[width=\linewidth]{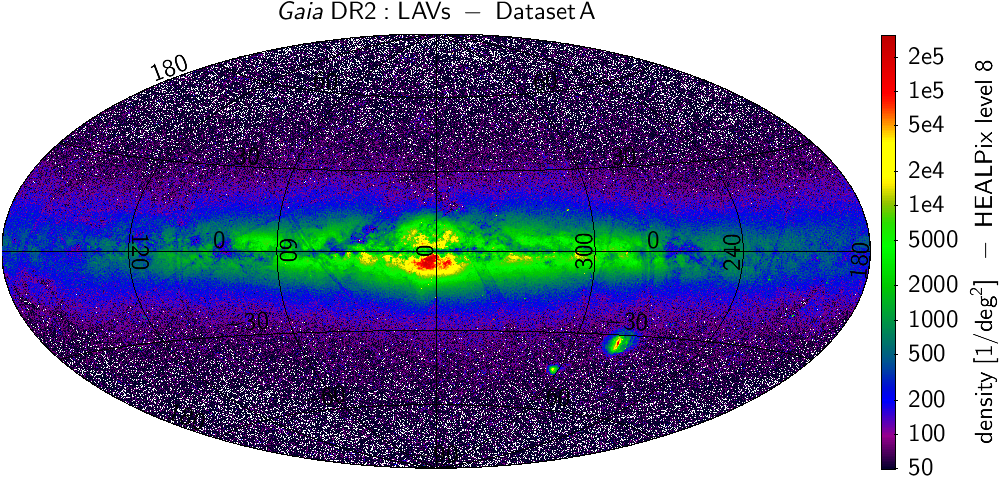}
	\caption{Sky density of all LAV candidates in Dataset~A (Galactic coordinates).
	         The density of sources is counted within each tile of a HEALPix level 8 sky division, and reported per unit square degree according to the logarithmic colour scale shown on the right of the panel.
	}
	\label{Fig:sky_datasetA_densityAll}
\end{figure}

The number of CCD observations in \gmag divided by nine, to get an average number of transits that is comparable to the number of observations in \gbp and \grp, is also shown in Fig.~\ref{Fig:histo_NobsBPRP_zoom_datasetA}.
It is seen that sources have, in general, fewer transit measurements in \gbp and \grp than in \gmag.
This is especially true for sources with fewer than 20 transits, where the peaks of the distributions are located at lower values for \gbp and \grp than for \gmag.
This shortness of \gbp and \grp measurements relative to \gmag is confirmed from the distribution of $\NobsG/\NobsBP$ and $\NobsG/\NobsRP$ shown in Fig.~\ref{Fig:histo_ratioNobsRPBPG_zoom_datasetA}.
These ratios peak between 8.5 and 11 rather than between 8 and 9.
This is most probably due to the window assignment and the fact that BP and RP windows are larger than \gmag windows.
This will cause more sources in dense areas not being assigned a BP/RP window.
On top of this, in crowded regions, more windows will be truncated, and truncated BP/RP windows have not been included in the DR2 (neither will in DR3) processing.
Dense regions on the sky are obviously expected to be most affected by these effects.
Figure~\ref{Fig:sky_datasetA_fracNobsBpRpOverG} confirms this expectation, where the sky regions having the largest fraction of sources without \gbp and {\grp} measurements are located in the densest regions of the sky (shown in Fig.~\ref{Fig:sky_datasetA_densityAll}).

\paragraph{b) Variability amplitude proxy in \gmag.}


\begin{figure}
	\centering
	\includegraphics[trim={0 81pt 0 0},clip,width=\linewidth]{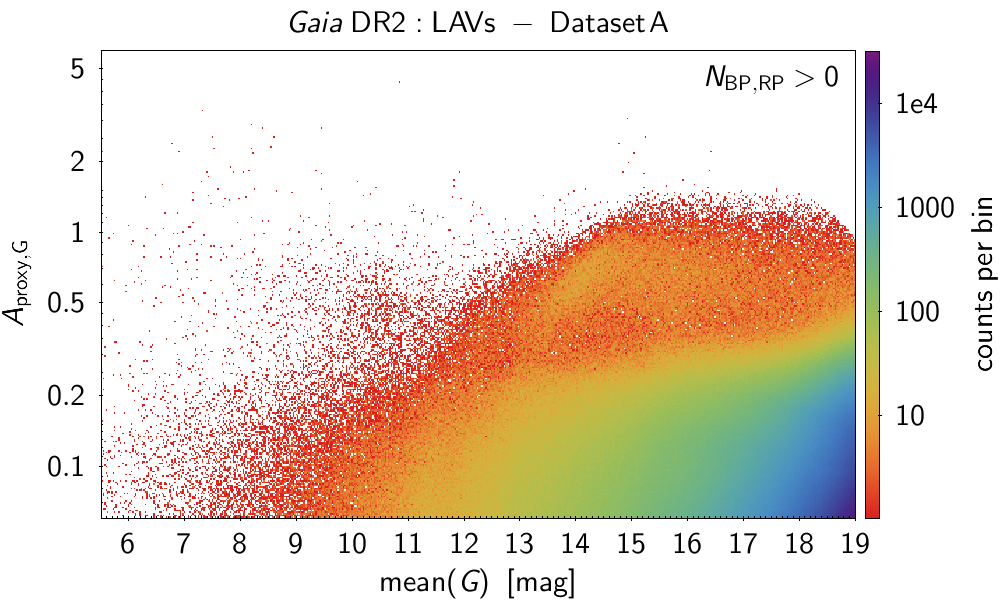}
	\vskip -0.5mm
	\includegraphics[trim={0 0 0 48pt},clip,width=\linewidth]{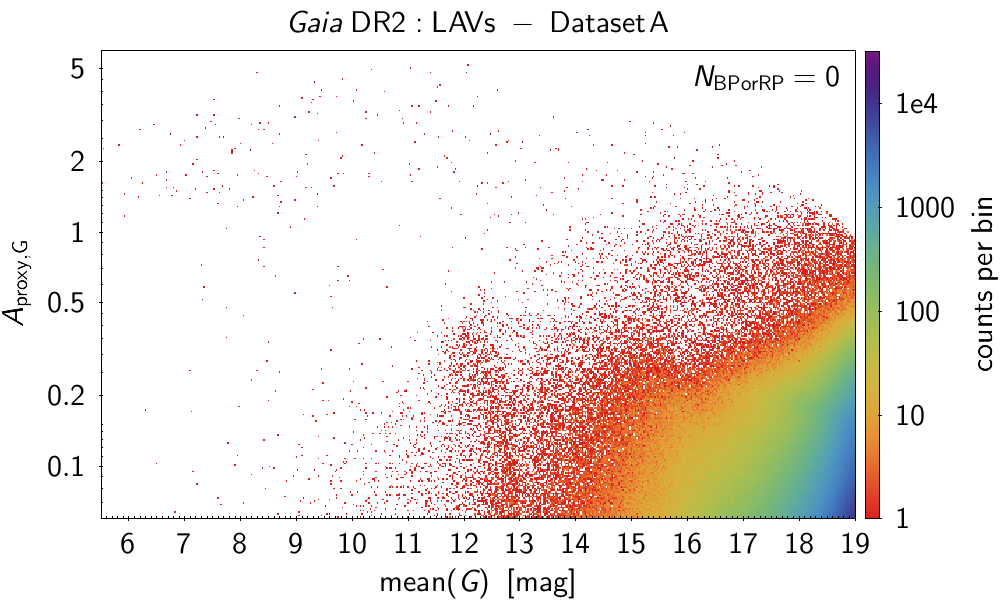}
	\caption{Density maps of the amplitude proxy \varProxyG versus mean \gmag magnitude of LAV candidates in Dataset~A with non-empty \gbp and \grp time series (upper panel), or having at least one of the two time series empty (lower panel).
	         Both panels have the same colour range, shown on the right of the each panel, giving the number of counts per bin.
	}
	\label{Fig:varProxyGvsG_datasetA}
\end{figure}

The distribution of \varProxyG versus \gmag for Dataset~A is shown in the upper panel of Fig.~\ref{Fig:varProxyGvsG_datasetA} for sources that have measurements in \gmag, \gbp and \grp.
A subset of sources with $0.3 \lesssim \varProxyG \lesssim 1$ is noticeable, reminiscent of what is expected from Miras.
The distribution of sources that lack measurements in at least one of the \gbp or \grp time series is also shown, in the bottom panel of Fig.~\ref{Fig:varProxyGvsG_datasetA}.
A density excess is seen around $\gmag \simeq 12.5$~mag for $\varProxyG \lesssim 0.5$, and around $\gmag \simeq 15.5$~mag for $\varProxyG \lesssim 0.3$.
These may be spurious variability detections due to photometric calibration issues.

\section{\fluxexcess}
\label{Appendix:BPRPexcess}

\begin{figure}
	\centering
	\includegraphics[width=\linewidth]{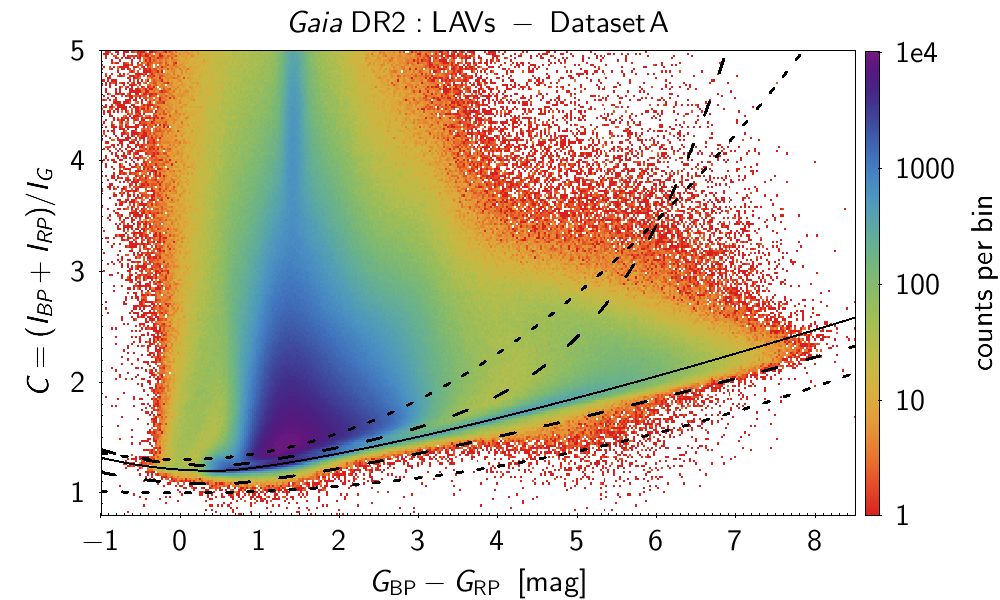}
	\caption{Density map of the {\fluxexcess} {\excessFactor} (Eq.~\ref{Eq:excessFactor}) versus $\BPminusRP$ colour for all sources in Dataset~A.
	         The solid line is the function given by Eq.~\ref{Eq:excessFactorFunction}, while the long-dashed thick lines are the limits given by Eq.~\ref{Eq:conditionGoodExcessFactor} outside of which \gbp and/or \grp are considered to be unreliable relative to \gmag.
	         For information, the limits proposed by \citet{ArenouLuriBabusiaux_etal18} are shown in short-dashed lines.
	}
	\label{Fig:excessFactor_datasetA}
\end{figure}

\begin{figure}
	\centering
	\includegraphics[width=\linewidth]{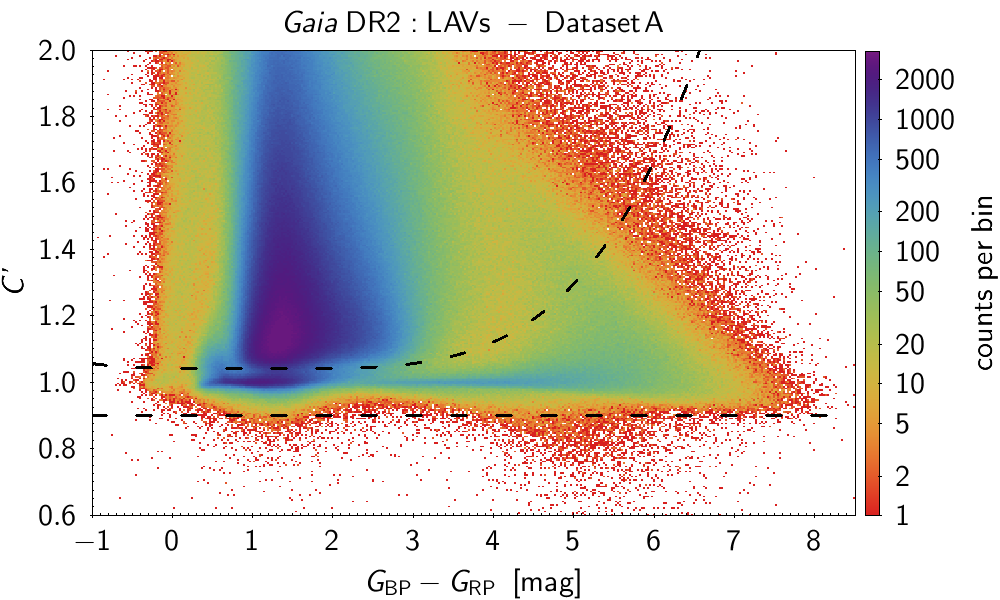}
	\caption{Same as Fig.~\ref{Fig:excessFactor_datasetA}, but for the normalized {\fluxexcess} {\excessFactorNormalized} defined by Eq.~\ref{Eq:normalizedExcessFactor}.
	         The upper dashed line is the function $1.04 + 0.001 \, (\BPminusRP-1)^3$ above which \gbp and/or \grp should not be reliable (see text).
	         The lower dashed line is the lower limit at $\excessFactorNormalized=0.9$.
	         The ordinate is zoomed compared to Fig.~\ref{Fig:excessFactor_datasetA} for better visibility.
	}
	\label{Fig:excessFactorNormalizedVsBPminusRP_datasetA}
\end{figure}

\begin{figure}
	\centering
	\includegraphics[width=\linewidth]{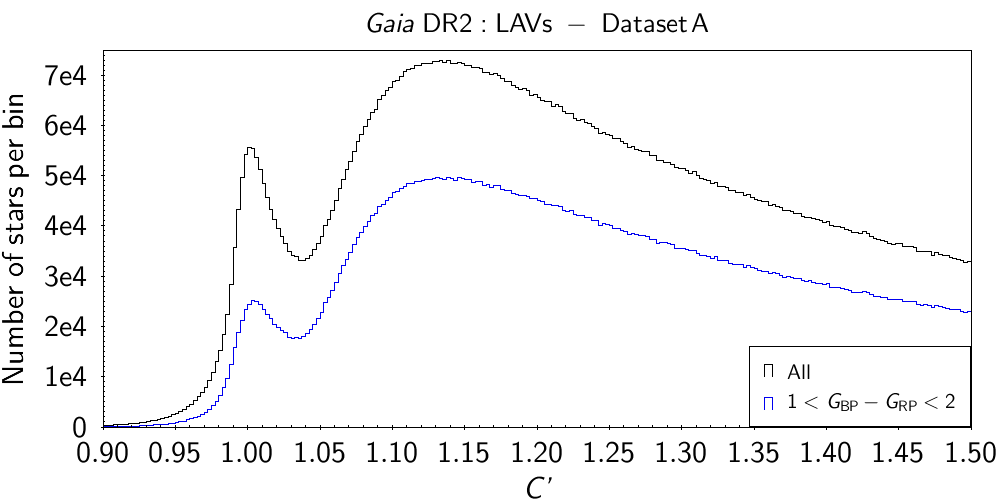}
	\caption{Histogram of the normalized \fluxexcess for all sources in Dataset~A (upper black histogram) and for sources in Dataset~A that have \BPminusRP colours between 1~mag and 2~mag (lower blue histogram).
	         The abscissa range hase been limited for better visibility.
	}
	\label{Fig:histo_Enormalized}
\end{figure}

\citet{EvansRielloDeAngeli_etal18} defined the {\fluxexcess} {\excessFactor} as
\begin{equation}
  \excessFactor = \frac{\fluxBP + \fluxRP}{\fluxG} \;.
\label{Eq:excessFactor}
\end{equation}
Its value should be close to one given the \gmag, \gbp and \grp transmission curves%
\footnote{
A combined figure of the calibrated DR2 passbands is provided in the \Gaia Image of the Week \textsl{IoW\_20180316} published on the ESA \Gaia web pages at \url{https://www.cosmos.esa.int/web/gaia/iow_20180316}, while the nominal pre-launch version is available in \cite{JordiGebranCarrasco10}.} \citep[see Sect.~8 and Figs.~20 and 21 in][]{EvansRielloDeAngeli_etal18}.

The value of \excessFactor versus \BPminusRP is shown in Fig.~\ref{Fig:excessFactor_datasetA} for all sources in Dataset~A.
The band of well-behaved single sources identified by \citet{EvansRielloDeAngeli_etal18} in their Fig.~17 is well visible in Fig.~\ref{Fig:excessFactor_datasetA}.
We define it more precisely with the following fiducial \fluxexcess function:
\begin{align}
\label{Eq:excessFactorFunction}
  \excessFactorFiducial & = \\
                &\hspace{-0.5cm} 1.2 + 0.06 \, (\BPminusRP-0.4)^2 \;\;\;\;\;\;\;\;\;\;\mathrm{if}\;(\BPminusRP<0.6)\;, \nonumber\\
                &\hspace{-0.5cm} 1.2024 + 0.1 \, (\BPminusRP-0.6)^{1.27} \;\,\;\;\mathrm{if}\;(\BPminusRP>0.6)\;. \nonumber
\end{align}
The function is shown by the solid line in Fig.~\ref{Fig:excessFactor_datasetA}.
For information, the limits $1+0.015*(\BPminusRP)^2  < \excessFactor < 1.3+0.06*(\BPminusRP)^2$ proposed by \citet[][their Eq.~2]{ArenouLuriBabusiaux_etal18} to select sources with acceptable \fluxexcesses are shown by the dashed lines in Fig.~\ref{Fig:excessFactor_datasetA}.

We now define for each source the normalized {\fluxexcess} {\excessFactorNormalized} by 
\begin{equation}
  \excessFactorNormalized = \frac{\excessFactor}{\excessFactorFiducial}
\label{Eq:normalizedExcessFactor}
\end{equation}
The resulting diagram is shown in Fig.~\ref{Fig:excessFactorNormalizedVsBPminusRP_datasetA}, and the histogram of \excessFactorNormalized is plotted in Fig.~\ref{Fig:histo_Enormalized}.
A local minimum is observed in the histogram around $\excessFactorNormalized \sim 1.04$, suggesting an upper limit around this value for well-behaved single sources.
We therefore adopt the colour-dependent upper limit \excessFactorNormalizedLimit identifying well-behaved sources as $\excessFactorNormalizedLimit = 1.04 + 0.001 \, (\BPminusRP-1)^3$.
This limit is shown by the upper dashed line in Figs.~\ref{Fig:excessFactor_datasetA} and \ref{Fig:excessFactorNormalizedVsBPminusRP_datasetA}.
The characteristics of LAVs with \excessFactorNormalized values larger than this limit are analysed in the next section.
Sources with too low \fluxexcesses are then checked in Sect.~\ref{Appendix:BPRPexcess_lowFactors}.

\subsection{Large \fluxexcesses}
\label{Appendix:BPRPexcess_groupL}

Several causes of the large \fluxexcesses have been presented in \citet{EvansRielloDeAngeli_etal18}, to which we refer.
Here, we check several properties that could be specific to LAVs for the origin of the large \fluxexcesses.
Indeed, 95\% of Dataset~A have too large \fluxexcesses.
For this purpose, we compare a representative subset of these sources, called subset L, with a subset of well-behaved sources around the fiducial line, called subset F.
Both subsets are restricted to sources with parallax uncertainties better than 10\%.
Subset L is defined with $1.11 \!<\! \excessFactorNormalized \!<\! 1.3$ and subset F with $\excessFactorNormalized \!<\! 1.02$, such that they have statistically comparable number of sources, of 99\,833 and 97\,540, respectively.

\paragraph{Number of \gmag, \gbp, \grp measurements.}

\begin{figure}
	\centering
	\includegraphics[width=\linewidth]{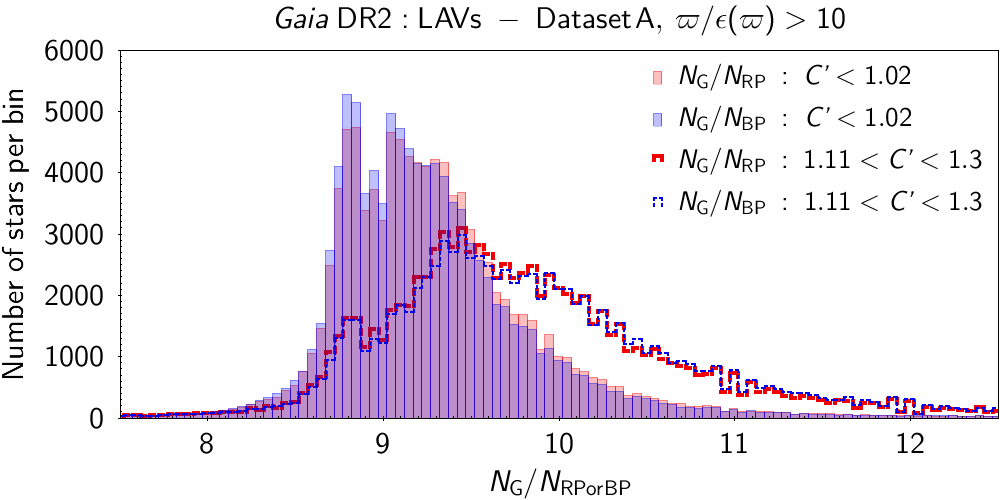}
	\caption{Same as Fig.~\ref{Fig:histo_ratioNobsRPBPG_zoom_datasetA}, but for the subsets of Dataset~A that have parallax relative uncertainties better than 10\% and either $\excessFactorNormalized \!<\! 1.02$ (filled histograms) or $1.11 \!<\! \excessFactorNormalized \!<\! 1.3$ (dashed lines).
	         Histograms of \gbp are shown in blue and those of \grp are shown in red.
	         Bins are 0.05 wide.
	} 
	\label{Fig:histo_ratioNobsRPBPG_datasetA_parOverErrGT10_subsetsFL}
\end{figure}

Because of the large amplitudes of LAVs, non-similar time sampling in \gmag, \gbp and \grp time series may lead to incompatible mean magnitudes in the three bands.
The distributions of $\NobsG / \NobsBP$ and $\NobsG / \NobsRP$ are shown in Fig.~\ref{Fig:histo_ratioNobsRPBPG_datasetA_parOverErrGT10_subsetsFL} for subsets~F (filled histograms) and L (dashed-line histograms).
They reveal, on the mean, slightly larger values of these ratios for subset~L than for subset~F, pointing to of lack of epoch measurements in \gbp and \grp time series relative to \gmag time series.
This is most probably due to the fact that these sources are located in dense regions of the sky (see below), leading to more difficult observation conditions in \bandBP and \bandRP spectrophotometry than in \gmag point-spread photometry.
$\NobsG/\NobsBP$ and $\NobsG/\NobsRP$ are, however, only 5\% to 10\% larger in subset~L than in subset~F.
It is improbable that it would be at the origin of the large \fluxexcesses observed in subset~L.

\paragraph{Location in the observational HRD.}
\begin{figure}
	\centering
	\includegraphics[trim={0 93pt 0 0},clip,width=\linewidth]{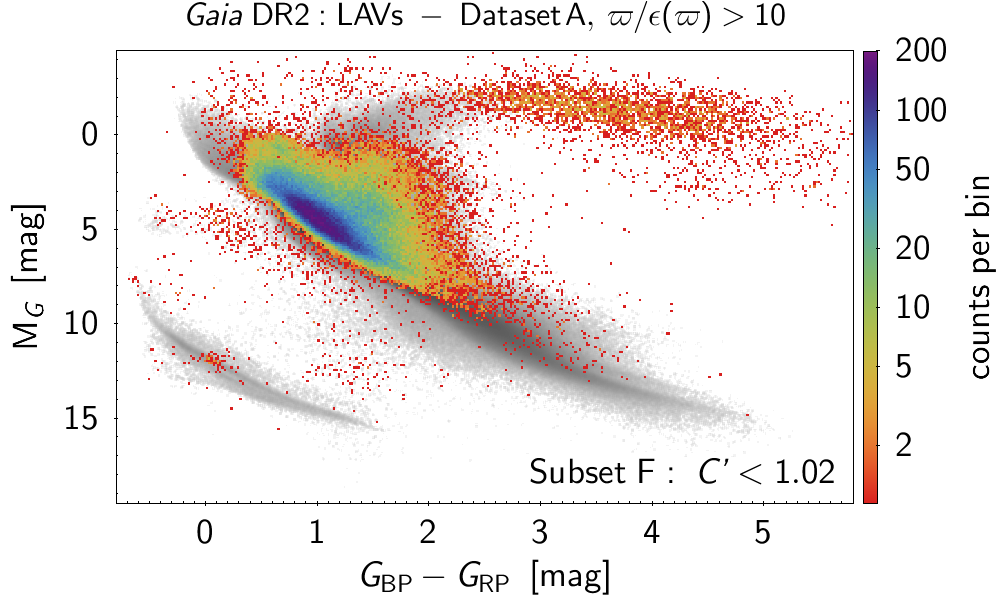}
	\vskip -0.5mm
	\includegraphics[trim={0 0 0 50pt},clip,width=\linewidth]{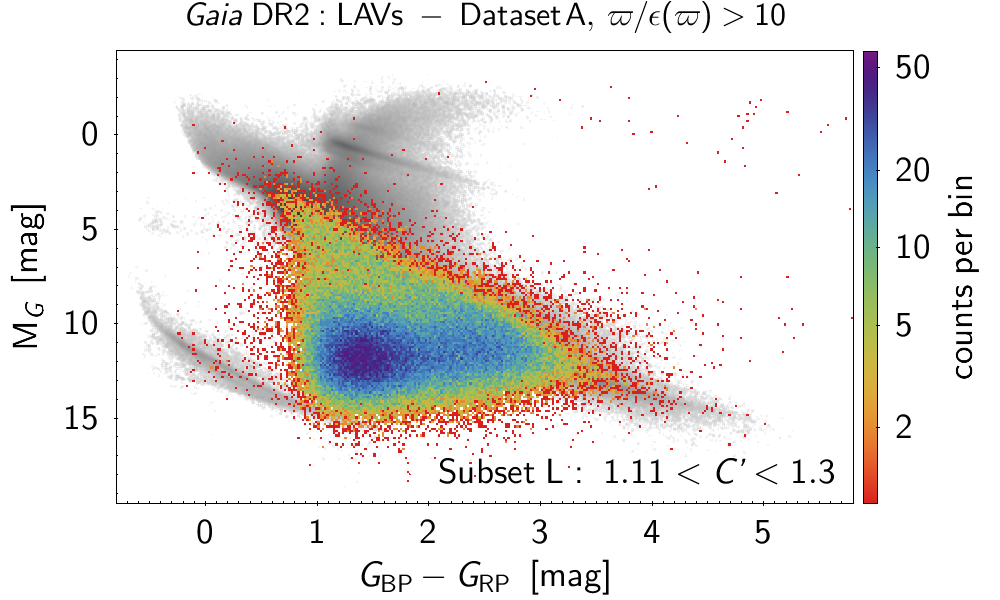}
	\caption{Density maps of the observational HRDs of two dataset~A sub-samples with relative parallax uncertainties better than 10\%: for sources with $\excessFactorNormalized \!<\! 1.02$ in the top panel (subset~F in the text) and for sources with $1.11 \!<\! \excessFactorNormalized \!<\! 1.3$ in the bottom panel (subset~L in the text).
	         The density maps are plotted on top of \Gaia DR2 sources (constant and variable) with relative parallax uncertainties better than 2.5\% (light grey in the background).
	         Density goes from low in red to high in black on a logarithmic scale.
	} 
	\label{Fig:CMabs_parallaxOverErrGT10_Enormalized}
\end{figure}

\begin{figure}
	\centering
	\includegraphics[width=\linewidth]{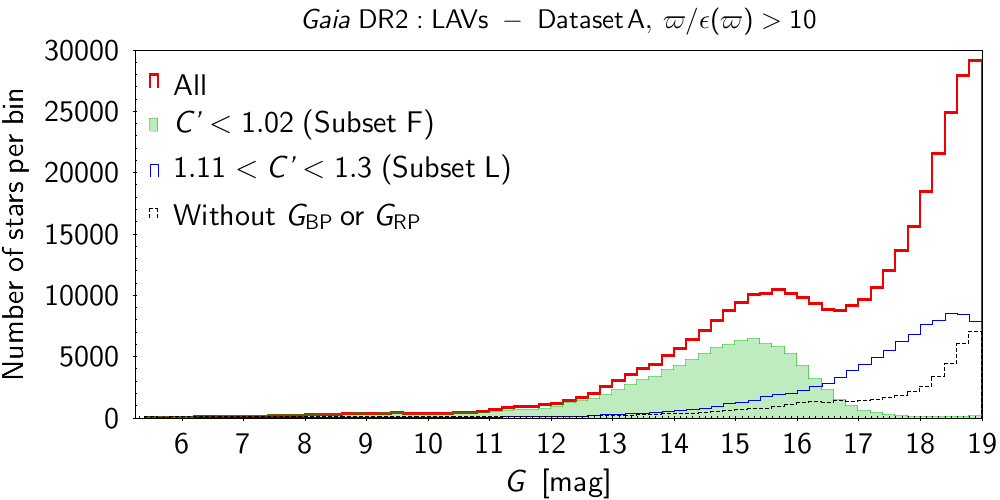}
	\caption{Histogram of \gmag magnitude for LAV candidates in Database~A that have parallax uncertainties better than 10\% (red thickline).
	         Subset~F (see text) therein that has $\excessFactorNormalized \!<\! 1.02$ is shown by the filled green histogram, and subset~L that has $1.11 \!<\! \excessFactorNormalized \!<\! 1.3$ is shown by the thin blue histogram.
	         Also shown in black dashed histogram is the subset that has no \gbp and/or \grp in the DR2 archive.
	         Bins are 0.2~mag wide.
	}
	\label{Fig:histo_magG_datasetA_parallaxOverErrorGT10}
\end{figure}

The distributions of the two subsets differ in the observational HRD, as shown in Fig.~\ref{Fig:CMabs_parallaxOverErrGT10_Enormalized}.
Subset~F populates the observational HRD in the expected regions of the diagram (top panel of Fig.~\ref{Fig:CMabs_parallaxOverErrGT10_Enormalized}).
In contrast, subset~L is mainly located in the region of the observational HRD between the MS and WD sequence where we do not expect to have many stars (bottom panel in the figure).
Subset~L sources have also fainter apparent magnitudes, on the mean, than subset~F sources (Fig.~\ref{Fig:histo_magG_datasetA_parallaxOverErrorGT10}), with $\gmag \lesssim 17.5$~mag for the majority of subset~L (blue histogram), while the bulk of subset~F has $12 \lesssim \gmag\,\mathrm{[mag]} \lesssim 17$ (green filled histogram).

\paragraph{Sky distribution.}

\begin{figure}
	\centering
	\includegraphics[trim={0 0 0 0},clip,width=\linewidth]{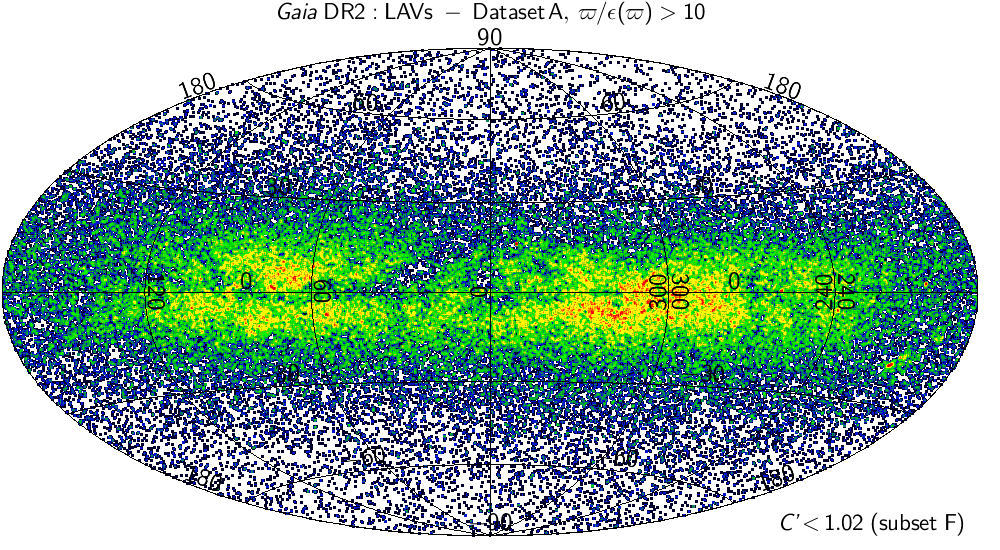}
	\vskip -0.5mm
	\includegraphics[trim={0 0 0 50pt},clip,width=\linewidth]{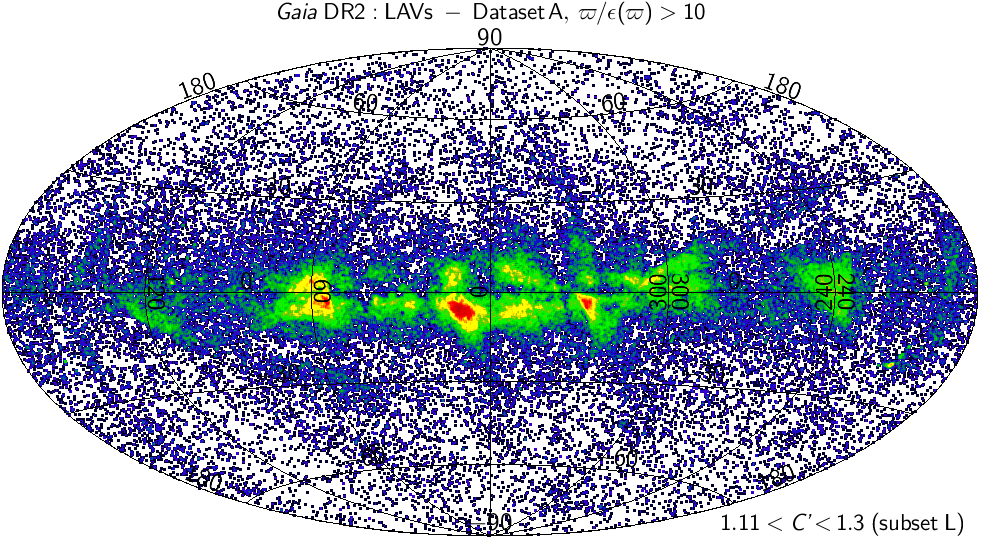}
	\caption{Sky density (Galactic coordinates) of LAV candidates in Database~A that have parallax uncertainties better than 10\% and either $\excessFactorNormalized \!<\! 1.02$ (top panel; subset~F in the text) or $1.11 \!<\! \excessFactorNormalized \!<\! 1.3$ (bottom panel; subset~L in the text).
	         Density goes from low in black to high in red on a logarithmic scale.
	}
	\label{Fig:skyDensity_datasetA_comparisonEnorm}
\end{figure}

The distributions of the two subsets also differ in the sky, as shown in Fig.~\ref{Fig:skyDensity_datasetA_comparisonEnorm}.
Subset~F (top panel) is relatively homogeneously distributed on the sky, with a predominance in the Galactic plane.
In contrast, subset~L sources (bottom panel) are mainly clumped towards specific regions of the Galactic plane, and predominantly towards the Bulge.
Their sky distribution points towards regions with dense regions and with large extinction, which is compatible with them being predominantly faint.

\paragraph{Astrometric solution.}

\begin{figure}
	\centering
	\includegraphics[trim={0 93pt 0 0},clip,width=\linewidth]{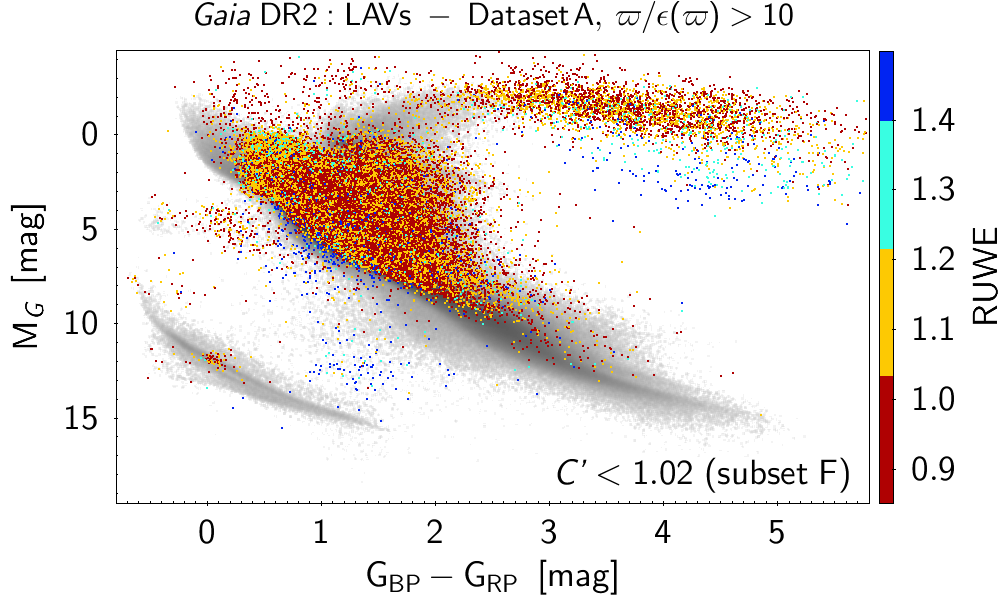}
	\vskip -0.5mm
	\includegraphics[trim={0 0 0 50pt},clip,width=\linewidth]{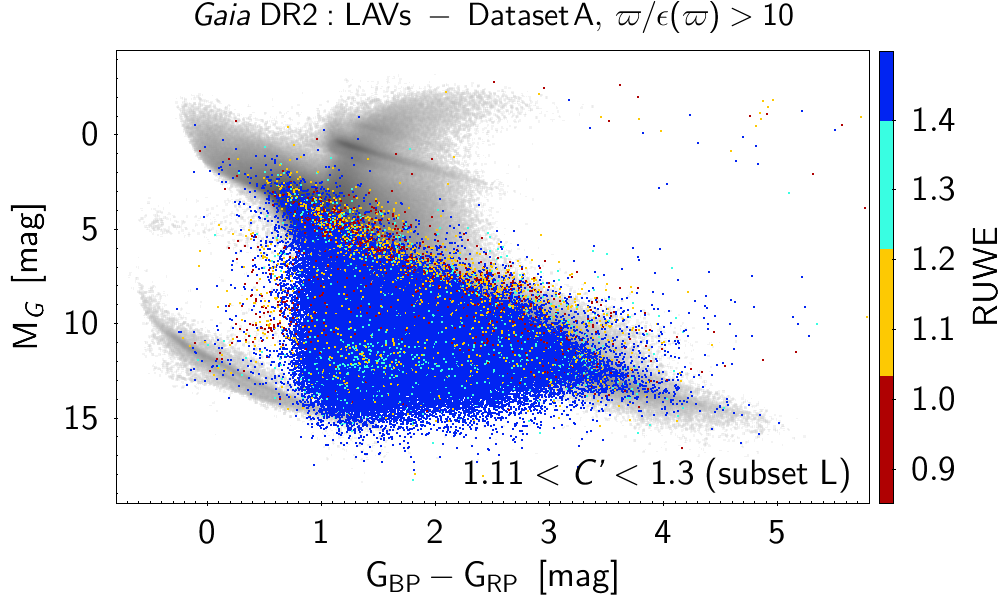}
	\caption{Same as Fig.~\ref{Fig:CMabs_parallaxOverErrGT10_Enormalized}, but colour coded with the value of RUWE according to the colour-scale shown on the right of the figure.
	         RUWE values larger than 1.5 are rendered in blue and values smaller than 0.85 are rendered in red.
	} 
	\label{Fig:CMabs_parallaxOverErrGT10_Enormalized_withRuwe}
\end{figure}

\begin{figure}
	\centering
	\includegraphics[trim={0 88pt 0 0},clip,width=\linewidth]{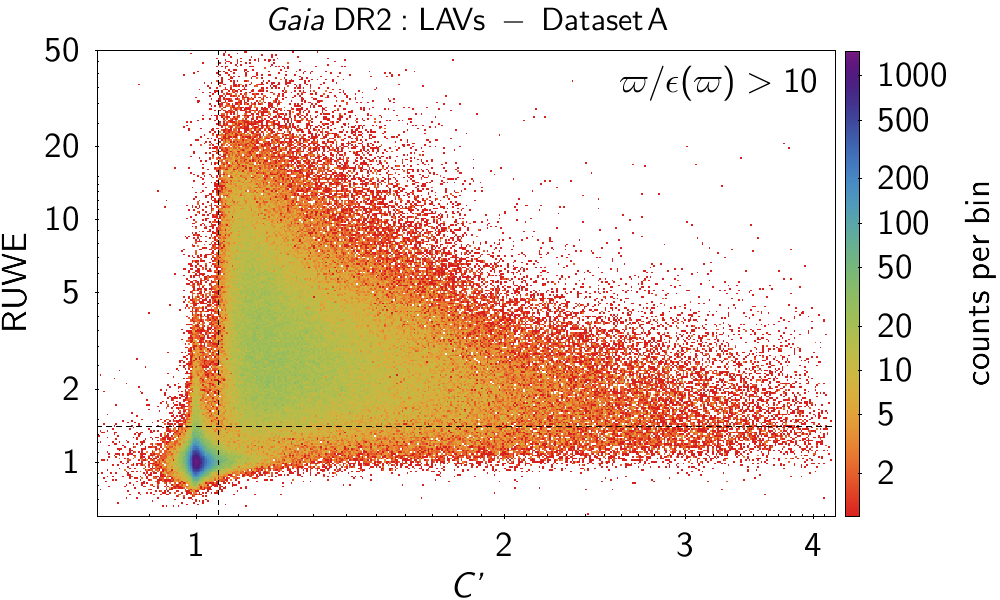}
	\vskip -0.5mm
	\includegraphics[trim={0 0 0 50pt},clip,width=\linewidth]{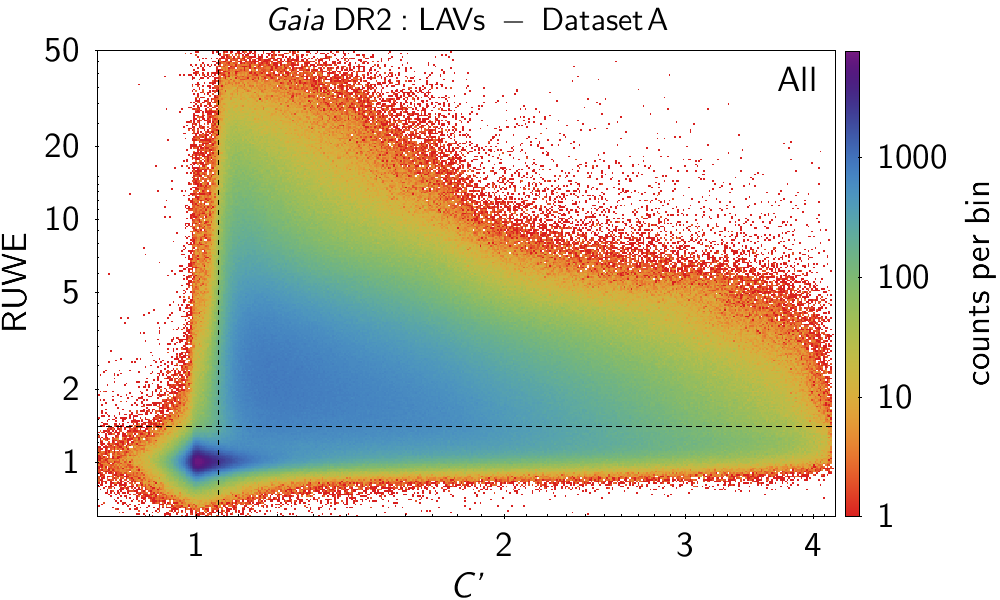}
	\caption{Density map of the Renormalized Unit Weight Excess versus normalized \fluxexcess for sources in Dataset~A that have parallax uncertainties better than 10\% (top panel).
	         The same figure, but for all sources in Dataset~A, is shown in the bottom panel.
	         Vertical and horizontal dotted lines have been drawn at $\excessFactorNormalized = 1.04$ and $\mathrm{RUWE} = 1.4$, respectively, as eye guides.
	         The axes ranges have been limited for better visibility.
	}
	\label{Fig:excessNormalizedVsRuwe_datasetA}
\end{figure}

The specific properties of subset~L, that is being faint, located in dense regions of the sky, and having large \fluxexcesses, raise the question of the validity of their astrometric solution.
One parameter to check in this respect is the Renormalized Unit Weight Excess (RUWE) from the astrometric solution.
It is a goodness of fit parameter that quantifies the departure from an astrometric single star model fit.
In this respect, it can be a very useful indicator of astrometric multiplicity in sources (e.g. binaries), as has been demonstrated by \citet{BelokurovPenoyreOh_etal20}.
These authors selected a high-quality sample of \Gaia DR2 stars located at high latitudes, with low extinction, and having low \fluxexcesses, among other filtering criteria.
In the vast majority of cases, however, large RUWE values can also be due to various uncalibrated effects in DR2.
Therefore, RUWE is commonly used in DR2 to identify inaccurate astrometric solutions, but at the expense of also removing potential binary sources.
In this paper, we do not use RUWE as a criterion to filter the photometric data, given that the relation between the two is not clearly established.
We explore this a little more below.

The RUWE is plotted colour-coded in the observational HRD shown in Fig.~\ref{Fig:CMabs_parallaxOverErrGT10_Enormalized_withRuwe} for both subsets~F (upper panel) and L (bottom panel).
Subset~F sources (upper panel) have for the great majority of them RUWE valus below 1.4, the limit proposed on the \Gaia DR2 known issues pages below which the astrometry is reliable.
In contrast, almost all subset~L sources (lower panel) that are between the MS and the WD sequence have RUWE values above 1.4, 
It is interesting in this respect to note that the few subset~F sources that are located in that same region of the observational HRD also have large RUWE values.
They are reminiscent of the DR2 problematic astrometric cases highlighted in the known issues on the \Gaia web pages \footnote{See presentation \url{https://www.cosmos.esa.int/documents/29201/1770596/Lindegren_GaiaDR2_Astrometry_extended.pdf/1ebddb25-f010-6437-cb14-0e360e2d9f09} mentioned in the \Gaia web page of known issues for DR2  \url{https://www.cosmos.esa.int/web/gaia/dr2-known-issues}}.

There is actually a bi-modal distribution in the \excessFactorNormalized versus RUWE plane, as shown in Fig.~\ref{Fig:excessNormalizedVsRuwe_datasetA}, with a first peak around $(\excessFactorNormalized,\mathrm{RUWE})=(1,1)$, and a second concentration of points in the region $\excessFactorNormalized \gtrsim 1.05$ and $\mathrm{RUWE}\gtrsim 1.4$.
The origin of this bi-modality is not known.
The two quantities could be linked with each other in dense regions.
However, RUWE values are also directly impacted by the \fluxexcess, because the normalization factor depends on the position of the colour-magnitude diagram.

It must also be noted that large RUWE values do not necessarily imply bad astrometry.
Small uncertainties on sky positions may lead to large RUWE values.
Caution must thus be taken when imposing a limit on RUWE, as this could remove good astrometric cases.
Here, we do not impose any limit on RUWE.

Finally, we checked that there is no specific correlation between RUWE and photometric variability amplitude.
This excludes an effect of photometric amplitude on astrometry.

\subsection{Low \fluxexcesses}
\label{Appendix:BPRPexcess_lowFactors}

\begin{figure}
	\centering
	\includegraphics[width=\linewidth]{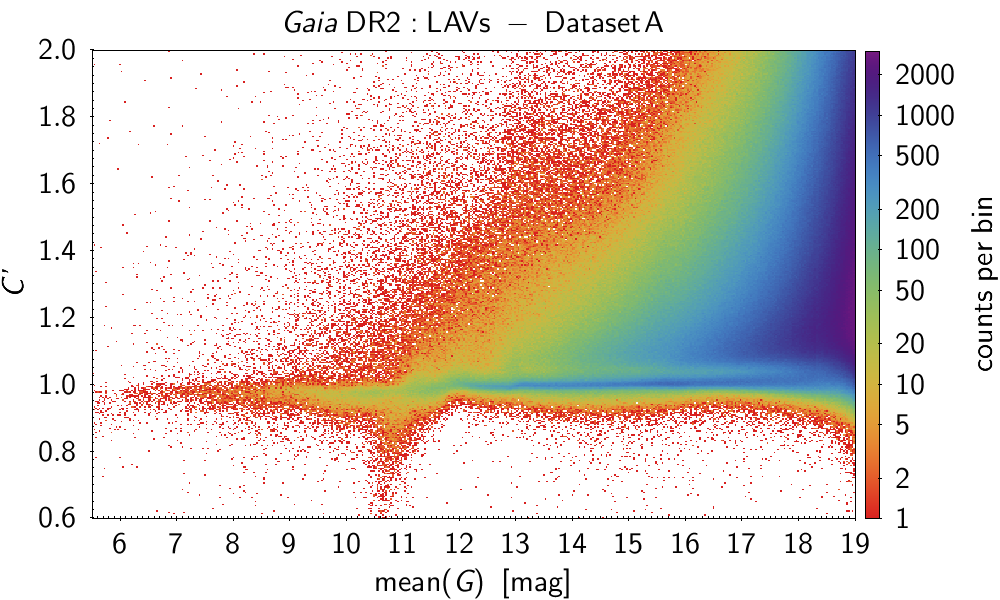}
	\caption{Same as Fig.~\ref{Fig:excessFactorNormalizedVsBPminusRP_datasetA}, but versus \gmag magnitude.
	}
	\label{Fig:excessFactorNormalizedVsG_datasetA}
\end{figure}

\begin{figure}
	\centering
	\includegraphics[width=\linewidth]{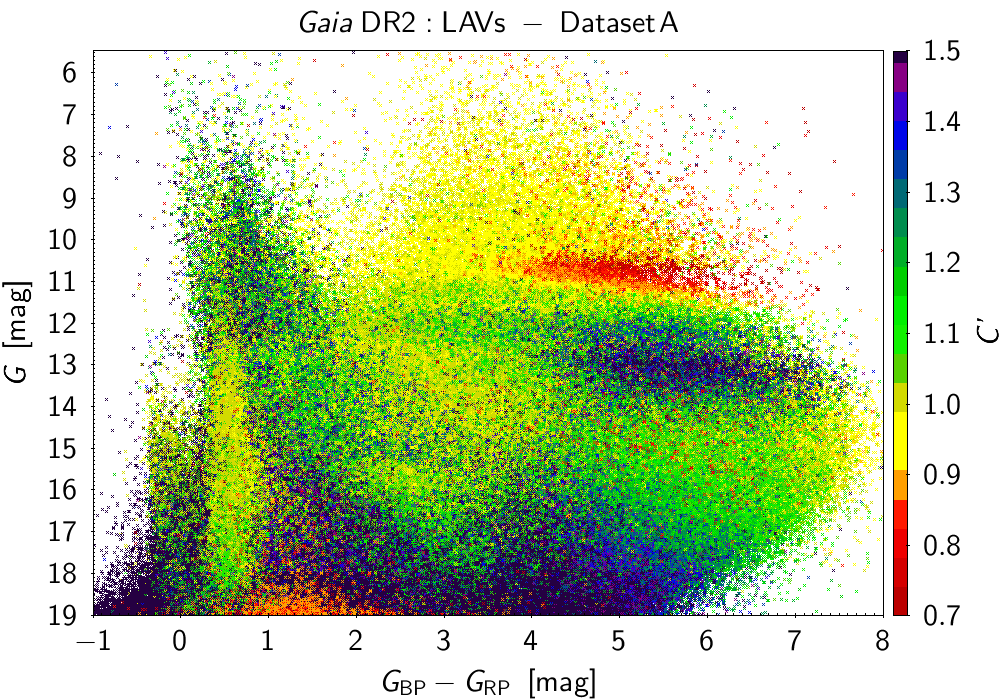}
	\caption{Colour-magnitude diagram of Dataset~A with the normalized \fluxexcess colour-coded according to the colour scale shown on the right of the figure.
	         The colour scale has been limited between 0.7 and 1.5, sources with \excessFactorNormalized values outside this range being rendered in the colour at the respective end of the scale.
	         Sources with $\excessFactorNormalized<0.9$ have been plotted on top of other sources for better visibility as they contain a very small number of sources compared to the size of Dataset~A.
	}
	\label{Fig:CM_withExcessFactorNormalized_datasetA}
\end{figure}

A (small) fraction of Dataset~A LAVs have too low \fluxexcesses compared to the fiducial values, especially for red sources (see Fig.~\ref{Fig:excessFactorNormalizedVsBPminusRP_datasetA}), a feature not observed (at least for red stars) in Fig.~17 of \citet{EvansRielloDeAngeli_etal18}.
It mainly occurs at magnitudes between $10 \lesssim \gmag\,\mathrm{[mag]} \lesssim 11.5$, as shown in Fig.~\ref{Fig:excessFactorNormalizedVsG_datasetA}.
Figure~\ref{Fig:CM_withExcessFactorNormalized_datasetA}, which colour-codes the normalized \fluxexcess across the colour-magnitude diagram, reveals that it significantly impacts bright red stars with $\BPminusRP \gtrsim 4$~mag in that magnitude range.

The too-low \fluxexcesses imply \fluxG fluxes that are too high with respect to \fluxBP$+$\fluxRP.
We adopt a lower limit of $\excessFactorNormalized=0.9$ below which \excessFactorNormalized is considered unreliable.
The number of sources in Dataset~A with $\excessFactorNormalized<0.9$ is only few thousand (see Table~\ref{Tab:datasetsSummary}).
However, they are at the origin of a significant shortage of the reddest LPVs at $\gmag\simeq$11~mag in Datasets~B and C defined in the main text of this article, well visible in CM diagrams like in Fig.~\ref{Fig:CM_density_DatasetC}.

\subsection{Summary}
\label{Appendix:BPRPexcess_summary}

In summary, our final condition to select sources with reliable (normalized) \fluxexcess is
\begin{equation}
  0.9 < \excessFactorNormalized < 1.04 + 0.001 \, (\BPminusRP-1)^3 \;\; .
\label{Eq:conditionGoodExcessFactor}
\end{equation}
It is used in the construction of Dataset~B, which results in the selection of only 5\% of LAV candidates from Dataset~A to have reliable \fluxexcesses.
For the 95\% remaining candidates, the photometric values are to be taken with care, mainly, but not only (see below), due to poor \gbp and \grp quality in DR2.

It must be noted that in the DR2 papers and documentation the \fluxexcess has been presented as a quality flag.
This parameter, however, simply informs on the consistency between \gmag, \gbp and \grp fluxes.
While for DR2 it is probably true that in many cases the BP/RP integrated fluxes are of poorer quality due mostly to lower-resolution background calibration and contamination/blend cases (see known DR2 issues listed on the \Gaia DR2 Web site%
\footnote{
\url{https://gea.esac.esa.int/archive/documentation/GDR2/index.html}
}%
),
similar problems (in particular the handling of close pairs and extended sources) will also affect \gmag-band measurements.
In the case of extended sources, for instance, the \gmag-band measurements may show larger variations than for BP/RP due to the smaller window and different scan directions, and we know already that there were some misclassifications of extended sources as RR~Lyrae in DR2.
The red LAVs with too-low \fluxexcesses described in Sect.~\ref{Appendix:BPRPexcess_lowFactors} give another example of cases where \gmag is in fault rather than \gbp or \grp.
The user should thus keep in mind that, in principle, outliers in the distribution of \fluxexcess could be due to problems in any of the bands.

\section{Amplitude proxy for BP$+$RP}
\label{Appendix:AproxyBPplusRP}

In this appendix, we derive the amplitude proxy \varProxyBPplusRP for the summed \bandBPplusRP magnitude.
The derivation is based on the variances $\sigma^2(\epochFluxBP)$ and $\sigma^2(\epochFluxRP)$ of the \epochFluxBP and \epochFluxRP time series and on the covariance $\mathrm{Cov}(\epochFluxBP,\epochFluxRP)$ between these two time series, defined by
\begin{align}
  \sigma^2(\epochFluxBP) &= \frac{1}{\NobsBP} \sum_i({\epochFluxBP}_{\!,\,i}-\fluxBP)^2 \label{Eq:varianceBP} \, , \\
  \sigma^2(\epochFluxRP) &= \frac{1}{\NobsRP} \sum_j({\epochFluxRP}_{\!,\,j}-\fluxRP)^2 \label{Eq:varianceRP} \, , \\
  \mathrm{Cov}(\epochFluxBP,\epochFluxRP) &= \frac{1}{\NobsRPandBP} \sum_k({\epochFluxBP}_{\!,\,k}-\fluxBP)({\epochFluxRP}_{\!,\,k}-\fluxRP) \, .
\label{Eq:covarianceBPRP}
\end{align}

The variance $\sigma^2(\epochFluxBP + \epochFluxRP)$ of the summed flux $\epochFluxBP + \epochFluxRP$ is then given by
\begin{equation}
  \sigma^2(\epochFluxBP + \epochFluxRP) = \sigma^2(\epochFluxBP) + \sigma^2(\epochFluxRP) + 2 \; \mathrm{Cov}(\epochFluxBP,\epochFluxRP) \, .
\label{Eq:varianceBPplusRP}
\end{equation}

It must be noted that covariance is usually defined for random variables.
In the case of \epochFluxBP and \epochFluxRP, they could be considered independent if we neglect the minimal frequency overlap between the two band passes.
They can, however, be correlated due, for example, to crowdedness simultaneously affecting \gbp  and \grp, or to correlated perturbation induced by stray light.
Aging of the optic/electronics and the occurence of solar storms, are other examples impacting both \gbp and \grp.
Besides, the physics behind variable stars leads, in most cases, to a coordinated variability of the flux throughout the optical spectrum.
All these effects lead to a non-zero covariance, as shown later in the next section (see in particular Fig.~\ref{Fig:histo_covProxyBPRP}).

\subsection{Definition}
\label{Appendix:AproxyBPplusRP_definition}
We define an amplitude proxy \varProxyBPplusRP for the summed \bandBPplusRP flux in the same way as we defined the amplitude proxies for \epochFluxG, \epochFluxBP and \epochFluxRP:
\begin{equation}
  \varProxyBPplusRPsquare = \frac{\sigma^2(\epochFluxBP + \epochFluxRP)}{(\fluxBP+\fluxRP)^2} \;,
\label{Eq:varProxyBPplusRPsquare}
\end{equation}
which becomes, using Eq.~\ref{Eq:varianceBPplusRP},
\begin{align}
  \varProxyBPplusRPsquare &= \frac{\fluxBP^{\,2}}{(\fluxBP+\fluxRP)^2}\;\varProxyBPsquare \nonumber\\
                          &\;\;\;\;\; + \frac{\fluxRP^{\,2}}{(\fluxBP+\fluxRP)^2}\;\varProxyRPsquare \nonumber\\
                          &\;\;\;\;\; + \frac{2 \; \mathrm{Cov}(\epochFluxBP,\epochFluxRP)}{(\fluxBP+\fluxRP)^2} \, .
\label{Eq:varProxyBPplusRPsquare_withFluxes}
\end{align}

We now define the last term in Eq.~\ref{Eq:varProxyBPplusRPsquare_withFluxes} as a 'covariance proxy'
\begin{equation}
  \varProxyCovBPplusRP = \frac{2 \; \mathrm{Cov}(\epochFluxBP,\epochFluxRP)}{(\fluxBP+\fluxRP)^2} \;,
\label{Eq:covarianceProxy}
\end{equation}
and Eq.~\ref{Eq:varProxyBPplusRPsquare_withFluxes} becomes
\begin{multline}
  \varProxyBPplusRP = \\
  \frac{\sqrt{\fluxBP^{\,2}\;\varProxyBPsquare + \fluxRP^{\,2}\;\varProxyRPsquare + (\fluxBP+\fluxRP)^2\;\varProxyCovBPplusRP}}
                           {(\fluxBP+\fluxRP)} \, .
\label{Eq:varProxyBPplusRPwithCov}
\end{multline}

\subsection{Computation}
\label{Appendix:AproxyBPplusRP_computation}

\begin{figure}
	\centering
	\includegraphics[width=\linewidth]{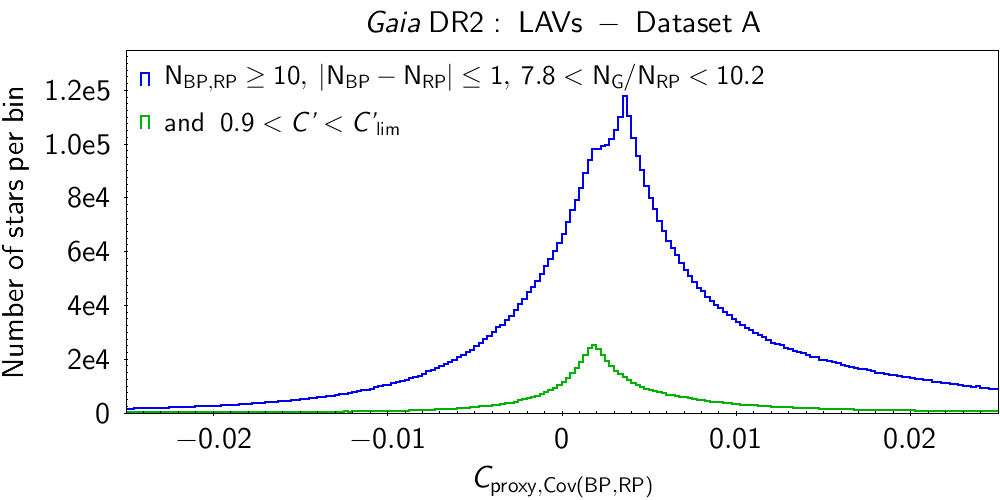}
	\caption{Histograms of the covariance proxy \varProxyCovBPplusRP defined by Eq.~\ref{Eq:varProxyBPplusRPwithCov} and computed from the information published in \Gaia DR2 using Eqs.~\ref{Eq:varProxyBPplusRPwithCov} and \ref{Eq:varProxyBPplusRP_varProxyG}.
	         The blue (upper) histogram represents the sample of Dataset~A satisfying the conditions $N_\mathrm{BP,RP}\ge10$, $|\NobsBP-\NobsRP| \le 1$, and $7.8 < N_{obs}(G)\,/\,N_{obs}(RP) < 10.2$.
	         The sub-sample therein having reliable \fluxexcesses according to condition \ref{Eq:conditionGoodExcessFactor} is shown by the green (lower) histogram.
	         Bins are $2.5 \, 10^{-4}$ wide.
	}
	\label{Fig:histo_covProxyBPRP}
\end{figure}

An estimate of the covariance proxy \varProxyCovBPplusRP (Eq.~\ref{Eq:covarianceProxy}) can be obtained under the assumption that
\begin{equation}
  \varProxyBPplusRP \simeq \varProxyG \; .
\label{Eq:varProxyBPplusRP_varProxyG}
\end{equation}
This assumption is debatable for cases with bad \fluxexcesses, but has the advantage to allow an estimate of \varProxyCovBPplusRP by combining Eqs.~\ref{Eq:varProxyBPplusRPwithCov} and \ref{Eq:varProxyBPplusRP_varProxyG}.
The histogram of the resulting \varProxyCovBPplusRP values is shown in Fig.~\ref{Fig:histo_covProxyBPRP} for the subset of Dataset~A that has similar number of transit measurements in \gmag, \gbp and \grp for each source (see figure caption), a condition that is necessary for a valid comparison of the properties of \epochFluxG, \epochFluxBP and \epochFluxRP time series given the large amplitudes considered here.

\begin{figure}
	\centering
	\includegraphics[trim={0 80pt 0 0},clip,width=\linewidth]{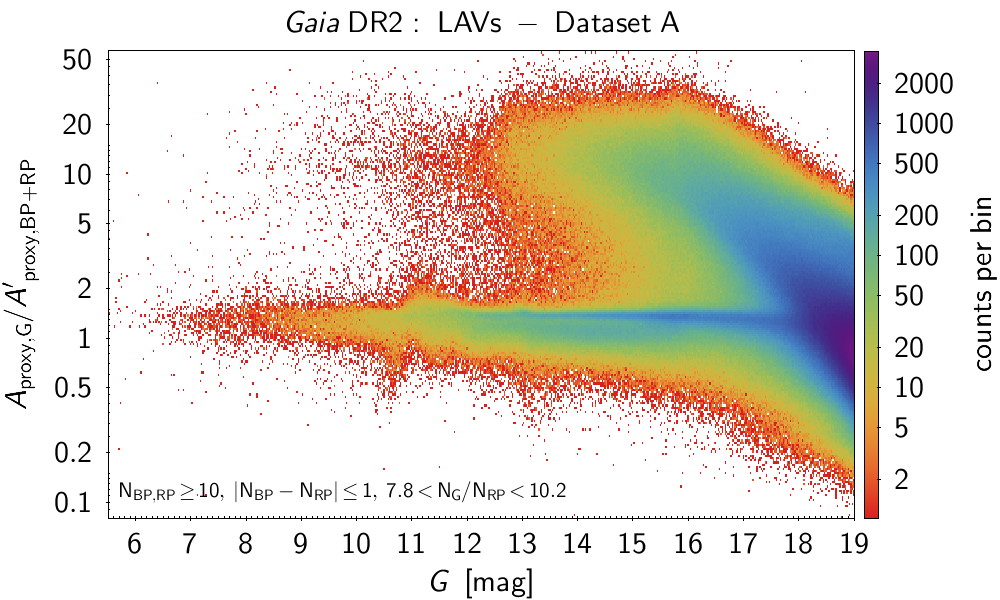}
	\vskip -0.5mm
	\includegraphics[trim={0 0 0 50pt},clip,width=\linewidth]{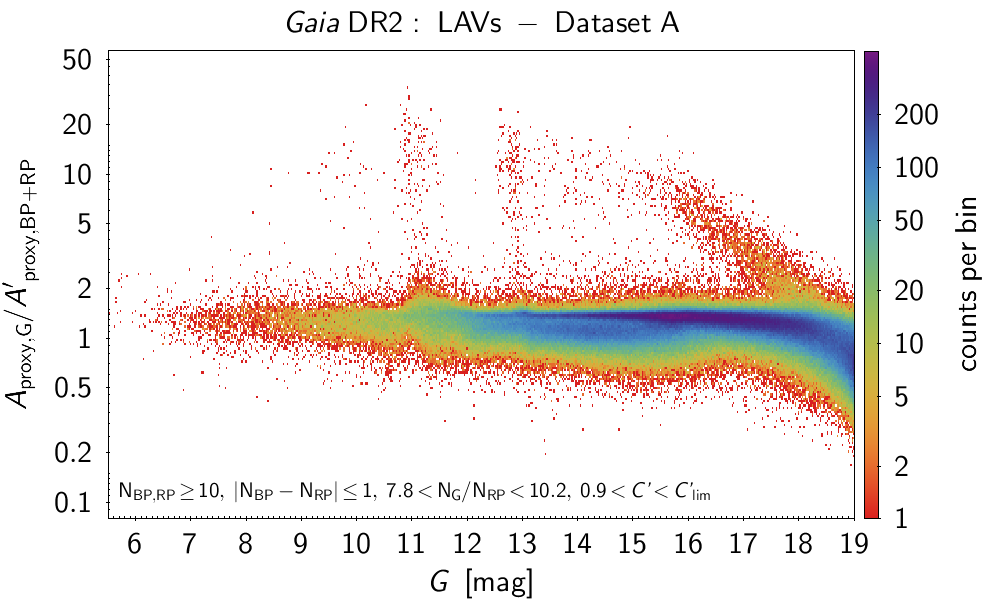}
	\caption{Density maps of the ratio between the variability proxy in \epochFluxG (Eq.~\ref{Eq:AproxyG}) and the variability proxy in $\epochFluxBP+\epochFluxRP$ neglecting the covariance between \epochFluxBP and \epochFluxRP (Eq.~\ref{Eq:varProxyBPplusRPwithoutCov}), versus mean \gmag magnitude.
	         The top panel shows the map for all sources in Dataset~A while the bottom panel shows the map for the sub-sample of Dataset~A that has reliable \fluxexcess according to Eq.~\ref{Eq:conditionGoodExcessFactor}.
	         The ordinate ranges are identical in both panels and have been limited for better visibility.
	}
	\label{Fig:varProxyGoverVarProxyBPplusRP_datasetA}
\end{figure}

\begin{figure}
	\centering
	\includegraphics[trim={0 80pt 0 0},clip,width=\linewidth]{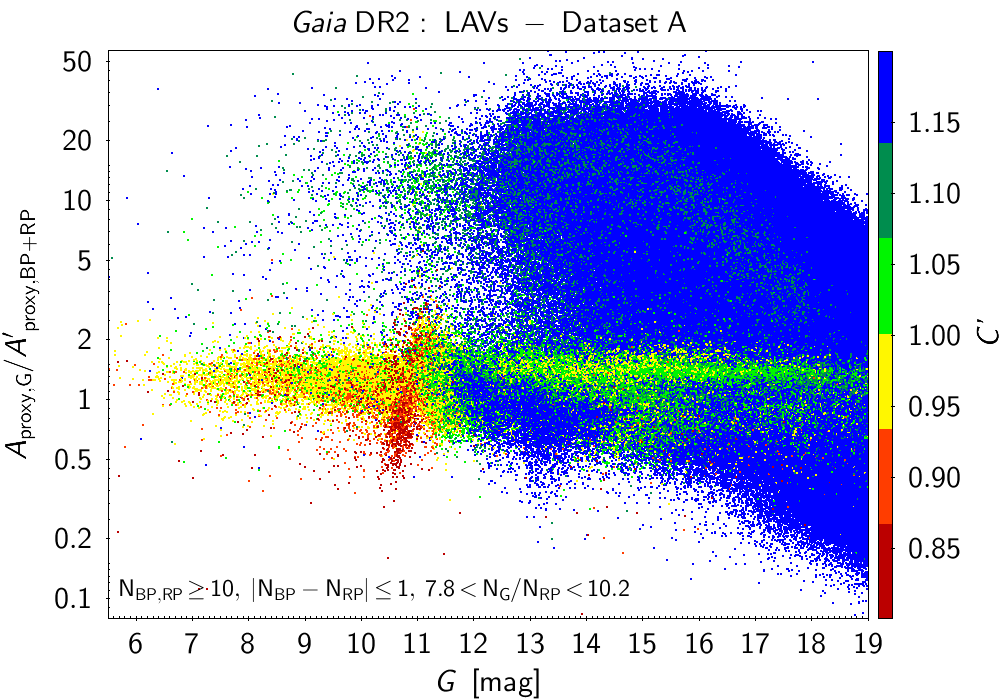}
	\vskip -0.5mm
	\includegraphics[trim={0 0 0 50pt},clip,width=\linewidth]{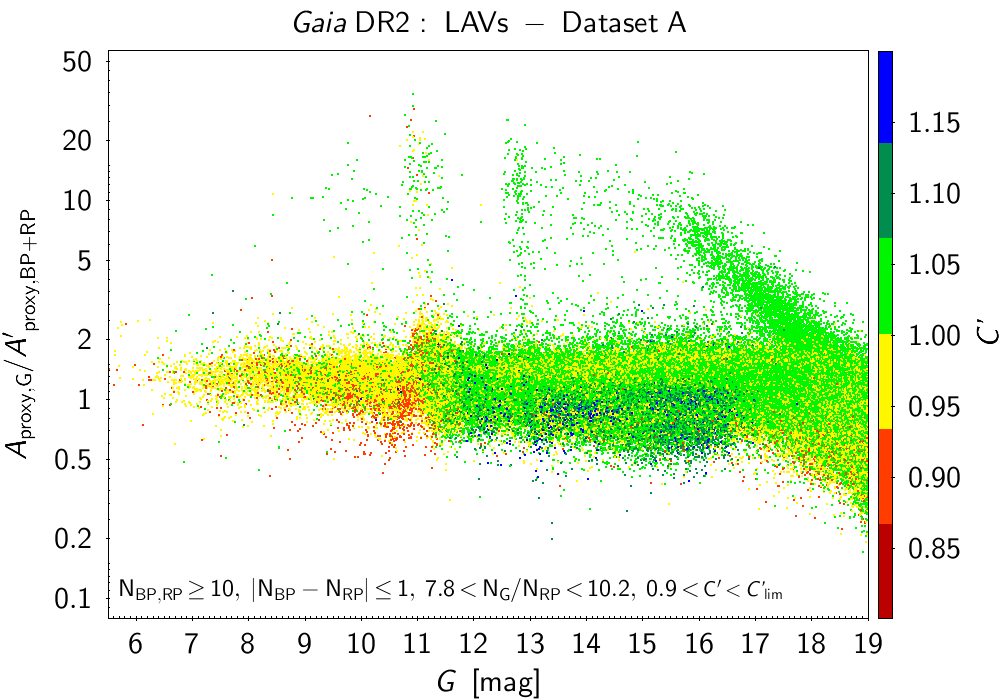}
	\caption{Same as Fig.~\ref{Fig:varProxyGoverVarProxyBPplusRP_datasetA}, but colour-coded with the value of the normalized \fluxexcess according to the colour scale shown on the right of the figure.
	         The scales are identical to those in Fig.~\ref{Fig:varProxyGoverVarProxyBPplusRP_datasetA}.
	}
	\label{Fig:varProxyGoverVarProxyBPplusRP_datasetA_colourCoded}
\end{figure}

Figure~\ref{Fig:histo_covProxyBPRP} shows a distribution of \varProxyCovBPplusRP centred on a positive value between 0.002 and 0.004.
These values are of the same order of magnitude as the values of $\varProxyGsquare > 0.0036$ considered in this study.
The covariance term is thus not negligible relative to the variances.
We indeed expect, for the great majority of variable stars, a concomitant increase (or decrease) in the red and blue filters.

A correct computation of the covariance between \epochFluxBP and \epochFluxRP time series should be done directly from the light curves using Eq.~\ref{Eq:covarianceBPRP}, rather than using Eq.~\ref{Eq:varProxyBPplusRPwithCov} with the approximation \ref{Eq:varProxyBPplusRP_varProxyG}.
The flux time series, however, are not available for the majority of sources in \Gaia DR2.
We therefore consider, as an approximation, the variability proxy \varProxyBPplusRPnullCov that neglects the covariance term.
It writes
\begin{equation}
  \varProxyBPplusRPnullCov = \frac{\sqrt{\fluxBP^{\,2}\;\varProxyBPsquare + \fluxRP^{\,2}\;\varProxyRPsquare}}
                                  {(\fluxBP+\fluxRP)} \; .
\label{Eq:varProxyBPplusRPwithoutCov}
\end{equation}
We have, in general,
\begin{equation}
  \varProxyBPplusRPnullCov < \varProxyBPplusRP
\end{equation}
since the covariance is, on the mean, positive.

\subsection{Analysis}
\label{Appendix:AproxyBPplusRP_analysis}

Since $\epochFluxG \simeq \epochFluxBP + \epochFluxRP$, we expect to have $\varProxyG \simeq \varProxyBPplusRP$.
The ratio $\varProxyG / \varProxyBPplusRPnullCov$ is shown versus \gmag magnitude in the top panel of Fig.~\ref{Fig:varProxyGoverVarProxyBPplusRP_datasetA} for all sources in Database~A.
For magnitudes brighter than 18~mag, the ratio peaks at $\varProxyG / \varProxyBPplusRPnullCov \simeq 1.35$.
The fact that it is larger than one reflects the omission of the (\epochFluxBP,\epochFluxRP) covariance term in the computation of \varProxyBPplusRPnullCov (Eq.~\ref{Eq:varProxyBPplusRPwithoutCov}).
At the fainter side of the catalogue ($\gmag>18$~mag), the ratio decreases with increasing magnitude for the bulk of the data, reaching values below one.
The likely cause of this effect is the residual astrophysical background, which becomes significative at faint magnitudes.
The fact that only a low-resolution background calibration was used in DR2 affects the amplitude proxies.
Source blending in the \bandBP and \bandRP spectrophotometers would also impact these proxies.
As the orientation of the spectra on the CCD varies with each transit, the blending is differently affected at different times in the photometric time series.

The top panel of Fig.~\ref{Fig:varProxyGoverVarProxyBPplusRP_datasetA} reveals the presence of a large number of sources having $\varProxyG / \varProxyBPplusRPnullCov$ ratios (much) larger than two, that is with variability amplitudes much larger in \gmag than in the combined \bandBPplusRP band.
These sources also have large (normalized) \fluxexcesses, as shown in the top panel of Fig.~\ref{Fig:varProxyGoverVarProxyBPplusRP_datasetA_colourCoded}, pointing to non-reliable \gbp and/or \grp time series.

If we limit the sample to sources with reliable \fluxexcesses using condition \ref{Eq:conditionGoodExcessFactor} of Appendix~\ref{Appendix:BPRPexcess}, the $\varProxyG / \varProxyBPplusRPnullCov$ versus \gmag diagram becomes much cleaner.
This is shown in the bottom panels of Figs.~\ref{Fig:varProxyGoverVarProxyBPplusRP_datasetA} and \ref{Fig:varProxyGoverVarProxyBPplusRP_datasetA_colourCoded}.
The filter on \excessFactorNormalized also cleans the faint side of the diagram, where the remaining departure from $\varProxyG / \varProxyBPplusRPnullCov \simeq 1$ at magnitudes fainter than 18~mag results from larger noise in BP and RP.

\begin{figure}
	\centering
	\includegraphics[width=\linewidth]{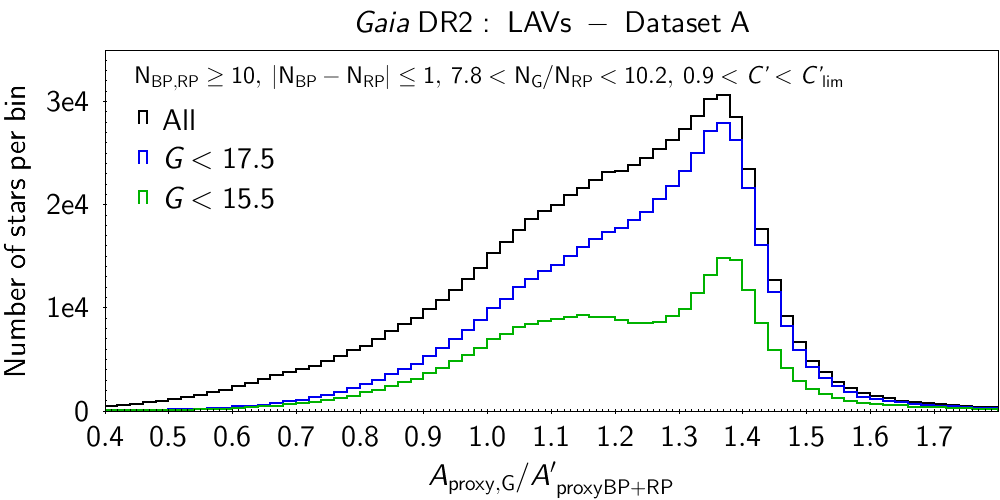}
	\caption{Histogram (in black) of $\varProxyG / \varProxyBPplusRPnullCov$ for the sample of Dataset~A with $N_\mathrm{BP,RP}\ge10$, $|\NobsBP-\NobsRP| \le 1$, $7.8 < N_{obs}(G)\,/\,N_{obs}(RP) < 10.2$, and $0.9<\excessFactorNormalized<\excessFactorNormalizedLimit$.
	         The blue and green histograms are for the subsets therein with $\gmag<17.5$~mag and $G<15.5$~mag, respectively.
	         The abscissa range has been limited for better visibility.
	}
	\label{Fig:histo_varProxyGoverBPRP}
\end{figure}

The patterns observed at $\varProxyG / \varProxyBPplusRPnullCov \gtrsim 1.5$ in the bottom panel of Fig.~\ref{Fig:varProxyGoverVarProxyBPplusRP_datasetA} are spurious.
From the histograms shown in Fig.~\ref{Fig:histo_varProxyGoverBPRP}, we take the limit of 1.5 above which we consider \varProxyBPplusRPnullCov to be unreliable.

At the small side of $\varProxyG / \varProxyBPplusRPnullCov$, we note that a small ratio can indicate larger-than-expected variability in \gbp and \grp, but that the variability in \gmag would still be good.
However, if variability in \gbp and \grp is to be studied, then a lower limit should also be applied, which we set at 0.8
This mainly impacts faint sources (see Fig.~\ref{Fig:varProxyGoverVarProxyBPplusRP_datasetA}, bottom panel).

Our final condition to select sources with reliable \varProxyBP and \varProxyRP amplitude proxies is summarize as:
\begin{equation}
  0.8 < \varProxyG / \varProxyBPplusRPnullCov < 1.5 \;\; .
\label{Eq:conditionVarProxyGBPRPRatio}
\end{equation}
It is used in the construction of Dataset~C.

\section{The electronic table}
\label{Appendix:electronicTable}

\begin{table*}
\caption{Attributes published in our electronically available catalogue of \Gaia DR2 LAVs (Dataset~A).
         The second column indicates the notation, if any, used in this paper for the attribute given in the first column.
         The third column provides a description of the attribute.
         An asterisc next to the attribute indicates that the data is a direct import from the \Gaia DR2 archive.
        }
\centering
\begin{tabular}{l l l}
\hline\hline
Attribute & Notation & Description \\
\hline
\texttt{source\_id$^*$}          & & \Gaia DR2 source ID\\
\texttt{dataset\_B}              & & (boolean) Is in Dataset~B\\
\texttt{dataset\_C}              & & (boolean) Is in Dataset~C\\
\texttt{l\_deg$^*$}              & & Galactic longitude (degree) \\
\texttt{b\_deg$^*$}              & & Galactic latitude (degree) \\
\texttt{parallax\_mas$^*$}       & & Parallax (milli arsec) \\
\texttt{parallax\_error\_mas$^*$}& & Parallax uncertainty (milli arsec) \\
\texttt{phot\_g\_n\_obs$^*$}     & \NobsG     & Number of points in \gmag \\
\texttt{phot\_g\_meanflux$^*$}   & \fluxG & Mean flux in the \gmag band (electron/sec) \\
\texttt{\small{phot\_g\_meanflux\_error$^*$}}
                                 & \fluxErrorG & Mean flux error in the \gmag band (electron/sec) \\
\texttt{phot\_g\_mean\_mag$^*$}  & \gmag & Mean \gmag magnitude (mag) \\
\texttt{phot\_bp\_n\_obs$^*$}    & \NobsBP & Number of points in \gbp \\
\texttt{phot\_bp\_meanflux$^*$}  & \fluxBP & Mean flux in \bandBP (electron/sec) \\
\texttt{\small{phot\_bp\_meanflux\_error$^*$}}
                                 & \fluxErrorBP & Mean flux error in \bandBP (electron/sec) \\
\texttt{phot\_bp\_mean\_mag$^*$} & \gbp & Mean \gbp magnitude (mag) \\
\texttt{phot\_rp\_n\_obs$^*$}    & \NobsRP & Number of points in \grp \\
\texttt{phot\_rp\_meanflux$^*$}  & \fluxRP & Mean flux in \bandRP (electron/sec) \\
\texttt{\small{phot\_rp\_meanflux\_error$^*$}}
                                 & \fluxErrorRP & Mean flux error in \bandRP (electron/sec) \\
\texttt{phot\_rp\_mean\_mag$^*$} & \grp & Mean \grp magnitude (mag) \\
\texttt{\small{phot\_bp\_rp\_excess\_factor$^*$}\!\!\!\!\!\!}
                                 & \excessFactor & \fluxexcess (Eq.~\ref{Eq:excessFactor}) \\
\texttt{\small{phot\_bp\_rp\_excess\_factor\_normalized}\!\!\!\!\!\!}
                                 & \excessFactorNormalized & Normalized \fluxexcess (Eq.~\ref{Eq:normalizedExcessFactor}) \\
\texttt{\small{isGoodBpRpExcessFactorNormalized}\!\!\!\!\!\!}
                                 &  & (boolean) Is \fluxexcess reliable ?\\
\texttt{DR2\_LPV}                & & (boolean) Is LPV candidate in DR2 \\
\texttt{DR2\_RRL\_SOS}           & & (boolean) Is RR\_Lyrae candidate in DR2 (sos table) \\
\texttt{DR2\_RRL\_Classif}       & & (boolean) Is RR\_Lyrae candidate in DR2 (classif table) \\
\texttt{DR2\_Cep\_SOS}           & & (boolean) Is Cepheid candidate in DR2 (sos table) \\
\texttt{DR2\_Cep\_Classif}           & & (boolean) Is Cepheid candidate in DR2 (classif table) \\
\texttt{DR2\_dSct\_SXPhe}        & & (boolean) Is $\delta$~Scuti \& SXPHE candidate in DR2 \\
\texttt{DR2\_RotMod}             & & (boolean) Is rotation modulation candidate in DR2 \\
\texttt{DR2\_STS}                & & (boolean) Is short time-scale candidate in DR2 \\
\texttt{amplProxyG}              & \varProxyG & Amplitude proxy in \gmag (Eq.~\ref{Eq:AproxyG}) \\
\texttt{amplProxyBP}             & \varProxyBP & Amplitude proxy in \gbp (Eq.~\ref{Eq:AproxyBP}) \\
\texttt{amplProxyRP}             & \varProxyRP & Amplitude proxy in \grp (Eq.~\ref{Eq:AproxyRP}) \\
\texttt{amplProxyBPplusRPwithoutCov}& \varProxyBPplusRPnullCov & Eq.~\ref{Eq:varProxyBPplusRPwithoutCov} \\
\texttt{group\_1\_in\_C}         & & (boolean) Is in Group~1 (for Dataset~C only, \texttt{NULL} otherwize)\\
\texttt{group\_2\_in\_C}         & & (boolean) Is in Group~2 (for Dataset~C only, \texttt{NULL} otherwize) \\
\texttt{group\_3\_in\_C}         & & (boolean) Is in Group~3 (for Dataset~C only, \texttt{NULL} otherwize) \\
\texttt{group\_4\_in\_C}         & & (boolean) Is in Group~4 (for Dataset~C only, \texttt{NULL} otherwize) \\
\texttt{group\_4a\_in\_C}        & & (boolean) Is in Subgroup~4a (for Dataset~C only, \texttt{NULL} otherwize) \\
\hline
\end{tabular}
\label{Tab:catalogueColumns}
\end{table*}

The electronic version of the catalogue will be made available through Vizier.
It is currently available at \url{http://obswww.unige.ch/~mowlavi}.
The list of attributes published in the table is given in Table~\ref{Tab:catalogueColumns}, with reference to equations given in either the main body of the paper or in one of the Appendixes.

\end{appendix}

\end{document}